\documentclass[a4paper, 12pt]{article}

\usepackage{tikz}
\usepackage{latexsym,amsmath,amsfonts,amssymb}
\usepackage{amsthm}
\usepackage{mathrsfs}
\usepackage[latin1]{inputenc}
\usepackage[american]{babel}
\usepackage{bbm}
\usepackage[nosort]{cite}
\usepackage[pdfencoding=auto]{hyperref}
\hypersetup{colorlinks, citecolor=[rgb]{.7,0,0}, linkcolor=[rgb]{0,0,0.7}, urlcolor=[rgb]{0,0,0.5}}

\renewcommand{\baselinestretch}{1.2}
\newcommand{\dvol}{d\mathrm{vol}}

\newcommand{\parfrac}[2]{\frac{\partial #1}{\partial #2}}

\newcommand{\ud}[2]{^{#1}_{\phantom{#1}#2}}
\newcommand{\du}[2]{_{#1}^{\phantom{#1}#2}}

\newcommand{\wt}{\widetilde}

\newcommand{\matht}[1]{\texorpdfstring{\ensuremath{\boldsymbol{#1}}}{#1}}
\newcommand{\ts}{\textstyle}

\newcommand{\eg}{\textit{e.g.}}

\newcommand{\ie}{\textit{i.e.}}

\newcommand{\first}{1\textsuperscript{st}}
\newcommand{\second}{2\textsuperscript{nd}}

\numberwithin{equation}{section}

\newcommand{\nn}{\nonumber}
\newcommand{\mat}[1]{\begin{pmatrix} #1 \end{pmatrix}}

\newcommand{\be}{\begin{equation}} \newcommand{\ee}{\end{equation}}
\newcommand{\bea}{\begin{equation} \begin{aligned}} \newcommand{\eea}{\end{aligned} \end{equation}}
\newcommand{\ba}{\begin{array}} \newcommand{\ea}{\end{array}}

\newcommand{\cA}{\mathcal{A}}
\newcommand{\cB}{\mathcal{B}}
\newcommand{\cC}{\mathcal{C}}

\newcommand{\cE}{\mathcal{E}}
\newcommand{\cF}{\mathcal{F}}
\newcommand{\cG}{\mathcal{G}}

\newcommand{\cI}{\mathcal{I}}
\newcommand{\cJ}{\mathcal{J}}

\newcommand{\cL}{\mathcal{L}}
\newcommand{\ccL}{\mathscr{L}}

\newcommand{\cN}{\mathcal{N}}
\newcommand{\cO}{\mathcal{O}}
\newcommand{\cP}{\mathcal{P}}
\newcommand{\cQ}{\mathcal{Q}}

\newcommand{\cU}{\mathcal{U}}

\newcommand{\cW}{\mathcal{W}}

\newcommand{\cZ}{\mathcal{Z}}

\newcommand{\bC}{\mathbb{C}}

\newcommand{\bP}{\mathbb{P}}

\newcommand{\bR}{\mathbb{R}}

\newcommand{\bZ}{\mathbb{Z}}

\newcommand{\fq}{\mathfrak{q}}

\newcommand{\mrE}{\mathrm{E}}
\newcommand{\mrg}{\mathrm{g}}
\newcommand{\unit}{\mathbbm{1}}
\newcommand{\mg}{\mathrm{g}}

\DeclareMathOperator{\Tr}{Tr}

\DeclareMathOperator{\re}{\mathbb{R}e}
\DeclareMathOperator{\im}{\mathbb{I}m}

\makeatletter
\def\blfootnote{\gdef\@thefnmark{}\@footnotetext}
\makeatother

\begin{document}

\thispagestyle{empty}
\begin{flushright}
SISSA  01/2021/FISI
\end{flushright}
\vspace{13mm}  
\begin{center}
{\huge  A gravity interpretation for the Bethe Ansatz \\[.5em] expansion of the $\cN=4$ SYM index}
\\[13mm]
{\large Ofer Aharony$^1$, Francesco Benini$^{2,3,4}$, Ohad Mamroud$^1$, Elisa Milan$^5$}
 
\bigskip
{\it
$^1$ Department of Particle Physics and Astrophysics, \\[.0em]
Weizmann Institute of Science, Rehovot 7610001, Israel \\[.2em]
$^2$ SISSA, Via Bonomea 265, 34136 Trieste, Italy \\[.2em]
$^3$ INFN, Sezione di Trieste, Via Valerio 2, 34127 Trieste, Italy \\[.2em]
$^4$ ICTP, Strada Costiera 11, 34151 Trieste, Italy \\[.2em]
$^5$ Department of Physics, Technion, Haifa, 32000, Israel \\[.2em]
}

\bigskip
\bigskip

{\parbox{16cm}{\hspace{5mm}
The superconformal index of the $\cN=4$ $SU(N)$ supersymmetric Yang-Mills theory counts the $1/16$-BPS states in this theory, and has been used via the AdS/CFT correspondence to count black hole microstates of $1/16$-BPS black holes. On one hand, this index may be related to the Euclidean partition function of the theory on $S^3\times S^1$ with complex chemical potentials, which maps by the AdS/CFT correspondence to a sum over Euclidean gravity solutions. On the other hand, the index may be expressed as a sum over solutions to Bethe Ansatz Equations (BAEs). We show that the known solutions to the BAEs that have a good large $N$ limit, for the case of equal chemical potentials for the two angular momenta, have a one-to-one mapping to (complex) Euclidean black hole solutions on the gravity side. This mapping captures both the leading contribution from the classical gravity action (of order $N^2$), as well as non-perturbative corrections in $1/N$, which on the gravity side are related to wrapped D3-branes. Some of the BA solutions map to orbifolds of the standard Euclidean black hole solutions (that obey exactly the same boundary conditions as the other solutions). A priori there are many more gravitational solutions than Bethe Ansatz solutions, but we show that by considering the non-perturbative effects, the extra solutions are ruled out, leading to a precise match between the solutions on both sides.
}}
\end{center}

\newpage
\pagenumbering{arabic}
\setcounter{page}{1}
\setcounter{footnote}{0}
\renewcommand{\thefootnote}{\arabic{footnote}}

{\renewcommand{\baselinestretch}{.88} \parskip=0pt
\setcounter{tocdepth}{2}
\tableofcontents}


\section{Introduction and Summary}
\label{sec: intro}

The AdS/CFT correspondence \cite{Maldacena:1997re, Gubser:1998bc, Witten:1998qj, Witten:1998zw} maps black holes in asymptotically anti-de-Sitter (AdS) spacetimes to coarse-grained descriptions of states in conformal field theories (CFTs), such that the Bekenstein-Hawking entropy of the black holes may be given a statistical mechanics interpretation as a counting of CFT microstates. For general non-supersymmetric black holes we do not have good methods to count these states in the strongly coupled theories which map under the AdS/CFT correspondence to weakly curved backgrounds (where the Bekenstein-Hawking computation is valid). In the last few years, considerable progress has been made \cite{Cabo-Bizet:2018ehj, Choi:2018hmj, Benini:2018ywd, Honda:2019cio, ArabiArdehali:2019tdm, Kim:2019yrz, Cabo-Bizet:2019osg, Amariti:2019mgp, Lezcano:2019pae, Lanir:2019abx, Cabo-Bizet:2019eaf, ArabiArdehali:2019orz, Cabo-Bizet:2020nkr, Murthy:2020rbd, Agarwal:2020zwm, Benini:2020gjh, GonzalezLezcano:2020yeb, Copetti:2020dil, Goldstein:2020yvj, Cabo-Bizet:2020ewf, Amariti:2020jyx, Choi:2021lbk, Amariti:2021ubd, Jejjala:2021hlt} in performing this counting for $1/16$-BPS black holes in type IIB string theory on AdS$_5\times S^5$, which map to $1/16$-BPS states in the ${\cal N}=4$ $SU(N)$ supersymmetric Yang-Mills (SYM) theory on $S^3$ (this followed related results in other backgrounds, starting with \cite{Benini:2015eyy} and including \cite{Hosseini:2016tor, Benini:2016hjo, Benini:2016rke, Hosseini:2016cyf, Cabo-Bizet:2017jsl, Azzurli:2017kxo, Hosseini:2017fjo, Benini:2017oxt, Hosseini:2018uzp, Crichigno:2018adf, Suh:2018tul, Hosseini:2018usu, Fluder:2019szh, Gang:2019uay, Kantor:2019lfo, Choi:2019zpz, Bobev:2019zmz, Nian:2019pxj, Benini:2019dyp}).

The counting of $1/16$-BPS states is based on computing the superconformal index \cite{Romelsberger:2005eg, Kinney:2005ej}, which is a sum over these states with chemical potentials for four of the five charges carried by the black holes (the black holes carry three global symmetry charges $R_{1,2,3}$ in the Cartan algebra of the $SU(4)_R$ symmetry, and two angular momenta $J_{1,2}$ in AdS$_5$). It is not possible to directly relate the index to the number of BPS states with given charges, both because the index is insensitive to one of the charges, and because it counts states with a factor of $(-1)^F$, so there can be cancellations between bosonic and fermionic states. Indeed, when the index was first computed in the large $N$ limit with real chemical potentials \cite{Kinney:2005ej}, a mismatch was found between the value of the index and the expectation from black holes. However, this discrepancy does not arise for generic complex chemical potentials \cite{Choi:2018hmj, Benini:2018ywd},%
\footnote{It was already shown in AdS$_4$ \cite{Benini:2016rke} how the index with complex chemical potentials captures the Bekenstein-Hawking entropy of BPS black holes.}
and in various limits the index was successfully matched (at leading order in the large $N$ limit) with the expectation from the black hole solutions.

More precisely, the superconformal index is a grand-canonical partition function and one would expect its Legendre transform \cite{Hosseini:2017mds} to agree --- at leading order --- with the black hole entropy. This is indeed the case, if one performs the Legendre transform with respect to the four available chemical potentials. This was shown both directly in the large $N$ limit \cite{Benini:2018ywd}, as well as going through a Cardy-like limit that retains states with charges much larger than the central charge \cite{Choi:2018hmj}.

In this paper we do not discuss the black hole microstate counting, but instead we aim to understand better the interpretation (on the gravity side) of the superconformal index as a Euclidean partition function, hoping that this better understanding will also be useful in the future for the microstate counting problem. In general, as we review in Section~\ref{part_to_index}, the superconformal index is related to the Euclidean partition function of the theory on $S^3\times S^1$, with specific background fields related to the chemical potentials appearing in the index (which can be real or complex), which ensure that supersymmetry is preserved \cite{Festuccia:2011ws, Dumitrescu:2012ha, Closset:2013vra}. The AdS/CFT correspondence tells us that this partition function should be described in the large $N$ limit on the gravity side in terms of a sum over all asymptotically AdS$_5\times S^5$ solutions of type IIB string theory that satisfy appropriate boundary conditions (related to the background fields). Up to now, only specific solutions, which give the leading contribution to the partition function for certain chemical potentials, were considered; however, the full answer on the gravity side includes all of these solutions, and also the fluctuations around these solutions, including both perturbative contributions in $1/N$ (from supergravity fields) and non-perturbative contributions (from wrapped D-branes). And moreover, different solutions dominate for different values of the chemical potentials.

We would like to ask whether the sum over Euclidean gravity solutions and the fluctuations around them can be reproduced by a direct computation of the index. There are several different methods that have been used to compute the index in the large $N$ limit.
One is the so-called Cardy limit \cite{Choi:2018hmj, Choi:2018vbz, Cassani:2021fyv, ArabiArdehali:2021nsx}, a sort of ``high-temperature'' limit taken on the chemical potentials in which the integral expression of the index considerably simplifies, and that can then be easily followed by the large $N$ limit. Another one is the Bethe Ansatz method \cite{Benini:2018ywd}, described below. A third method is the saddle-point approximation applied to a non-analytic extension of the index integral formula \cite{Cabo-Bizet:2019eaf, Cabo-Bizet:2020nkr, Cabo-Bizet:2020ewf}. A fourth method is an expansion of the index integral formula around a complexified Gross-Witten-Wadia model \cite{Copetti:2020dil, Choi:2021lbk}.

In this paper we focus on the Bethe Ansatz method \cite{Benini:2018ywd}, which is valid when the chemical potentials for the angular momenta obey specific relations; we discuss in this paper only the simplest case when this method applies, which is for equal chemical potentials for the two angular momenta. In this method the index is written as a sum over solutions to Bethe Ansatz Equations (BAEs) (at least when these solutions are discrete). Some of the known solutions to these equations give contributions which in the large $N$ limit resemble our expectations from the gravity side; there is a leading term (in the logarithm of the contribution to the partition function) of order $N^2$, then power-law corrections in $1/N$, and then non-perturbative corrections of order $e^{-N}$. This raises a natural question --- can we identify the sum over Bethe Ansatz solutions in the CFT computation of the index, with the sum over gravitational solutions? A priori there is no reason for such a matching of individual terms in the sum (as opposed to the full partition function); in particular, it has not been shown on the gravity side that the partition function localizes to a discrete sum over solutions, analogous to the Bethe Ansatz result. However, we will show that there is in fact a precise matching between the two sides. 

In one direction, we show that every known solution to the Bethe Ansatz equations, which has a good large $N$ limit, may be identified with a specific Euclidean solution on the gravity side. These solutions are either Euclidean black holes (belonging to one of the two branches of known supersymmetric black hole solutions), or $\bZ_m$ orbifolds of Euclidean black holes which satisfy the same boundary conditions. The matching involves both the leading contribution of order $N^2$ to the logarithm of the partition function, as well as the form of the non-perturbative corrections, which on the gravity side arise from wrapped D3-branes.

In the other direction, we note that a priori there are many more gravitational solutions with the correct boundary conditions than Bethe Ansatz solutions, and that if we naively sum over all solutions on the gravity side we obtain divergences, since the real part of their actions is not bounded from below. However, our supersymmetric Euclidean solutions are all complex-valued, and it is not obvious what the rules for including such solutions in the gravitational theory are. We use (as previous works do) a specific prescription for evaluating the action of the complex solutions (at leading order in $1/N$) by analytic continuation. However, it is not clear when this analytic continuation is valid, since various poles may be encountered when one tries to explicitly continue the relevant contours in the gravitational path integral (to the extent that such a path integral description makes sense to begin with). And moreover, when we analytically continue only the leading contribution to some solution, we risk having subleading terms become dominant over the leading ones (for instance, a non-perturbative correction of the form $e^{-N a^2}$ is negligible for real $a$, but becomes dominant when $\re (a^2) < 0$). We suggest a specific prescription for which solutions should be included in the path integral, namely that solutions for which the non-perturbative contributions from wrapped D3-branes are exponentially large (rather than exponentially suppressed) should not be included. With this (reasonable but not rigorously derived) prescription, we find a precise one-to-one match between the solutions on both sides.

The wrapped D3-brane solutions mentioned above, which give non-perturbative corrections on the gravity side matching the Bethe Ansatz solutions, wrap an $S^3$ inside the $S^5$ and an $S^1$ in the AdS$_5$ coordinates. In addition to these, there are also other Euclidean wrapped D3-branes that preserve supersymmetry, which wrap an $S^1$ inside the $S^5$ and an $S^3$ in the AdS$_5$ coordinates. These additional D3-branes are not related to non-perturbative terms in the Bethe Ansatz contribution of the Hong-Liu solutions, but they still give corrections to the gravitational partition function. The interpretation of these corrections in the Bethe Ansatz expansion will be discussed elsewhere \cite{ABMMinprogress}. In this paper, whenever we mention wrapped D3-branes without indicating their type, we will always be referring to the first type of D3-brane solutions.

Our analysis in this paper focuses on the Eucllidean path integral, but along the way, we have also found new supersymmetric Lorentzian probe D3-brane configurations in the BPS black hole backgrounds (and complexification thereof), that generalize the giant and dual giant graviton configurations in AdS$_5$ \cite{McGreevy:2000cw, Grisaru:2000zn, Hashimoto:2000zp}. It would be interesting to analyze how these configurations affect the counting of black hole microstates, and in particular if they may be related to any multicenter black hole solutions in AdS, but this lies beyond the scope of this work.

After reviewing the superconformal index, the partition function, and their relation in Section~\ref{part_to_index}, we review in Section~\ref{sec: Field theory analysis} the Bethe Ansatz method and the known solutions (found in \cite{Hosseini:2016cyf, Hong:2018viz}), and we compute their perturbative and non-perturbative contributions to the index in the large $N$ limit. In Section~\ref{sec: black holes 5d} we review the known black hole solutions using the consistent truncation to a 5d supergravity theory, and in particular the Euclidean supersymmetric black hole solutions. We show that the action of some of these solutions agrees (at leading order in $1/N$) with the Bethe Ansatz results, but that more solutions exist on the gravity side. In Section~\ref{sec: Branes} we lift the Euclidean black hole solutions to 10d, and compute the non-perturbative contributions to the partition function from wrapped Euclidean D3-branes in these backgrounds. In Section~\ref{comparison} we describe the prescription mentioned above for keeping only the solutions whose non-perturbative corrections from wrapped D3-branes are small, and we show that these solutions match precisely to a subclass of the Bethe Ansatz solutions, including both the leading term and the form of the non-perturbative corrections. Finally, in Section~\ref{sec: Orbifolds} we show that some specific orbifolds of Euclidean black hole solutions also contribute to the same Euclidean partition function (since they have the same boundary conditions), and that the orbifold solutions which satisfy our criterion precisely match with the remaining Bethe Ansatz solutions. Appendix~\ref{app: functions} contains reference material, Appendices~\ref{app: computation}--\ref{app: Three charges} contain details on field theory and gravity computations, while in Appendix~\ref{app: giant gravitons} we present the new Lorentzian D3-brane giant graviton configurations.

\subsection{Future directions}

There are many open questions left by our analysis. Some of them are:

\begin{itemize}
\item The rules of the AdS/CFT correspondence in the presence of complex sources (such as the background metric or gauge fields) are not clear. A commonly used method is to take results for the classical action on the gravity side computed for real sources, and to analytically continue them to complex sources, where the result may also be interpreted as related to a complex gravity solution (with complex boundary conditions related to the complex sources). Analytic continuations of this type may be dangerous, both because poles or other features may be encountered when shifting the contours where the gravitational fields are valued, and because subleading contributions for real sources may become dominant for some complex-valued sources. In this paper we discuss in detail an example of the latter problem, and suggest a prescription in which only (complex) gravitational solutions which do not have any instabilities related to non-perturbative contributions becoming large should be included. It would be useful to find a rigorous justification for this prescription, by carefully analyzing the analytic continuations that are involved. 

\item Our results suggest that, as the Bethe Ansatz prescription does on the CFT side, the gravitational computation of the supersymmetric partition function on $S^3\times S^1$ may perhaps be localized to a discrete sum over supersymmetric gravity solutions. It would be interesting to perform such a localization on the gravity side, at least in the supergravity approximation and perhaps along the lines of \cite{Dabholkar:2010uh, Dabholkar:2011ec, Dabholkar:2014wpa}, and to see if it reproduces the results we found in this paper.

\item In this paper we only compared the leading exponents in the contributions of various solutions to the path integral --- both for the classical gravity solutions (contributing to the logarithm of the partition function at order $N^2$), and for the non-perturbative corrections coming from D3-branes. From the Bethe Ansatz we can also find the subleading contributions to the logarithm of the partition function, of order $\log(N)$ (already considered in \cite{GonzalezLezcano:2020yeb}) and of order $1$, and we seem to find no further perturbative corrections in $1/N$. It would be interesting to reproduce these corrections from one-loop fluctuations around our gravitational solutions, and to confirm that higher-loop corrections are absent on the gravity side. In the case of the D3-branes, we can again compute on the field theory side also the coefficients of the exponentials that we matched, and it would be interesting to match those with multiplicities of D3-brane solutions on the gravity side.

\item So far there is no classification of all solutions to the Bethe Ansatz equations for the $SU(N)$ $\cN=4$ SYM theories, and for finite $N$ there are certainly additional solutions beyond the ones we analyzed, including continuous families of solutions \cite{ArabiArdehali:2019orz, Lezcano:2021qbj, Benini:2021ano}. It would be interesting to classify all the solutions, and to understand whether some of the other solutions might have an interpretation as classical gravity solutions as well.

\end{itemize}

There are also many possible generalizations of our analysis:

\begin{itemize}

\item We analyzed only the case of equal chemical potentials for the two angular momenta, since in this case the Bethe Ansatz computation of the index is the simplest. The Bethe Ansatz computation works whenever the ratio between the two angular momenta is rational \cite{Benini:2018mlo, Benini:2020gjh}, but for general rational values, extra parameters appear in the sum over Bethe Ansatz solutions, and the number of these parameters is of order $N$, so that in the large $N$ limit there is an exponentially large number of terms appearing in the sum. On the other hand, on the gravity side nothing drastic seems to change when the chemical potentials are not equal (or even when their ratio is not rational), though some of the details of our analysis may be modified. This suggests that the extra parameters should be summed over before matching to the gravity side, in order to still have a matching between the two sides. It would be interesting to perform these sums and to match the Bethe Ansatz results to a sum over gravity solutions also for other ratios of chemical potentials. Note that it is not obvious if this should be possible or not; even if the gravity computation localizes as mentioned above, it is not guaranteed that the same sum over solutions will appear on the gravity side as on the field theory side (only the full partition function has to match). However, our results suggest that such a mapping may be possible also for general ratios of the angular momenta chemical potentials.

\item Many of our computations, both on the field theory side (using the Bethe Ansatz) and on the gravity side, may be generalized to general theories with 4d $\cN=1$ superconformal symmetry. Indeed, the leading contributions from both sides have been matched in some other cases \cite{Lanir:2019abx,Benini:2020gjh}. It would be interesting to check if also in these other cases, the full sum over Bethe Ansatz solutions may be matched in the large $N$ limit to a sum over gravitational solutions, as we found for the $\cN=4$ SYM theories. It may also be possible to generalize the analysis to supersymmetric $S^{d-1}\times S^1$ partition functions of $d$-dimensional SCFTs in other dimensions (perhaps starting from \cite{Cassani:2019mms, Kantor:2019lfo, Bobev:2020pjk}).

\item Non-perturbative corrections to the Euclidean partition function coming from wrapped branes, similar to the ones we found, presumably exist for all black hole solutions in any dimension. It would be interesting to evaluate these contributions and, where possible, to match them to the field theory side.%
\footnote{Non-perturbative corrections to the contribution to the superconformal index (or other supersymmetric partition functions) coming from AdS$_5$ have been studied, \eg, in \cite{Biswas:2006tj, Arai:2019xmp, Arai:2020qaj}.}
Of course, this requires lifting these black hole solutions --- which are often found in truncations of the gravitational theory to a lower-dimensional theory --- to 10d or 11d (depending on the case). It would also be interesting to understand if there are other cases where these contributions ``destabilize'' some complexified solutions of the gravity equations of motion, as we found.

\end{itemize}

Last but not least, it would be interesting to understand the implications of our analysis (which is purely in Euclidean signature) to the counting of $1/16$-BPS black hole microstates.

\section{The sphere partition function and the index}
\label{part_to_index}

\subsection{The sphere thermal partition function}
\label{sphere_part_func}

In this paper we would like to match the partition function of $\cN = 4$ $SU(N)$ super Yang-Mills (SYM) on $S^3 \times S^1$, which is closely related to the superconformal index, with the gravitational partition function of its holographic dual, type IIB string theory on a spacetime which is asymptotically AdS$_5\times S^5$. We will begin in this section with the field theory side, by describing the precise relation between the partition function and the superconformal index.

Using $\cN=1$ notation, the field content of the $\cN=4$ SYM theory consists of a vector multiplet and three chiral multiplets $X$, $Y$, $Z$, all in the adjoint representation of the gauge group, with a superpotential proportional to $W = \Tr \bigl( X[Y,Z] \bigr)$. The R-symmetry is $SU(4)_R$: going to the Cartan $U(1)^3$, we choose a basis of generators $R_{1,2,3}$ each giving R-charge 2 to a single chiral multiplet and zero to the other two, in a symmetric way. In addition, local operators (or, equivalently, states in the theory on $S^3$) are labeled by two angular momenta $J_{1,2}$, which are half-integer and each rotates an $\bR^2 \subset \bR^4$ in which $S^3$ is embedded. We define the fermion number as $F = 2J_1$. Note that all fields in the theory (and thus all states) have integer charges under $R_{1,2,3}$ and obey
\be
\label{modone}
F = 2J_{1,2} = R_{1,2,3} \pmod{2} \;.
\ee

We can add chemical potentials for each of these five charges, and the thermal partition function of the theory on $S^3$ can then be expressed as a trace over the Hilbert space of the theory as
\be
\label{part_func}
Z= \Tr{}\left( e^{-\beta H + \sum_i \beta \, \Omega_i J_i + \frac{1}{2} \sum_a \beta \, \Phi_a R_a }\right) \;,
\ee
where $\beta$ is the inverse temperature, $H$ is the Hamiltonian on $S^3$ in the absence of background fields (related to the operator dimension by the state/operator correspondence),  and $\Omega_i$ (with $i=1,2$) and $\Phi_a$ (with $a=1,2,3$) are chemical potentials. The $\tfrac{1}{2}$ for the R-charges comes from our convention that the $J_i$ are half-integers, while the $R_a$ are integers. With this convention, given \eqref{modone}, the partition function \eqref{part_func} is periodic under shifts
\be
\label{shift symmetry}
\{ \Omega_1, \Omega_2, \Phi_1, \Phi_2, \Phi_3 \} \to \Bigl\{\Omega_1+\frac{2\pi i n_1}{\beta}, \Omega_2+\frac{2\pi i n_2}{\beta}, \Phi_1+\frac{2\pi i m_1}{\beta}, \Phi_2+\frac{2\pi i m_2}{\beta}, \Phi_3+\frac{2\pi i m_3}{\beta} \Bigr\}
\ee
for any integers $n_1,n_2,m_1,m_2,m_3$ whose sum is even.

As usual, we can think of \eqref{part_func} as a Euclidean partition function%
\footnote{Actually, there could be a proportionality constant $Z_{S^1\times S^3} = e^{-C(\beta, \Omega_i, \Phi_a)} Z$, where $Z$ is as in \eqref{part_func}, coming from background dependent counterterms. For our purposes this distinction can be ignored, as explained below.}
on $S^1_\beta \times S^3$ where the Euclidean time direction $t_\mrE$ has periodicity $\beta$ (up to a zero-point energy that we will discuss below). The chemical potentials $\Omega_i$ and $\Phi_a$ can be thought of as background holonomies (on the thermal circle) of gauge fields coupled to the conserved currents (whose charges are the angular momenta and the R-charges). For example, the $\Phi_a$ are equivalent to background fields $A^a_{t_\mrE} = \frac{i}{2} \Phi_a$ for the gauge fields coupled to the currents of $R_a$. In principle, the chemical potentials in \eqref{part_func} can be real or complex, as long as the partition function \eqref{part_func} converges; the inverse temperature $\beta$ can also be complex, though then the geometrical Euclidean interpretation is less clear.

For the chemical potentials related to the angular momenta, the background fields in question are really off-diagonal components of the metric; for an angular momentum involving a shift in an angular coordinate $\phi$ of $S^3$, we can think of them as $g_{t_\mrE \phi}$ components of this metric, proportional to $(i \Omega_i)$. This can also be thought of as a different complex structure for the fibration of the Euclidean time circle over the sphere \cite{Closset:2013vra}. When $\Omega$ is purely imaginary, the modification in the metric is real, while otherwise the metric becomes complex. 

In this case we can also implement the chemical potentials in a different way. Since wavefunctions of states with angular momentum $J$ have angular dependence $e^{i J \phi}$ on an angular variable $\phi$, we can alternatively keep the original $S^3$ metric on the sphere, but use a different identification around the time circle,
\be
\label{shifted ident}
(t_\mrE, \phi) \sim (t_\mrE + \beta, \phi - i \beta \Omega) \;,
\ee
which leads to the very same factor of $e^{\beta \Omega J}$ in the trace \eqref{part_func}. We will call this the ``coordinate shift realization'' of the chemical potentials, while the previous implementation is the ``metric realization''.

When $\Omega$ is purely imaginary, using \eqref{shifted ident} is a reasonable alternative description, equivalent to modifying the metric, but otherwise we need to be careful in how to interpret the identification \eqref{shifted ident} (together with the standard $(2\pi)$ periodicity of $\phi$). A consistent way to implement this realization for general complex $\Omega$ is to define a new angular coordinate
\be
\label{tildephi}
{\tilde \phi} = \phi + i \Omega t_\mrE
\ee
such that the new coordinate $\tilde \phi$ still has a $2\pi$ periodicity, and it can be taken to be real (even when $\Omega$ is complex), while  the other identification is simply
$(t_\mrE, \tilde\phi) \sim  (t_\mrE + \beta, \tilde\phi)$.
This defines the coordinate range of $\phi$ and $t_\mrE$ for the ``coordinate shift realization'' of the chemical potentials.
Note that in general the original coordinate $\phi$ of the $S^3$ is complex now (in particular since $\beta$ may also be complex, so that $t_\mrE$ which runs from $0$ to $\beta$ is also complex), but $\tilde\phi$ is real. The metric in the new coordinate $\tilde\phi$ is generally complex. Indeed, the metric in this new coordinate is essentially the same as the metric in the first approach discussed above. In this framework we know the range of the coordinates even when $\Omega$ and $\beta$ are complex, and this will be useful when we discuss holography below.

\subsection{The superconformal index}

The superconformal index is defined for general 4d ${\cal N}=1$ superconformal field theories; in the ${\cal N}=4$ SYM theory it counts (with sign) 1/16 BPS states on $S^3$ preserving one complex supercharge $\cQ$, which we choose to be associated with a specific $U(1)_R$ symmetry with charge $r=(R_1+R_2+R_3)/3$. The superconformal index keeps track of some combinations of the R-charges and the two angular momenta $J_{1,2}$. It is useful to  introduce two flavor generators \mbox{$\fq_{1,2} = \frac12 (R_{1,2} - R_3)$} that commute with $\cQ$ and the $\cN=1$ superconformal subalgebra containing it. 
The superconformal index \cite{Romelsberger:2005eg, Kinney:2005ej} is then defined by the trace%
\footnote{Often the powers in the index are written as $p^{J_1+\frac12 r} \, q^{J_2+\frac12 r}$. Compared to this convention, we have swallowed a power of $pq$ into $y_1$ and $y_2$ in order to obtain a single-valued function. The relation of our variables to those of \cite{Kinney:2005ej} is $p = t^3x \big|_\text{there}$, $q = t^3/x \big|_\text{there}$, $y_1 = t^2v \big|_\text{there}$, $y_2 = t^2 w/v \big|_\text{there}$.}
\be
\label{index Hamiltonian definition}
\cI(p,q,y_1, y_2) = \Tr^\prime{} \left[ (-1)^F \, e^{-\beta\{\cQ, \bar\cQ \}} \, p^{J_1 + \frac12 R_3} \, q^{J_2 + \frac12 R_3} \, y_1^{\fq_1} \, y_2^{\fq_2} \right]
\ee
over the Hilbert space on $S^3$, where by $\Tr^\prime{}$ we mean the trace in the convention where the contribution of the vacuum is 1. Here $p$, $q$, $y_{1,2}$ are fugacities, and it is convenient to introduce chemical potentials $\sigma$, $\tau$, $\Delta_{1,2}$ such that
\be \label{fugacities}
p = e^{2\pi i \sigma} \;,\qquad\qquad q = e^{2\pi i \tau} \;,\qquad\qquad y_{1,2} = e^{2\pi i \Delta_{1,2}} \;.
\ee
The index is well-defined for $|p|, |q|<1$, namely for $\im(\sigma),\, \im(\tau) > 0$. Given \eqref{modone}, the index is a single-valued function of the fugacities \eqref{fugacities}, \ie, it is periodic under integer shifts of the chemical potentials $\sigma$, $\tau$, $\Delta_1$ and $\Delta_2$. By standard arguments 
\cite{Witten:1982df}, the index only counts states annihilated by $\cQ$ and $\bar\cQ$ and is thus independent of $\beta$.

For the purpose of relating the partition function \eqref{part_func} to the superconformal index, note that in order to
preserve the specific supercharge $\cQ$ mentioned above on $S^1_\beta\times S^3$, the chemical potentials in \eqref{part_func} have to satisfy \cite{Cabo-Bizet:2018ehj}
\be \label{susy_cond}
\beta \, \bigl( 1 + \Omega_1 + \Omega_2 - \Phi_1 - \Phi_2 - \Phi_3 \bigr) = 2\pi i n \;,\qquad n\in \bZ \;.
\ee
In addition, in order for \eqref{part_func} to have the interpretation of computing a thermal partition function, we must have anti-periodic boundary conditions for the fermions (and in particular for the supercharge) around the thermal cycle, and this requires $n$ to be odd \cite{Cabo-Bizet:2018ehj}. We will choose $n=1$, but this does not affect anything because of the periodicity \eqref{shift symmetry} of the chemical potentials.

Our specific conserved supercharge has the algebra \cite{Kinney:2005ej}
\be
\{\cQ, \bar\cQ\} = H - J_1 - J_2 - \frac{1}{2} \bigl( R_1 + R_2 + R_3 \bigr) \;.
\ee
Thus, we can write the partition function \eqref{part_func} as
\be
Z = \Tr{} \left(e^{-\beta\{\cQ, \bar\cQ\}  + \sum_i \beta(\Omega_i - 1) J_i + \frac{1}{2}\sum_a \beta (\Phi_a - 1) R_a }  \right) \;,
\ee
and using the relation \eqref{susy_cond} between the chemical potentials one finds
\be
Z = \Tr{} \left(e^{-\beta\{\cQ, \bar\cQ\}} \, e^{\pi i R_3} \, e^{\sum_i \beta(\Omega_i - 1) \left( J_i + \frac{1}{2}R_3 \right)} \, e^{\beta (\Phi_1 - 1) \frac{R_1 - R_3}2} \, e^{\beta (\Phi_2 - 1) \frac{R_2 - R_3}2 } \right) \;.
\ee
For the charge assignments of our theory, the factor $e^{\pi i R_3}$ is equivalent to $(-1)^F$.  Note also that in the new expression, since $(R_1-R_3)$ and $(R_2-R_3)$ are even and $J_i + \frac{1}{2}R_3$ is an integer, the periodicities of the remaining four chemical potentials are reduced to $(2\pi i / \beta)$.
Defining new chemical potentials for the four independent charges that, given \eqref{susy_cond}, we have access to, namely
\be  \label{param_relation}
\sigma = \frac{\beta \, (\Omega_1-1)}{2\pi i} \;,\qquad \tau = \frac{\beta \, (\Omega_2-1)}{2\pi i} \;,\qquad \Delta_a = \frac{\beta \, (\Phi_a-1)}{2\pi i} \quad\text{for } a=1,2 \;,
\ee
one finally finds that
\bea
\label{eq: partition function index relation}
Z &= \Tr{} \left( (-1)^F \, e^{-\beta\{\cQ, \bar\cQ\}} \, e^{2 \pi i \sigma \left( J_1 + \frac{1}{2}R_3 \right)} \, e^{2 \pi i \tau \left( J_2 + \frac{1}{2}R_3 \right)} \, e^{2 \pi i \Delta_1 \fq_1 } \, e^{2 \pi i \Delta_2 \fq_2} \right) \\ 
	&= e^{-\beta E_0} \; \cI(\sigma,\tau, \Delta_1,\Delta_2) \;,
\eea
where $\cI$ is the superconformal index \eqref{index Hamiltonian definition} defined above, and $E_0$ (which is related to the charges of the ground state on $S^3$) is called here the Casimir energy.%
\footnote{If our renormalization scheme requires additional counterterms, they would appear similarly to $E_0$.}
This relation between the thermal partition function \eqref{part_func} (when \eqref{susy_cond} holds) and the index will enable us to relate the index that we compute in the field theory, to computations of the  $S^1 \times S^3$ partition function using the gravitational dual.

Note that, in general, $E_0$ is scheme dependent. The authors of \cite{Assel:2015nca} argued that supersymmetry dictates (if a supersymmetric regularization is used) a specific choice of renormalization scheme which fixes the value of $E_0$ to a specific value $E_\text{SUSY}$ depending on the chemical potentials, which is then called the supersymmetric Casimir energy. In \cite{Cabo-Bizet:2018ehj} a localization computation verified \eqref{eq: partition function index relation} when \eqref{susy_cond} applies and found the supersymmetric Casimir energy.

Both of these papers relied on coupling the theory to curved space using the R-multiplet. Recently, it was argued that for theories with an anomalous $U(1)_R$ symmetry (in the sense of non-conservation of the current in the presence of a background field, as in $\cN=4$ SYM) this coupling induces supercurrent anomalies, see \cite{Papadimitriou:2017kzw, An:2017ihs, Papadimitriou:2019gel, Papadimitriou:2019yug, Kuzenko:2019vvi, Katsianis:2020hzd, Bzowski:2020tue}. They suggest that additional counterterms should be added to the action in order to restore supersymmetry. These counterterms could change the value of $E_0$. A similar argument was made in \cite{Closset:2019ucb}, where it was suggested that in order to regularize the theory in a supersymmetric and diffeomorphism invariant way one should add an additional counterterm to the action. This changes \eqref{eq: partition function index relation} to $Z = e^{-\beta W} e^{-\beta E_\text{SUSY}} \cI$, where $W$ is the additional counterterm and $E_\text{SUSY}$ is the same as before.

We will avoid this discussion by considering the ratio between the partition function and the contribution of the vacuum, which equals the index $\cI$  in every consistent scheme. On the gravity side we realize this through ``background subtraction'', namely regularizing the gravitational action by subtracting the contribution coming from empty thermal AdS, and we compare the result of this regularization to the computation of the index.%
\footnote{Note that background subtraction is believed to give the same results for the ratio we compute as a more precise computation using holographic renormalization in the bulk, as in \cite{Cassani:2019mms}. With the specific regularization used there, $E_0$ matches the usual Casimir energy in a free CFT, see also Appendix~A of \cite{Assel:2015nca} (and our footnote~\ref{foo: gravitational casimir energy}).}


\section{Field theory analysis}
\label{sec: Field theory analysis}

We are interested in the superconformal index \eqref{index Hamiltonian definition} of the four-dimensional $\cN=4$ supersymmetric Yang-Mills theory with gauge group $SU(N)$. 

\paragraph{Bethe Ansatz formulation.} The index is independent of the gauge coupling, and thus can be computed exactly \cite{Sundborg:1999ue, Aharony:2003sx, Romelsberger:2005eg, Kinney:2005ej} in terms of a certain contour integral. However, in order to extract the large $N$ limit, it is convenient to consider the so-called Bethe Ansatz formula \cite{Closset:2017bse, Benini:2018mlo}. In this paper we will mainly be interested in the case of two equal angular fugacities:
\be
p= q \qquad\Leftrightarrow\qquad \sigma = \tau \;.
\ee
In this case the Bethe Ansatz formula reads%
\footnote{This expression was derived in \cite{Benini:2018mlo} assuming that the Bethe Ansatz equations have isolated solutions. It has been noticed in \cite{ArabiArdehali:2019orz, Lezcano:2021qbj, Benini:2021ano} that for the $\cN=4$ SYM theory, the equations can also have continuous families of solutions. In this case the expression of their contributions should likely be modified. In this paper we will ignore the contributions of those continuous families, leaving their study to future work.
\label{foo: continuous sols}}
\be
\label{BA formula}
\cI(q, y_1, y_2) = \kappa_N \sum_{\hat u \,\in\, \text{BAEs}} \cZ(\hat u; \Delta, \tau) \, H(\hat u; \Delta, \tau)^{-1} \;.
\ee
The sum is over the solution set $\{\hat u\}$ to a system of transcendental equations, dubbed here Bethe Ansatz Equations (BAEs). Defining the Bethe Ansatz (BA) operators
\be
\label{BA operators}
Q_i(u; \Delta, \tau) \,\equiv\, e^{6\pi i \sum_j u_{ij}} \prod_{j=1}^N \frac{\theta_0( u_{ji} + \Delta_1; \tau)\, \theta_0 (u_{ji} + \Delta_2; \tau) \, \theta_0(u_{ji} - \Delta_1 - \Delta_2; \tau) }{ \theta_0( u_{ij} + \Delta_1; \tau)\, \theta_0 (u_{ij} + \Delta_2; \tau) \, \theta_0(u_{ij} - \Delta_1 - \Delta_2; \tau) }
\ee
for $i=1, \dots, N$, the BA equations are given by
\be
\label{BAEs}
1 = \frac{Q_i}{Q_N} \qquad\qquad\text{for}\quad i = 1, \dots, N-1 \;.
\ee
The quasi-elliptic function $\theta_0$ is defined in Appendix~\ref{app: functions}, while $u_{ij} = u_i - u_j$. The unknowns are the ``complexified $SU(N)$ holonomies'' $u_i$ living on a torus of modular parameter $\tau$, namely with identifications
\be
u_i \,\sim\, u_i + 1 \,\sim\, u_i + \tau \qquad\qquad\text{for}\quad i=1, \dots, N-1 \;,
\ee
while $u_N$ is fixed by the constraint $\sum_{i=1}^N u_i = 0$. The BAEs (\ref{BAEs}) are invariant under such shifts.
In fact, a stronger property holds: $Q_i$ are invariant under shifts of the components of the antisymmetric tensor $u_{ij}$ by $1$ or $\tau$, even relaxing the condition that $u_{ij} = u_i - u_j$.

It was proven in \cite{Benini:2018mlo} that only the solutions that are not invariant under any non-trivial element of the Weyl group of $SU(N)$ (namely, only solutions with all $u_i$ different on the torus) actually contribute to the sum in (\ref{BA formula}). Moreover, it was shown in \cite{Benini:2021ano} that only the solutions to a more restrictive set of equations,
\be
\label{reduced BAEs}
Q_i = (-1)^{N-1} \qquad\qquad\text{for } i = 1, \dots, N
\ee
and dubbed ``reduced Bethe Ansatz equations'', actually contribute to the index. Notice that $\prod_{i=1}^N Q_i = 1$ identically. We stress that from each solution in terms of $u_{ij}$ on the torus, one gets up to $N^2$ solutions in terms of $u_i$, related by a shift of the ``center of mass'' of the first $N-1$ components. Some of those, however, could be equivalent up to the Weyl group action, and thus the exact multiplicity should be determined by a case-by-case analysis.

The prefactor in (\ref{BA formula}) is
\be
\kappa_N = \frac1{N!} \left( \frac{ (q;q)_\infty^2 \, \wt\Gamma(\Delta_1; \tau,\tau) \, \wt\Gamma(\Delta_2; \tau,\tau) }{ \wt\Gamma(\Delta_1 + \Delta_2; \tau,\tau) } \right)^{N-1}
\ee
where $(z; q)_\infty$ is the $q$-Pochhammer symbol while $\wt\Gamma(u; \sigma, \tau)$ is the elliptic gamma function, defined in Appendix~\ref{app: functions}. The function $\cZ$ is given by
\be
\cZ(u; \Delta, \tau) = \prod_{i\neq j}^N \frac{ \wt\Gamma(u_{ij} + \Delta_1; \tau,\tau) \, \wt\Gamma(u_{ij} + \Delta_2; \tau, \tau) }{ \wt\Gamma(u_{ij} + \Delta_1 + \Delta_2; \tau,\tau) \, \wt\Gamma(u_{ij}; \tau,\tau) }
\ee
and it coincides (using \eqref{Gamma product}) with the integrand of the standard integral formula for the superconformal index. Finally, $H$ is a Jacobian defined as
\be
\label{def Jacobian}
H = \det\left[ \frac1{2\pi i} \, \parfrac{\log(Q_i / Q_N)}{u_j} \right]_{i,j=1, \dots, N-1} \,\equiv\, \det(\cA_{ij}) \;.
\ee
Here we introduced the $(N-1)\times(N-1)$ Jacobian matrix $\cA_{ij}$ for later convenience.

Notice that $Q_i$, $\kappa_N$, $\cZ$ and $H$ are all invariant under integer shifts of $\tau$, $\Delta_1$ and $\Delta_2$, implying that the superconformal index (\ref{BA formula}) is a single-valued function of the fugacities. This is also apparent from the Hamiltonian definition (\ref{index Hamiltonian definition}), as already noted.

The superconformal index of the $\cN=4$ SYM theory with gauge group $U(N)$ is related to the one for gauge group $SU(N)$ by the simple relation
\be
\cI_{U(N)} = \cI_{U(1)} \, \cI_{SU(N)} \;,
\ee
where
\be
\label{index U(1)}
\cI_{U(1)}(p,q,y_1, y_2) = (p;p)_\infty \, (q;q)_\infty \, \frac{\wt\Gamma(\Delta_1; \sigma, \tau) \, \wt\Gamma(\Delta_2; \sigma, \tau) }{ \wt\Gamma(\Delta_1 + \Delta_2; \sigma, \tau) }
\ee
is the index of the free $\cN=4$ $U(1)$ theory.

\paragraph{Hong-Liu (HL) solutions.} The full set of solutions to (\ref{reduced BAEs}) is not known (see also footnote~\ref{foo: continuous sols}), however, a large set was found in \cite{Hosseini:2016cyf, Hong:2018viz} and we will refer to them as HL solutions. They are labelled by three integers:
\be
\{m,n,r\} \qquad\text{such that}\qquad N = m \cdot n \;,\qquad 0 \leq r < n \;.
\ee
The solutions are
\be
\label{HL solutions u_j}
u_j \equiv u_{\hat\jmath \hat k} = \bar u + \frac{\hat\jmath}m + \frac{\hat k}n \left( \tau + \frac rm \right) \;.
\ee
Here we have decomposed the index $j= 0, \dots, N-1$ into the indices $\hat\jmath = 0, \dots, m-1$ and $\hat k = 0, \dots, n-1$. Morever, $\bar u$ is a constant chosen in such a way to solve the $SU(N)$ constraint (but recall that what enters in all formulas are the differences $u_{ij} = u_i - u_j$). Notice that these solutions are in one-to-one correspondence with subgroups of $\bZ_N \times \bZ_N$ of order $N$. It turns out that each HL solution has multiplicity $N$ (besides the $N!$ coming from permutations of $u_i$'s), corresponding to the inequivalent solutions to the $SU(N)$ constraint.

The BAEs (\ref{reduced BAEs}) are invariant under $SL(2,\bZ)$ modular transformations of the torus, namely under the generators
\be
T : \begin{cases} \tau \mapsto \tau + 1 \\ u \mapsto u \end{cases} \qquad
S : \begin{cases} \tau \mapsto - 1/\tau \\ u \mapsto u/\tau \end{cases} \qquad
C : \begin{cases} \tau \mapsto \tau \\ u \mapsto -u \;. \end{cases}
\ee
It follows that the HL solutions form orbits under $PSL(2,\bZ)$, completely classified by the integer $d = \gcd(m,n,r)$. The action of $PSL(2,\bZ)$ is given by
\be
T: \{m,n,r\} \mapsto \{m, n, r+m\} \;,\qquad
S: \{m,n,r\} \mapsto \left\{ \gcd(n,r) \,,\, \frac{m\, n}{\gcd(n,r)} \,,\, \frac{m(n-r)}{\gcd(n,r)} \right\}
\ee
where the last entry of $\{m',n',r'\}$ is understood mod $n'$.

\subsection{Contributions of Hong-Liu solutions}
\label{sec: HL contrib}

We will compute the contribution of each of the HL solutions (\ref{HL solutions u_j}) to the sum in (\ref{BA formula}). While $H$ will remain somehow implicit, we will be able to obtain a very explicit expression for $\cZ$.

It is convenient to rewrite $u_{ij}$ as
\be
\label{HL solution u_ij}
u_{ij} \equiv u_{(j_1k_1)(j_2k_2)} = v_{j_1j_2} + w_{k_1k_2}
\ee
with $j_1, j_2 = 0, \dots, m-1$ and $k_1, k_2 = 0, \dots, n-1$, as well as
\be
v_{j_1 j_2} = v_{j_1} - v_{j_2} = \frac{j_1 - j_2}m \;,\qquad\qquad w_{k_1k_2} = w_{k_1} - w_{k_2} = \frac{k_1 - k_2}n \Bigl( \tau + \frac rm \Bigr) \;.
\ee
We also define
\be
\xi_j = e^{2\pi i v_j} \;,\qquad\qquad \zeta_k = e^{2\pi i w_k} \;.
\ee

\subsubsection{Elliptic gamma functions}

To evaluate the contribution to the index, we need to compute
\bea
\label{def gamma Delta}
\gamma_\Delta &= \sum_{i\neq j}^N \log \left(\wt\Gamma(u_{ij} + \Delta; \tau, \tau) \right) \Big|_{(\ref{HL solution u_ij})} \\
&= \sum_{j_1, j_2=0}^{m-1} \sum_{k_1 \neq k_2}^{n-1} \log \left(\wt\Gamma\bigl( v_{j_1j_2} + w_{k_1k_2} + \Delta; \tau , \tau \bigr)\right) + n \sum_{j_1 \neq j_2}^{m-1} \log \left(\wt\Gamma\bigl( v_{j_1 j_2} + \Delta; \tau , \tau \bigr) \right)
\eea
for $\Delta\in\{0, \Delta_1, \Delta_2, \Delta_1+\Delta_2\}$. Notice that while $\exp(\gamma_\Delta)$ is well defined, $\gamma_\Delta$ is only defined modulo $2\pi i$. We set
\be
y = e^{2\pi i \Delta} \;.
\ee
When $\Delta$ is a generic complex number, one easily verifies the shift property
\be
\label{shift property of gamma}
\gamma_{\Delta + \frac1m} - \gamma_\Delta = - N \log \left(\frac{\wt\Gamma\bigl( \Delta + \frac1m; \tau, \tau \bigr)}{\wt\Gamma(\Delta; \tau, \tau)} \right) \;.
\ee
We will use this relation in the following. It follows that $\gamma_{\Delta+1} = \gamma_\Delta$.

\paragraph{The case \matht{\Delta \neq 0}.}
Our strategy is to expand the functions appearing in (\ref{def gamma Delta}) inside a common domain of convergence, and then manipulate and resum the expansions. As long as we obtain exact expressions, the latter will be valid on the full domain of analyticity, possibly up to ambiguities by $2\pi i$ from the logarithms (such ambiguities disappear when considering $\exp(\gamma_\Delta)$). The details of the computation are in Appendix~\ref{app: case Delta neq 0}. Defining
\be
\label{def: tau check}
\check\tau = m\tau + r \qquad\qquad\text{and}\qquad\qquad \check q = e^{-2\pi i /\check\tau} \;,
\ee
we obtain the exact expression%
\footnote{Taking the exponential of the right-hand side, one verifies that it is invariant under $m\Delta \to m\Delta + 1$, performed in all terms but the first one in the second line (while if we do not take the exponential, we are left with $-\pi i mn^2 - N\log(-1)$ plus possible multiples of $2\pi i$ from the branch cuts of the logs). It follows that the right-hand side satisfies the shift property (\ref{shift property of gamma}).
\label{foo: shift}}
\be
\label{gamma_Delta resummed}
\begin{aligned}
\quad \gamma_\Delta &= \pi i \frac{N^2}m \, \frac{(\check\tau-m\Delta) \bigl(\check\tau - m\Delta - \tfrac12 \bigr) (\check\tau - m\Delta-1)}{3\check\tau^2} - \frac{ \pi i m}6 \bigl( \check\tau - m\Delta - \tfrac12 \bigr) \quad\rule{0pt}{2em} \\
&\quad -N \log\left( \wt\Gamma(\Delta; \tau,\tau) \right) - N \log\left( \theta_0 \Bigl( \frac{N\Delta}{\check\tau}; - \frac n{\check\tau} \Bigr) \right) + m \sum_{k=0}^\infty \log \left( \frac{ \psi\bigl( \frac{k+1+m\Delta}{\check\tau/n} \bigr) }{ \psi \bigl( \frac{ k-m\Delta }{\check\tau/n} \bigr) } \right) \;.
\end{aligned}
\ee
The function $\psi(u)$ appearing here is defined in Appendix~\ref{app: functions}.

The expression we obtained will be convenient when taking the large $N$ limit, however, if we are interested in an explicit exact expression, even the remaining infinite sum can be computed exploiting the modular formula (\ref{degenerate modular formula}). We obtain:%
\footnote{One can verify that under $r \to r + n$, which corresponds to $\check\tau \to \check\tau + n$, the quantity $\gamma_\Delta$ gets shifted by $m\pi i \frac{n(n+1)(n-1)}3$ (which is always a multiple of $2\pi i$) plus possible multiples of $2\pi i$ from the branch cuts of the logs. It follows that $\exp(\gamma_\Delta)$ is invariant, as it should.}
\bea
\gamma_\Delta &= \pi i \frac{N^2}m \, \frac{(\check\tau-m\Delta) \bigl(\check\tau - m\Delta - \tfrac12 \bigr) (\check\tau - m\Delta-1)}{3\check\tau^2} - \frac{ \pi i m}6 \bigl( \check\tau - m\Delta - \tfrac12 \bigr) \\
&\quad + \pi i m \, \cQ \bigl( m\Delta; \check\tau/n, \check\tau/n \bigr) - (N-m) \log \left( \theta_0 \Bigl( \frac{N\Delta}{\check\tau}; - \frac n{\check\tau} \Bigr) \right) \\
&\quad  - N \log \left( \wt\Gamma(\Delta; \tau,\tau) \right) + m \log\left( \wt\Gamma\Bigl( m\Delta; \frac{\check\tau}n, \frac{\check\tau}n \Bigr) \right) \;.
\eea
Here $\cQ(u; \tau, \tau)$ is a cubic polynomial in $u$, defined in (\ref{Q polynomial degenerate}).

\paragraph{The case \matht{\Delta = 0}.}
This case requires a separate treatment, which can be found in Appendix~\ref{app: case Delta = 0}. We obtain
\be
\label{gamma_0 resummed}
\gamma_0 = \pi i \frac{N^2}m \, \frac{\check\tau \bigl( \check\tau - \frac12 \bigr) (\check\tau -1) }{ 3 \check\tau^2} - \frac{\pi i m \check\tau}6 + N \log \left( \frac{\check\tau}N \right) + 2N \log \left[ \frac{ (q;q)_\infty }{ (\check q^n; \check q^n)_\infty } \right] \;.
\ee
Notice that we could have obtained this expression directly as the $\Delta\to0$ limit of (\ref{gamma_Delta resummed}). To do that, we need the following asymptotic behaviors, easy to derive:
\bea
\theta_0\left( \frac{N\Delta}{\check\tau} ; - \frac n{\check\tau} \right) &\;\stackrel{\Delta \to 0}{\sim}\; \bigl( 1 - e^{2\pi i N\Delta/\check\tau} \bigr) \, (\check q^n; \check q^n)_\infty^2 \\
\wt\Gamma(\Delta; \tau, \tau) &\;\stackrel{\Delta\to0}{\sim}\; \frac1{1-e^{2\pi i \Delta}} \, \frac1{(q;q)_\infty^2} \;.
\eea
It follows that
\be
-N \lim_{\Delta\to0} \log \left[ \wt\Gamma(\Delta; \tau,\tau) \, \theta_0\left( \frac{N\Delta}{\check\tau} ; - \frac n{\check\tau} \right) \right] = 2 N \log \left[ \frac{(q;q)_\infty }{ (\check q^n; \check q^n)_\infty} \right] + N \log \left( \frac{\check\tau}N \right) \;.
\ee
The last sum in (\ref{gamma_Delta resummed}) is regular at $\Delta=0$: most terms cancel out, and one uses $\psi(0)= e^{\pi i /12}$.

\subsubsection{The Jacobian}

The Jacobian was defined in (\ref{def Jacobian}). It is convenient to introduce an auxiliary chemical potential $\Delta_3$ defined by
\be
\label{constraint def Delta3}
\sum_{a=1}^3 \Delta_a - \sigma - \tau \in \bZ
\ee
(and then set $\sigma = \tau$), in order to make manifest the action of the Weyl group of $SU(3)$ --- the global symmetry contained in $SU(4)_R$ that commutes with the preserved supercharges $\cQ, \bar\cQ$. We can then rewrite the BA operators (\ref{BA operators}) as
\be
\label{BA operators alternative}
Q_i = e^{-2\pi i \sum_j u_{ij}} \prod_{a=1}^3 \prod_{j=1}^N \frac{ \theta_0(u_{ji} + \Delta_a; \tau) }{ \theta_0(u_{ij} + \Delta_a; \tau) } \;.
\ee
With a few manipulations that we report in Appendix~\ref{app: Jacobian}, the $(N-1)\times(N-1)$ Jacobian matrix $\cA_{ij}$ can be compactly written as
\be
\label{Aij resummed}
\cA_{ij} = N \, b_N \, \cB_{ij} + \cC_{ij} \;.
\ee
Here
\bea
\label{elements of Jacobian}
b_N &= - 4 + \frac1{\check\tau} \sum_{a=1}^3 \left[ 2m\Delta_a +1 + \cG \left(0, \frac{N\Delta_a}{\check\tau}; - \frac n{\check\tau} \right) \right] \\
\cB_{ij} &= \delta_{ij} + 1 \\
\cC_{ij} &= \sum_{a=1}^3 \biggl[ \cG(u_{iN}, \Delta_a; \tau) + \cG(u_{jN}, \Delta_a; \tau) - \cG(u_{ij}, \Delta_a; \tau) - \cG(0, \Delta_a; \tau) \biggr]
\eea
with $i,j=1, \dots, N-1$, where the function $\cG(u,\Delta; \tau)$ is defined in Appendix~\ref{app: functions}. Notice that $\det (\cB) = N$ and $(\cB^{-1})_{ij} = \delta_{ij} - 1/N$.  To compute the determinant, we factorize $\cA$ and obtain
\be
\log (H) \,\equiv\, \log \bigl( \det(\cA) \bigr) = N\log (N) + (N-1)\log (b_N) + \log \left[ \det \left( \unit + \frac{b_N^{-1}}N \cB^{-1}\cC \right) \right] \;.
\ee
The last term of this expression is still rather implicit, however this will be good enough to compute the large $N$ limit.

\subsubsection{Total contribution}

We denote the total contribution of the HL solution $\{m,n,r\}$ to the index as
\be
\cI_{\{m,n,r\}} = N \cdot N! \cdot \kappa_N \, \cZ(u; \Delta, \tau) \, H(u; \Delta, \tau)^{-1} \Big|_{(\ref{HL solution u_ij})} \;.
\ee
The factor $N \cdot N!$ takes into account the multiplicity of the solution $\{m,n,r\}$: indeed $N!$ comes from the Weyl group action that permutes the eigenvalues $u_i$, while $N$ comes from the inequivalent choices of sets of points $\{u_i\}$ on the torus that give rise to the same $u_{ij}$'s, as discussed after (\ref{HL solutions u_j}). Such $N$ inequivalent choices correspond to shifts of the center of mass of the first $N-1$ holonomies, and can be parametrized, for $(\hat\jmath, \hat k) \neq (m,n)$, as
\be
u_{\hat\jmath \hat k}^{(\ell)} = \bar u + \frac{\hat\jmath}m + \frac{\hat k}n \left( \tau + \frac rm \right) + \frac{\ell_1\tau + \ell_2}N
\ee
with $\ell_1 = 0, \dots, m-1$ and $\ell_2 = 0, \dots, n-1$.
As discussed before, all these solutions give exactly the same contribution to $\cZ$ and $H$.

Using that $\wt\Gamma(u + \Delta_1 + \Delta_2; \sigma, \tau)^{-1} = \wt\Gamma(-u + \Delta_3; \sigma, \tau)$ and putting everything together, we eventually find
\be
\label{contribution (m,n,r)}
\boxed{\begin{aligned} \, \rule{0pt}{2em}
& \log \bigl( \cI_{\{m,n,r\}} \bigr) = \cP(m\Delta, \check\tau) + \log (N) + 2\log \left[ \frac{ \bigl( \check q^n; \check q^n)_\infty^N }{ (q;q)_\infty} \right] - N \log(\check\tau) - (N-1) \log (b_N) \\
&\;\; + \sum_{a=1}^3 \left[ - N \log \left[ \theta_0 \biggl( \frac{N \Delta_a}{\check\tau}; - \frac{n}{\check\tau} \biggr) \right] - \log \left( \wt\Gamma(\Delta_a; \tau, \tau) \right) + m \sum_{k=0}^\infty \log \left( \frac{ \psi\left( \frac{k+1+m\Delta_a}{\check\tau/n} \right) }{ \psi\left( \frac{k - m\Delta_a}{\check\tau/n} \right) } \right) \right] \; \\
&\;\; + \frac{\pi i m}6 \left( \sum_a m \Delta_a - 2\check\tau + \frac32 \right) - \log\left[ \det \left( \unit + \frac{b_N^{-1}}N \cB^{-1}\cC \right) \right] \;.
\rule[-1.8em]{0pt}{1em} \end{aligned}}
\ee
Recall that $\check\tau = m\tau +r$ and $\check q = e^{-2\pi i / \check\tau}$.
Here $\cP(m\Delta,\check\tau)$ is the contribution of order $N^2$:
\be
\cP(m\Delta,\check\tau) = \frac{\pi i N^2}{3m} \left[ \sum_{a=1}^3 \frac{(\check\tau - m\Delta_a) \bigl( \check\tau - m\Delta_a - \frac12 \bigr) ( \check\tau - m \Delta_a - 1) }{ \check\tau^2} - \frac{\check\tau \bigl( \check\tau - \frac12 \bigr)(\check\tau - 1) }{ \check\tau^2} \right] \;.
\ee

\subsection[The large $N$ limit]{The large \matht{N} limit}
\label{sec: large N}

We proceed to compute the large $N$ limit of (\ref{contribution (m,n,r)}). Clearly, this depends on how we scale $m,n,r$ with $N$. In this paper we consider BA solutions $\{m,n,r\}$ with $m,r$ fixed and $n \to \infty$.

There are various terms in (\ref{contribution (m,n,r)}) that are exponentially suppressed in the limit and go to zero. For instance, $N \log \left( (\check q^n; \check q^n)_\infty \right) \sim - N \check q^n$ for $n \to \infty$, and thus this term is of order $\cO(N e^{-N})$ and goes to zero. It turns out that when the electric chemical potentials $\Delta_{a=1,2,3}$ are in a particular range, namely
\be
\label{range of validity}
0 < \im \left( \frac{m\Delta_a}{\check\tau} \right) < \im \left( - \frac1{\check\tau} \right) \;,
\ee
many other terms are exponentially suppressed as $\cO(N e^{-N})$. We have:
\begin{align}
\label{large N suppressions}
N \log \left[ \theta_0 \left( \frac{N\Delta_a}{\check\tau} ; - \frac n{\check\tau} \right) \right] &\,\sim\, - N \bigl( \tilde y_a^n + (\check q/ \tilde y_a)^n \bigr) \to 0\,, \nn \\[.5em]
m \sum_{k=0}^\infty \log \left( \frac{ \psi\left( \frac{k+1+m\Delta_a}{\check\tau/n} \right) }{ \psi\left( \frac{k - m\Delta_a}{\check\tau/n} \right) } \right) &\,\sim\, - N \sum_{k=0}^\infty \left[ \frac{k+1+m\Delta_a}{\check\tau} \, (\check q/\tilde y_a)^n - \frac{k-m\Delta_a}{\check\tau} \, \tilde y_a^n \right] \check q^{nk} \,\to\, 0\,, \nn \\[.5em]
N \, \cG\left( 0, \frac{N\Delta_a}{\check\tau}; - \frac n{\check\tau} \right) &\,\sim\, 2N \bigl( \tilde y_a^n - (\check q / \tilde y_a)^n \bigr) \,\to\, 0 \;.
\end{align}
On the other hand, the quantity $\log \bigl[ \det\bigl( \unit + \cB^{-1} \cC/ b_N N \bigr) \bigr]$ is of order $\cO(1)$. To see that, notice from (\ref{elements of Jacobian}) that the matrix $\cC$ has entries of order 1, the matrix $\cB^{-1}\cC$ has entries of order 1, $b_N$ is of order 1, and hence the matrix $\cE \equiv \cB^{-1} \cC/b_N N$ has entries of order $1/N$. It follows that all traces $\Tr (\cE^\ell)$ are of order 1 (or higher powers of $1/N$), and so is $\log \bigl[ \det(1+\cE) \bigr] = \sum_{\ell=1}^\infty \frac1\ell (-1)^{\ell+1} \Tr (\cE^\ell)$. We will study the Jacobian in more detail in Section~\ref{sec: pert corrections} below.

\begin{figure}[t]
\centering
\begin{tikzpicture}
\filldraw [yellow!15!white] (-2.16,-1) to (-.66,-1) to (1.33,2) to (-.16,2) to cycle;
\filldraw [purple!10!white] (-.66,-1) to (1.33,2) to (2.82,2) to (.83, -1) to cycle;
\draw [->] (-3,0) to (3,0);
\draw [->] (0,-1) to (0,2);
\draw [very thick, blue!80!black] (-.66,-1) to (1.33,2);
\draw [very thick, blue!80!black] (-2.16,-1) to (-.16,2);
\draw [very thick, blue!80!black] (.83, -1) to (2.82,2);
\filldraw [red!80!black] (0, 0) circle [radius=.07] node [below right, black] {\small $0$};
\filldraw [red!80!black] (1, 1.5) circle [radius=.07] node [below right, black] {\small $\tau$};
\filldraw [red!80!black] (-1.5, 0) circle [radius=.07] node [above left, black] {\small $-1$};
\filldraw [red!80!black] (1.5, 0) circle [radius=.07] node [below right, black] {\small $1$};
\filldraw [red!80!black] (-.5, 1.5) circle [radius=.07] node [above left, black] {\small $\tau-1$};
\filldraw [red!80!black] (2.5, 1.5) circle [radius=.07] node [below right, black] {\small $\tau + 1$};
\end{tikzpicture}
\caption{Fundamental strips for $[\Delta]_\tau$ and $[\Delta]'_\tau$. The function $[\Delta]_\tau$ is the restriction of $\Delta$ mod $1$ to the region $\im(-1/\tau) > \im(\Delta/\tau)>0$ (in yellow, on the left), while $[\Delta]'_\tau$ is the restriction of $\Delta$ mod $1$ to the region $0 > \im(\Delta/\tau) > \im(1/\tau)$ (in purple, on the right). We dubbed $\ccL_\tau$ the blue line through $0$ and $\tau$.
\label{fig: strip}}
\end{figure}
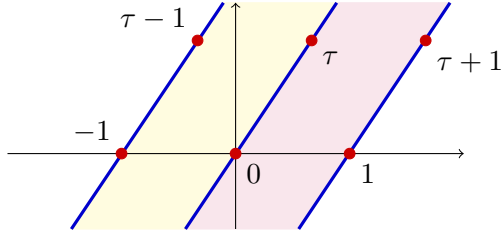

We conclude that in the range (\ref{range of validity}) it is relatively simple to take the large $N$ limit --- and the only term that contributes at order $\cO(N^2)$ is $\cP$. The range (\ref{range of validity}) corresponds to $m\Delta_a$ lying within an open strip in the complex plane, bounded by the line $\ccL_{\check\tau}$ through $0$ and $\check\tau$ on the right, and by the line through $-1$ and $\check\tau-1$ on the left (see Figure \ref{fig: strip}). From (\ref{gamma_Delta resummed}) and as noticed in footnote~\ref{foo: shift}, $\gamma_\Delta$ is invariant if we shift $m\Delta \to m\Delta+1$ in all terms but $N\log\left(\wt\Gamma(\Delta; \tau,\tau)\right)$; similarly $b_N$ in (\ref{elements of Jacobian}) is invariant under that shift. We can exploit this fact to push $m\Delta_a$ towards the range (\ref{range of validity}).

To that purpose, we introduce the periodic discontinuous function $[\Delta]_\tau$ defined as
\be
[\Delta]_\tau = \left( z \;\middle|\; z = \Delta \text{ mod } 1 \,,\quad \im\Bigl( - \frac1\tau \Bigr) > \im\Bigl( \frac{z}\tau \Bigr) > 0 \right) \qquad\text{for }\;\; \im\left( \tfrac\Delta\tau \right) \not\in \bZ \times \im\left( -\tfrac1\tau \right) \,.
\ee
This gives the image of $\Delta$ under an integer shift that sets it between $\ccL_\tau -1$ on the left and $\ccL_\tau$ on the right (see Figure~\ref{fig: strip} for a picture of this domain, in yellow), while it remains undefined if $\Delta \in \ccL_\tau + \bZ$. In other words, $[\Delta]_\tau$ is defined by the conditions
\be
\label{properties square bracket function}
[\Delta]_\tau = \Delta \mod 1 \;,\qquad \im\Bigl( - \frac1\tau \Bigr) > \im\Bigl( \frac{[\Delta]_\tau}\tau \Bigr) > 0 \;.
\ee
We also define
\be
[\Delta]'_\tau = [\Delta]_\tau + 1 \;,
\ee
which gives the image of $\Delta$ under an integer shift that sets it between $\ccL_\tau$ on the left and $\ccL_\tau + 1$ on the right, whenever $\Delta$ does not lie on $\ccL_\tau + \bZ$. These functions satisfy
\be
[\Delta+1]_\tau = [\Delta]_\tau \;,\qquad [\Delta + \tau]_\tau = [\Delta]_\tau + \tau \;,\qquad [-\Delta]_\tau = - [\Delta]_\tau - 1 = - [\Delta]_\tau' \;.
\ee
In particular, $\sum_{a=1}^3 \Delta_a - 2\tau \in \bZ$ implies $[\Delta_3]_\tau = 2\tau - 1 - [\Delta_1 + \Delta_2]_\tau$, and multiplying by $m$ it also implies
\be
[m\Delta_3]_{\check\tau} = 2 \check\tau -1 - [m\Delta_1 + m\Delta_2]_{\check\tau} \;.
\ee

The strategy is thus to perform integer shifts of the $m\Delta_a$'s in the arguments of $\gamma_\Delta$ and $b_N$, in such a way that we land inside the range (\ref{range of validity}). This is always possible as long as none of the $m\Delta_a$ is exactly on the line $\ccL_{\check\tau}$ or one of its images under integer shifts. This allows us to write the contribution from the HL solution $\{m,n,r\}$ as
\begin{align}
& \log (\cI) = \cP\bigl( [m\Delta]_{\check\tau}, \check\tau \bigr) + \log (N) + 2\log\Biggl[ \frac{ \bigl( \check q^n; \check q^n)_\infty^N }{ (q;q)_\infty} \Biggr] \!- N \log(\check\tau) - (N \!-\! 1) \log \Bigl[ b_N\bigl( [m\Delta]_{\check\tau}, \check\tau \bigr) \Bigr] \nn\\
&\; + \sum_{a=1}^3 \left[ - N \log \left[ \theta_0 \left( \frac{n [m\Delta_a]_{\check\tau}}{\check\tau}; - \frac{n}{\check\tau} \right) \right] - \log \left( \wt\Gamma(\Delta_a; \tau, \tau) \right) + m \sum_{k=0}^\infty \log \left( \frac{ \psi\left( \frac{k+1+[m\Delta_a]_{\check\tau}}{\check\tau/n} \right) }{ \psi\left( \frac{k - [m\Delta_a]_{\check\tau}}{\check\tau/n} \right) } \right) \right] \nn\\
&\; + \frac{\pi i m}6 \left( \sum_a [m \Delta_a]_{\check\tau} - 2\check\tau + \frac32 \right) - \log \left[ \det \left( \unit + \frac{b_N\bigl( [m\Delta]_{\check\tau}, \check\tau \bigr)^{-1}}N \cB^{-1}\cC \right) \right] \;.
\label{contribution (m,n,r) shifted}
\end{align}
Note that no shift has been done within the matrix $\cC$ in the last term.
Because of the constraint (\ref{constraint def Delta3}) between $\Delta_a$ and $\tau$ that defines $\Delta_3$, the functions $\cP$ and $b_N$ simplify. Notice that, in general, $\im (-2/\check\tau) > \im \bigl( ( [m\Delta_1]_{\check\tau} + [m\Delta_2]_{\check\tau})/ \check\tau \bigr) > 0$ and therefore one should distinguish between two cases in parameter space:
\bea
\label{def 1st 2nd case}
\text{1$^\text{st}$ case}: & & \qquad\qquad [m\Delta_1]_{\check\tau} + [m\Delta_2]_{\check\tau} &= [m\Delta_1 + m\Delta_2]_{\check\tau} \\[.5em]
\text{2$^\text{nd}$ case}: & & \qquad\qquad [m\Delta_1]_{\check\tau} + [m\Delta_2]_{\check\tau} &= [m\Delta_1 + m\Delta_2]_{\check\tau} - 1 \;.
\eea
The second case is equivalent to $[m\Delta_1]'_{\check\tau} + [m\Delta_2]'_{\check\tau} = [m\Delta_1 + m\Delta_2]'_{\check\tau}$.
In a more permutation-invariant way, they can be written as
\bea
\label{def 1st 2nd case symmetric}
\text{1$^\text{st}$ case}: & & \qquad\qquad {\ts \sum_{a=1}^3} [m\Delta_a]_{\check\tau} - 2 \check\tau + 1 &= 0 \\[.5em]
\text{2$^\text{nd}$ case}: & & \qquad\qquad {\ts \sum_{a=1}^3} [m\Delta_a]'_{\check\tau} - 2 \check\tau - 1 &= 0 \;. \hspace{2.2cm}
\eea
The second case is also equivalent to $\sum_{a=1}^3 [m\Delta_a]_{\check\tau} - 2\check\tau+2 = 0$.

In the first case, with a little bit of algebra we find the large $N$ limit
\begin{equation}
\label{large N result case 1}
\boxed{\;\;\rule[-2.6em]{0pt}{5.7em}\begin{aligned}
\log \bigl( \cI_{\{m,n,r\}} \bigr) &= - \frac{\pi i N^2}m \, \frac{ [m\Delta_1]_{\check\tau} [m\Delta_2]_{\check\tau} [m\Delta_3]_{\check\tau} }{ \check\tau^2} + \log (N) - \log \bigl( \cI_{U(1)} \bigr) \qquad\quad\;\;\;\text{1$^\text{st}$ case} \\
&\quad -\log(\check\tau) + \frac{\pi i m}{12} - \log \left[ \det \left( \unit + \frac{b_N^{-1}}N \cB^{-1}\cC \right) \right] + \cO(N e^{-N}) \;.
\end{aligned}\;\;}
\end{equation}
Here $\cI_{U(1)}$ is the index of the free $U(1)$ theory (\ref{index U(1)}) specialized to $\sigma=\tau$, that we can also write as
\be
\cI_{U(1)}\Big|_{\sigma=\tau} = (q;q)_\infty^2 \prod\nolimits_{a=1}^3 \wt\Gamma(\Delta_a; \tau,\tau) \;.
\ee
In (\ref{large N result case 1}) the first term is of order $\cO(N^2)$, the second term of order $\cO\bigl( \log (N) \bigr)$, all other terms%
\footnote{In this paper we have not explicitly computed the determinant, which we leave for future work, however see a partial analysis in Section~\ref{sec: pert corrections}.}
are of order $\cO(1)$, and the corrections are exponentially small.
In the second case we similarly find%
\footnote{Notice that both in (\ref{large N result case 1}) and (\ref{large N result case 2}), $b_N$ stands for $b_N\bigl( [m\Delta]_{\check\tau}, \check\tau \bigr)$.}
\begin{equation}
\label{large N result case 2}
\boxed{\;\;\rule[-2.6em]{0pt}{5.7em}\begin{aligned}
\log \bigl( \cI_{\{m,n,r\}} \bigr) &= - \frac{\pi i N^2}m \, \frac{ [m\Delta_1]'_{\check\tau} [m\Delta_2]'_{\check\tau} [m\Delta_3]'_{\check\tau} }{ \check\tau^2} + \log (N) - \log \bigl( \cI_{U(1)} \bigr) \qquad\quad\;\;\;\text{2$^\text{nd}$ case} \\
&\quad -\log(\check\tau) + \pi i - \frac{\pi i m}{12} - \log \left[ \det \left( \unit + \frac{b_N^{-1}}N \cB^{-1}\cC \right) \right] + \cO(N e^{-N}) \;.
\end{aligned}\;\;}
\end{equation}
Here we used the $2\pi i $ ambiguity of the logarithm and the fact that $e^{\pi i m n^2 - \pi i m n} = 1$.

Our result is compatible with previous works on the subject. In particular, the leading contribution from special sub-families of HL solutions already appeared in \cite{Cabo-Bizet:2019eaf, ArabiArdehali:2019orz}, while the correction at order $\cO\bigl( \log(N) \bigr)$ was already found in \cite{GonzalezLezcano:2020yeb}, both for the basic solution as well as for the $\{m, n, 0\}$ sub-family of HL solutions.

\subsection{Non-perturbative corrections}
\label{non_pert}

The terms that we reported in the large $N$ limit expressions (\ref{large N result case 1}) and (\ref{large N result case 2}) are the leading $\cO(N^2)$ contribution and the perturbative corrections (notice that the determinant gives rise to both perturbative and non-perturbative corrections). Here we would like to estimate the non-perturbative corrections $\cO(N e^{-N})$. More precisely, we will compute the leading exponent --- or instanton action --- that controls the non-perturbative corrections, while leaving a more detailed analysis for future work.

In (\ref{large N result case 1}) and (\ref{large N result case 2}), the missing terms are
\begin{multline}
\label{non-perturbative corrections}
\log \bigl( \cI_{\{m,n,r\}} \bigr) = \Bigl( \begin{matrix} \ref{large N result case 1} \\[-.4em] \ref{large N result case 2} \end{matrix} \Bigr)
+ 2N \log \bigl( (\check q^n; \check q^n)_\infty \bigr) - (N-1) \log \left[ 1 \pm \sum_{a=1}^3 \cG \left( 0, \frac{[m\Delta_a]_{\check\tau}}{\check\tau/n}; - \frac n{\check\tau} \right) \right] \\
 - \sum_{a=1}^3 \left[ N \log \Biggl( \theta_0 \left( \frac{[m\Delta_a]_{\check\tau}}{\check\tau/n}; - \frac{n}{\check\tau} \right) \Biggr) - m \sum_{k=0}^\infty \log \left( \frac{ \psi\left( \frac{k+1+[m\Delta_a]_{\check\tau}}{\check\tau/n} \right) }{ \psi\left( \frac{k - [m\Delta_a]_{\check\tau}}{\check\tau/n} \right) } \right) \right] \;,
\end{multline}
where the two equation numbers and the $\pm$ signs in the first line correspond to the \first{} and \second{} case, respectively.
These non-perturbative corrections are needed in order to make the right-hand sides of (\ref{large N result case 1}) and (\ref{large N result case 2}) an analytic function of the chemical potentials for finite $N$, even though $[m\Delta_a]_{\check\tau}$ are not analytic functions. By expanding the various functions above using the formulas in Appendix~\ref{app: functions}, we find that the non-perturbative corrections have three kinds of exponents:
\be
\exp\left( \frac{2\pi i N}{m} \, k \, \frac{[m\Delta_a]_{\check\tau} - \ell}{\check\tau} \right) \;,\quad
\exp\left( \frac{2\pi i N}{m} \, k \, \frac{- [m\Delta_a]_{\check\tau} - 1 - \ell}{\check\tau} \right) \;,\quad
\exp\left( - \frac{2\pi i N}{m} \, \frac{k}{\check\tau} \right)\;,
\ee
for integers $k\geq 1$ and $\ell \geq 0$. All these exponents can be written as products of
\be
\exp\left( \frac{2\pi i N}{m} \, \frac{[m\Delta_a]_{\check\tau}}{\check\tau} \right) \;,\qquad
\exp\left( - \frac{2\pi i N}{m} \, \frac{[m\Delta_a]_{\check\tau} + 1}{\check\tau} \right) \;,\qquad
\exp\left( - \frac{2\pi i N}{m} \, \frac1{\check\tau} \right) \;.
\ee
In the \first{} case of parameter space, taking into account the relation (\ref{def 1st 2nd case symmetric}) among the chemical potentials, one finds that in fact all exponents can be written as products of
\be
\exp\left( \frac{2\pi i N}{m} \, \frac{[m\Delta_a]_{\check\tau}}{\check\tau} \right) \qquad\quad\text{for } a= 1,2,3.
\ee
Notice that because of the property (\ref{properties square bracket function}) of the function $[\;\;]_\tau$, all non-perturbative corrections are exponentially small in the large $N$ limit. In the \second{} case of parameter space, instead, all exponents can be written as products of
\be
\exp\left( - \frac{2\pi i N}{m} \, \frac{[m\Delta_a]'_{\check\tau}}{\check\tau} \right) \qquad\quad\text{for } a = 1,2,3.
\ee
Again, all non-perturbative corrections are exponentially small.

\subsection{Perturbative corrections}
\label{sec: pert corrections}

The perturbative corrections (\ie, the perturbative expansion in $1/N$) to the leading $\cO(N^2)$ contribution to $\log\bigl( \cI_{\{m,n,r\}}\bigr)$ come from the terms in (\ref{large N result case 1}) and (\ref{large N result case 2}). We were not able to evaluate the determinant explicitly, however we collected some evidence that --- at least in the case $m=1$ --- the determinant is of order $\cO(1)$ plus non-perturbative corrections of order $\cO(e^{-N})$, but without further perturbative corrections of order $\cO(1/N)$ or smaller. We thus conjecture that the series of perturbative corrections stops at order $\cO(1)$.

Our strategy is to expand
\be
- \log \biggl[ \det \biggl( \unit + \frac{\cB^{-1}\cC}{b_N N} \biggr) \bigg] = \sum_{\ell=1}^\infty \frac{(-1)^\ell }{\ell} \Tr \biggl[ \biggl( \frac{\cB^{-1}\cC}{b_N N} \biggr)^{\!\!\ell\;} \biggr]
\ee
and compute the first few terms on the right-hand side. The details of the computation are in Appendix~\ref{app: pert corrections}.
At first order, in the large $N$ limit we find
\be
\frac{\Tr\bigl( \cB^{-1} \cC \bigr) }{ b_N N} = 1 \pm \check\tau \Biggl( 1 - \sum_{a=1}^3 \cG\bigl( 0, \Delta_a; \tau \bigr) \Biggr) + \cO(e^{-N})
\ee
where the $\pm$ signs refer to the \first{} and \second{} case of parameter space, respectively. At second order the computation becomes considerably more involved, and we only performed it in the case $m=1$, \ie, for the T-transformed solutions. Setting $\check\tau = \tau + r$, in the large $N$ limit we find
\be
\frac{ \Tr\bigl( \cB^{-1}\cC \cB^{-1} \cC \bigr) }{ b_N^2 N^2 } = \frac12 \sum_{a,b=1}^3 \Biggl[ \wt\gamma \biggl( \frac{[\Delta_a]_{\check\tau} + [\Delta_b]_{\check\tau}}{ \check\tau}; - \frac1{\check\tau} \biggr) - \wt\gamma\biggl( \frac{ [\Delta_a]_{\check\tau} - [\Delta_b]'_{\check\tau}}{ \check\tau}; - \frac1{\check\tau} \biggr) \Biggr] + \cO(e^{-N}) \;.
\ee
Here we introduced the function
\be
\wt\gamma(u; \tau) \,\equiv\, \frac1{2\pi i} \, \partial_u \log \Bigl( \wt\Gamma(u; \tau, \tau) \Bigr) \;,
\ee
see also (\ref{expansion gamma tilde}). Both at first and second order (for $m=1$), the perturbative series in $1/N$ stops at order $\cO(1)$, the only large $N$ corrections being non-perturbative.

\subsection{The universal superconformal index}
\label{universal}

It is interesting to consider also a special case of the superconformal index (\ref{index Hamiltonian definition}), which is a simpler and more universal object. This is constructed by only using charges in the $\cN=1$ superconformal subalgebra,%
\footnote{All choices are physically equivalent.}
and it is defined as
\be
\cJ(p,q) = \Tr\, \left[ e^{-\pi i r} \, e^{-\beta\{\cQ, \bar\cQ\}} \, e^{2\pi i \sigma \left( J_1 + \frac r2 \right)} \, e^{2\pi i \tau \left( J_2 + \frac r2 \right)} \right]
\ee
where $r$ is the superconformal R-symmetry in the $\cN=1$ subalgebra. Since different states in the same supermultiplet have R-charges that differ by integers and the supercharge $\cQ$ has charge $r=1$, $\cJ$ has all the good properties of a supersymmetric index. When the R-charges are rational, this index will have some periodicity in the $\sigma,\tau$ plane.

For $\cN=4$ SYM, this special case corresponds to choosing
\be
\label{equal chemical potentials}
\Delta_1 = \Delta_2 = \Delta_3 \equiv \Delta = \frac{\sigma + \tau -1}3 \;.
\ee
To simplify even further we will take $\sigma = \tau$, which means $2\tau - 3\Delta = 1$. The corresponding index
\be
\cJ(q) = \Tr\, \left[ e^{-\pi i r} \, e^{-\beta\{\cQ, \bar\cQ \}} \, e^{2\pi i \tau ( J_1 + J_2 + r)} \right]
\ee
is periodic under $\tau \to \tau +3$, and it is defined for $\im(\tau) >0$.

As in Section~\ref{sec: large N} we are interested in the contribution%
\footnote{The contribution from the T-transformed solutions $\{1,N,r\}$ was already analyzed in \cite{Benini:2018ywd}.}
from HL solutions $\{m,n,r\}$, in a large $N$ limit in which $m,r$ are kept fixed while $n \to \infty$. This means applying the restriction (\ref{equal chemical potentials}) to the large $N$ expressions (\ref{large N result case 1}) and (\ref{large N result case 2}) in the first and second region of parameter space, respectively. One easily computes
\be
[m\Delta]_{\check\tau} = \begin{cases} \text{undefined} &\text{if } m+2r = 0 \mod 3 \\[1em] \dfrac{2\check\tau-1}3 &\text{if } m+2r=1 \mod 3 \\[1em] \dfrac{2\check\tau-2}3 &\text{if } m+2r = 2 \mod 3 \;. \end{cases}
\ee
The undefined case arises because for these chemical potentials $m\Delta$ happens to sit precisely on one of the images of $\ccL_{\check\tau}$ under integer shifts. For $m+2r = 1$ mod $3$ the solution sits in the 1$^\text{st}$ case of parameter space, according to (\ref{def 1st 2nd case}), while for $m+2r=2$ mod $3$ it sits in the 2$^\text{nd}$ case. It follows that the leading behavior of the index is
\be
\label{large N index special case final}
\log \left( \cI_{\{m,n,r\}} \right) \simeq \begin{cases}
\text{undefined} \qquad &\text{if } m + 2r = 0 \mod 3 \\[1em]
- \dfrac{\pi i N^2}{27 m} \, \dfrac{(2\check\tau-1)^3}{\check\tau^2} &\text{if } m+2r=1 \mod 3 \qquad (\text{\first{} case}) \\[1em]
- \dfrac{\pi i N^2}{27 m} \, \dfrac{(2\check\tau+1)^3}{\check\tau^2} &\text{if } m+2r=2 \mod 3 \qquad (\text{\second{} case}) \;.
\end{cases}
\ee
The undefined case should be interpreted as the fact that there are two exponentially-large competing contributions to $\cI_{\{m,n,r\}}$ (coming from the first and second case solutions when we slightly shift the chemical potentials in one direction or the other) with the same absolute value, but with a relative phase that oscillates very rapidly.


\section{Five-dimensional black holes}
\label{sec: black holes 5d}

In this section we review the gravitational solutions that contribute to the  $S^1 \times S^3$ partition function of the $\cN=4$ SYM theory, for supersymmetric values of the chemical potentials, and compare their contributions to what we obtained from the index.
Since we work at large $N$ on the field theory side, the gravitational side is semi-classical and its partition function should be a sum over all the classical solutions that satisfy the appropriate boundary conditions for our chemical potentials.%
\footnote{With perturbative corrections given by the fluctuations around them, which we will not consider in this work, and with non-perturbative corrections that we will discuss in the next section.}
Our field theory computation is independent of the 't~Hooft coupling constant; on the gravity side it will be useful to work at large 't~Hooft coupling, where we can approximate type IIB string theory by type IIB supergravity,%
\footnote{The field theory analysis tells us that the result should be the same for any value of the 't~Hooft coupling, and thus it should be independent of the string scale, but we do not know how to see this directly on the gravity side.}
and just look for appropriate solutions of this theory, which are asymptotically AdS$_5\times S^5$.

The rules of the AdS/CFT correspondence tell us how to find gravitational solutions when the chemical potentials in \eqref{part_func} correspond to real boundary conditions for the bulk fields in Euclidean space. Unfortunately, as discussed below, for generic values of the chemical potentials this is not consistent with \eqref{susy_cond}, so for most supersymmetric configurations some of the chemical potentials are complex, and we do not know what are the rules for constructing solutions dual to such complex chemical potentials. We will begin by dealing with this issue in the most naive possible way, by assuming that one can analytically continue the results computed for real fields (with real boundary conditions) to complex fields (with complex boundary conditions).%
\footnote{More precisely, the results are computed for a real metric, but for Euclidean charged black holes there is an imaginary gauge field. This still leads to a real action.}
We will find that this will lead to some problems, and we will suggest how to resolve them in the following sections.

In general, finding ten-dimensional black hole solutions is complicated.
Luckily, type IIB supergravity has several known consistent truncations, with known black hole solutions that satisfy our boundary conditions. The most general black hole solution has two angular momenta $J_1, J_2$ and three electric charges $Q_1, Q_2, Q_3$, where we defined $Q_a = R_a/2$. The simplest consistent truncation is to 5d $\cN=2$ minimal gauged supergravity, which contains just a single gauge field, and which can be used to describe solutions with $Q_1=Q_2=Q_3 \equiv Q$ (or, equivalently, with equal chemical potentials $\Phi_1 = \Phi_2 = \Phi_3$). In this section we use this truncation to consider these solutions, further limiting ourselves, for simplicity, to the case of equal angular momenta $J_1 = J_2 \equiv J$ (or, equivalently, $\Omega_1=\Omega_2$).
A detailed discussion of generalizations to black holes with more generic charges (or chemical potentials) is left to Appendices~\ref{app: Two angular momenta} and \ref{app: Three charges}. The analysis of this section mostly follows \cite{Cabo-Bizet:2018ehj}.

\subsection{Black hole solutions}

The bosonic fields of 5d $\cN=2$ minimal gauged supergravity are the metric $g_{\mu\nu}$ and a $U(1)$ gauge field $A_\mu$. The bosonic Lagrangian is%
\footnote{The Lagrangian used in \cite{Cvetic:2004hs, Chong:2005hr} is related to the one used here by $\lambda_\text{CLP} = - g^2$ and $A_\text{CLP} = \frac2{\sqrt3} A$. The normalization used here has the advantage that $A$ couples canonically to the superconformal R-symmetry current, however recall that the R-charges of scalar gauge-invariant operators are multiples of $2/3$.
\label{foo: gauge normalization}}
\bea
\label{5d action}
\cL &= \bigl( R + 12g^2 \bigr) *1 - \frac23 \, F \wedge *F + \frac8{27} \, F \wedge F \wedge A \;,
\eea
such that the radius of the vacuum AdS$_5$ solution is $1/g$. In the following we will set the dimensionful constant $g$ to 1. Specializing to the case of equal angular momenta $a=b$, the black hole solutions found in \cite{Cvetic:2004hs} in asymptotically global AdS$_5$ (with boundary $\mathbb{R}\times S^3$) are given by
\begin{align}
	\label{BH metric 1}
	ds^2 &= - \frac{(1 + r^2)}{\Xi_a}dt^2 - \frac{2q}{\Xi_a \, \rho^2} \nu\, \varpi + \frac{f_t}{\Xi_a^2 \, \rho^4} \varpi^2 + \frac{\rho^2}{\Delta_r} dr^2 + \frac{\rho^2}{\Xi_a} \Big( d\theta^2 + \sin^2(\theta) \, d\phi^2 + \cos^2 (\theta) \, d\psi^2 \Big) \nn \\
	A &= \frac{3q}{2\rho^2 \, \Xi_a} \varpi - \alpha \, dt \;,
\end{align}
where we defined
\bea
\label{BH metric 2}
\nu &= a \, \bigl( \sin^2(\theta) \, d\phi + \cos^2(\theta) \, d\psi \bigr) \;, \qquad\qquad\qquad\qquad & \varpi &= dt - \nu \;, \\
\Delta_r &= \frac{\rho^4 (1 + r^2) + q^2 + 2a^2q}{r^2} - 2m \;, & \rho^2 &= r^2 + a^2 \;, \\
f_t &= 2(m+a^2 q )\rho^2 - q^2 \;, & \Xi_a &= 1 - a^2 \;.
\eea
The coordinates $\phi, \psi$ have period $2\pi$ (with the fermions anti-periodic under this shift), while $\theta \in [0,\pi/2]$. The solution depends on three parameters $(a,q,m)$ (and on $\alpha$, which is arbitrary in Lorentzian signature).  

The black hole has an outer horizon at the largest positive root of $\Delta_r(r_+) = 0$, denoted by $r_+$. One can then solve for $m$ and find
\be
\label{from m to r+}
m = \frac{(r_+^2 + a^2)^2 \, (1+ r_+^2) + q^2 + 2a^2q }{ 2r_+^2} \;,
\ee
which can be used to eliminate $m$ and use $r_+$ instead.

The horizon is a Killing horizon generated by the Killing vector field
\be
\label{null Killing vector}
V = \parfrac{}{t} + \Omega \, \parfrac{}{\phi} + \Omega \, \parfrac{}{\psi} \;,
\ee
where
\be
\label{Omega}
\Omega = \frac{a \, (r_+^2 + a^2)(1 + r_+^2) + aq}{(r_+^2 + a^2)^2 + a^2q} \;.
\ee
Evaluating the surface gravity, the Hawking temperature is
\be
\label{Temperature}
T \equiv \frac1\beta = \frac{r_+^4 \big[ 1 + 2 (r_+^2 + a^2) \big] - (a^2+q)^2}{2\pi r_+ \bigl[ (r_+^2+a^2)^2 + a^2q\bigr]} \;.
\ee
The electrostatic potential at the horizon is
\be
\label{Phi}
\Phi \,\equiv\, \imath_V A \big|_{r_+} - \imath_V A \big|_\infty = \frac{ 3q r_+^2}{2 \bigl[ (r_+^2 + a^2)^2 + a^2q \bigr]} \;.
\ee

For our purposes we would like to continue these solutions to Euclidean signature and to compactify the Euclidean time direction, such that the boundary is $S^1_\beta\times S^3$. Following \cite{Cabo-Bizet:2018ehj}, these solutions can be analytically continued to real Euclidean solutions by rotating $t \to -i t_\mathrm{E}$, if we also take $a \to i \tilde a$. In terms of the original $a$, which is now purely imaginary, and the real-valued $m$ and $q$, one finds that the Euclidean metric is real, $A$ is purely imaginary (if $\alpha \in \bR$), $\imath_V A$ is real, $\beta$ is real, and $\Phi$ is real, while $\Omega$ is purely imaginary. We assume that we can rotate the integration over the gauge fields from real to imaginary fields without encountering any problems.

\paragraph{Expansion near the horizon.}
In the Euclidean solutions we need to periodically identify the Euclidean time direction $t_\mathrm{E}$, and make sure that the solutions are smooth when this cycle shrinks at the horizon. We expand the metric around $r \gtrsim r_+$, setting $R^2 = r - r_+$. The near-horizon metric takes the following form:
\begin{multline}
	\label{near-horizon expansion}
	ds^2 = h_{RR} \Bigl( dR^2 + R^2 \bigl( \tfrac{2\pi}{\beta} \, dt_\mrE \bigr)^2 \Bigr) + h_{\theta\theta} \, d\theta^2 + h_{\phi\phi} \bigl( d\phi + i \Omega\, dt_\mrE \bigr)^2 + h_{\psi\psi} \bigl( d\psi + i\Omega\, dt_\mrE \bigr)^2 \\
	+ 2h_{\phi\psi} \bigl( d\phi + i \Omega\, dt_\mrE \bigr) \bigl( d\psi + i\Omega\, dt_\mrE \bigr) \;.
\end{multline}
Here $h_{RR}$, $h_{\theta\theta}$, $h_{\phi\phi}$, $h_{\psi\psi}$, $h_{\phi\psi}$, $\Omega$ are functions of $R$ and $\theta$ with non-vanishing limits for $R \to 0$, which depend on the parameters $a,q,r_+$. The limit of $\Omega$ is the value given in (\ref{Omega}).

We see that the geometry is smooth around $r_+$ if we make the following identifications of the coordinates:
\be
\label{twisted identifications}
(t_\mathrm{E}, \phi, \psi) \,\cong\, \bigl( t_\mathrm{E} + \beta,\; \phi - i \Omega \beta,\; \psi - i \Omega \beta \bigr) \,\cong\, (t_\mathrm{E}, \phi + 2\pi, \psi) \,\cong\, (t_\mathrm{E}, \phi, \psi+2\pi) \;.
\ee
Moreover, the Killing vector field $V$ in (\ref{null Killing vector}) is the one that generates rotations of the circle that shrinks. The identifications \eqref{twisted identifications} take the same form as \eqref{shifted ident}, so we anticipate that the chemical potentials for the angular momenta will be realized, in our coordinate system for the black hole solutions, by modifying the coordinate identifications rather than by changing the CFT metric.

Regularity of the gauge field at $r=r_+$ requires (in a gauge that is regular at the horizon)
\be
\label{regularity condition gauge field}
\imath_V A \big|_{r_+} = 0 \qquad\Rightarrow\qquad \alpha = \Phi \;,
\ee
because the Wilson line around the shrinking circle should be trivial at $r=r_+$.

Now that we have smooth Euclidean solutions with the identification \eqref{twisted identifications}, we can compute their on-shell action.
As usual in holography, this requires a regularization for large $r$, and the computation was done in \cite{Cabo-Bizet:2018ehj} using background subtraction%
\footnote{The on-shell action was computed more carefully using holographic renormalization in \cite{Cassani:2019mms}, and was found to have an extra constant term $I_0 = \beta E_0 = \tfrac{3\pi\beta}{32g^2G_5}$ where $I_0$ is the action of thermal AdS$_5$.
\label{foo: gravitational casimir energy}}
(which assigns vanishing action to AdS$_5$). It has the form
\be \label{on_shell_action}
I_\text{SUGRA} = \frac{\pi \beta}{4 \, \Xi_a^2} \biggl[ m - (r_+^2 + a^2)^2 - \frac{ q^2 \, r_+^2 }{ (r_+^2 + a^2)^2 + a^2q } \biggr] \;.
\ee
Note that $I_\mathrm{SUGRA}$ satisfies the so-called ``quantum statistical relation''
\be
\label{quantum statistical relation}
I_\mathrm{SUGRA} = \beta E - S - 2 \beta \Omega J - \beta \Phi Q \;,
\ee
where $E$ is the energy and $S$ is the entropy of the black hole.

As mentioned above, for supersymmetric solutions we generally need the action for complex values of the chemical potentials, and we assume that we can extend the results above by analyticity to the full space of complex parameters $(a, q, r_+)$. As discussed in Section~\ref{sphere_part_func}, in the continued solutions $\phi$ and $\psi$ are complex, but their tilded versions \eqref{tildephi} obey standard identifications and remain real (but with a complex metric). A priori it is not clear whether the path integral with complex chemical potentials indeed receives contributions from such analytically-continued bulk solutions or not, and we will return to this issue below.

\paragraph{Expansion near the boundary.}
In order to precisely identify which chemical potentials the Euclidean solutions above correspond to, we need to analyze the behavior of the solutions near the boundary. The asymptotic $r\to\infty$ limit of the metric (\ref{BH metric 1}) can be written as
\be
ds^2 = \frac{d\hat r^2}{\hat r^2} + \hat r^2 \, ds^2_\text{bdry} + \cO(\hat r^0) \;,
\ee
where $\Xi_a \, \hat r^2 = r^2$ and
\be
ds^2_\text{bdry} = dt_\mathrm{E}^2 + d\theta^2 + \sin^2(\theta) \, d\phi^2 + \cos^2 (\theta) \, d\psi^2
\ee
is the round metric on $S^1_{\beta} \times S^3$, but still with the twisted identifications \eqref{twisted identifications}.
So, as expected from the discussion of Section~\ref{sphere_part_func} on the ``coordinate shift realization'' of the chemical potentials, the CFT metric remains the same, and the parameter $\Omega$ affects the boundary only through the identifications \eqref{twisted identifications}. In addition, we find that the gauge field near the boundary $r\to \infty$ is
\be \label{boundary_gauge}
A_\text{bdry} = i\Phi\, dt_\mathrm{E} \;,
\ee
corresponding to a chemical potential for the $U(1)$ global symmetry.

\paragraph{Shifted chemical potentials.} Note that shifts of $\Omega$ by $\tfrac{2\pi i}{\beta} \, \bZ$ do not change the global identification:
\begin{align}
\label{eq:shifted chemical potentials identification}
(t_\mathrm{E}, \phi, \psi) &\,\cong\, \Bigl( t_\mathrm{E} + \beta,\; \phi - i \bigl( \Omega + \tfrac{2\pi i \bZ}{\beta} \bigr) \beta,\; \psi - i \bigl( \Omega + \tfrac{2\pi i \bZ}{\beta} \bigr) \beta \Bigr) \\ 
&\,=\, (t_\mathrm{E} + \beta,\, \phi - i\Omega \beta + 2\pi \bZ,\, \psi - i\Omega \beta + 2\pi \bZ)
\,\cong\, (t_\mathrm{E}+\beta,\, \phi - i\Omega \beta,\, \psi - i\Omega \beta) \;, \nn
\end{align}
and thus they do not change the behavior of our solutions near the boundary. Thus, we actually found a family of black hole backgrounds with the same boundary metric and topology, but with different bulk metrics (corresponding to the shifted values of $\Omega$). Noting that solutions with angular momenta $J_1$ and $J_2$ have wavefunctions $e^{-i J_1 \phi - i J_2 \psi}$ and thus acquire a factor $e^{J_1 \Omega \beta + J_2 \Omega \beta}$ under the identification \eqref{eq:shifted chemical potentials identification}, and comparing to \eqref{part_func}, we see that our black hole configurations contribute to the partition function with
\be
\Omega_1=\Omega_2=\Omega \;.
\ee
The CFT partition function itself is periodic under shifts of $\Omega$ by  $\tfrac{2\pi i}{\beta}$, and this arises on the gravity side from summing over all these shifted solutions.

Similarly, if we consider a configuration with some charges $R_1, R_2, R_3$, recalling that the gauge field couples to $r=(R_1+R_2+R_3)/3$, the boundary condition \eqref{boundary_gauge} gives us a factor of $e^{\Phi \beta (R_1 + R_2 + R_3) / 3}$ when we go around the circle. So these black holes contribute to the partition function \eqref{part_func} with
\be
\Phi_1=\Phi_2=\Phi_3= \frac23 \, \Phi \;.
\ee
Note that since scalar operators in the bulk (which are periodic under the shift around the Euclidean circle) have R-charges which are multiples of $2/3$, shifts of $\Phi$ by $\tfrac{3\pi i}{\beta}\, \bZ$ correspond to large gauge transformations near the boundary (this is consistent also with the behavior of fermions, which are anti-periodic under the shift). 
So, as in our discussion above of the chemical potentials for angular momenta, configurations that differ by these shifts all contribute to the same partition function \eqref{part_func}.

All in all, if we consider the partition function \eqref{part_func} with these values of $\Omega_i$ and $\Phi_a$, it gets contributions from all the solutions with $\Omega = \Omega_1 + \tfrac{2\pi i}{\beta} \bZ$ and with \mbox{$\Phi = \frac{3}{2}\Phi_1 + \tfrac{3\pi i}{\beta} \bZ$}. In Appendices~\ref{app: Two angular momenta} and \ref{app: Three charges} we consider more general bulk solutions, with not necessarily equal angular momenta or $U(1)$ charges (and corresponding chemical potentials). In this context, contributions to \eqref{part_func} for specific values (equal or not) of the chemical potentials come from a 5-parameter infinite series of gravitational solutions, in which each of $\Omega_1$, $\Omega_2$, $\Phi_1$, $\Phi_2$ and $\Phi_3$ may be independently shifted by $\tfrac{2\pi i}{\beta} \bZ$.
Assuming that all of these complexified solutions contribute to the partition function, in the supergravity approximation we can write
\begin{multline}
Z_{S^1\times S^3}(\Omega_1, \Omega_2, \Phi_1, \Phi_2, \Phi_3) = {} \\
{} = \sum_{n_1, n_2, m_1, m_2, m_3 \,\in\, \bZ} e^{- I_\mathrm{SUGRA} \bigl( \Omega_1+\frac{2\pi i}{\beta} n_1,\; \Omega_2+\frac{2\pi i}{\beta} n_2 ,\; \Phi_1 + \frac{2\pi i}{\beta} m_1 ,\; \Phi_2 + \frac{2\pi i}{\beta} m_2 ,\; \Phi_3+\frac{2\pi i}{\beta}  m_3 \bigr)} \;,
\end{multline}
where the sum runs over all integers $n_1,n_2,m_1,m_2,m_3$ whose sum is even, in order for the periodicity to be consistent also with the behavior of fermions in the bulk, as in the QFT discussion around \eqref{shift symmetry}.

\subsection{Supersymmetry}

The solutions described above, as well as their analytic continuations to complex parameters, are supersymmetric if their parameters are related as
\be
\label{SUSY condition}
q = \frac{m}{1+2a} \;.
\ee
This gives a two-parameter family of solutions in terms of $(a,m)$. In terms of the equivalent set of parameters $(a,r_+)$, the supersymmetry condition becomes
\be
q = -a^2 + (1+2a) \, r_+^2 \mp i r_+ (r_+^2 - r_*^2) \;,
\ee
where $r_*^2 = 2a+a^2$ and the $\mp$ signs correspond to the two branches of a square root. For the upper sign, the expression above equals
\be
\label{first SUSY branch}
q = -(a-ir_+)^2(1-i r_+) \;.
\ee
We refer to the corresponding family of SUSY solutions as the first branch. On the other hand, for the lower sign the expression is obtained by sending $i \to -i$, and we call it the second branch. In the following we will focus on the first SUSY branch (\ref{first SUSY branch}), but we will sometimes mention also the results for the second branch.

In the supersymmetric cases, parametrized by $(a, r_+)$, the chemical potentials become
\bea
\label{SUSY chemical potentials}
\Phi &= \frac{3i r_+(1-i r_+)}{2(r_*^2 + i r_+)} \;,\qquad\qquad\qquad \beta = \frac{2\pi i a (a-ir_+)(r_*^2 + i r_+) }{ (r_*^2 - 3iar_+)(r_*^2 - r_+^2) } \;,\\[.5em]
\Omega &= \frac{ (r_*^2 + iar_+)(1-i r_+) }{ (r_*^2+ir_+)(a-ir_+) } \;.
\eea
They satisfy
\be
\label{SUSY relation chemical potentials}
\beta \, \bigl( 1 + 2\Omega - 2 \Phi \bigr) = 2\pi i \;.
\ee
In the Euclidean solutions there is a globally-well-defined Killing spinor, which changes sign when going around the Euclidean time cycle, whenever we have $(\pm 2\pi i)$ on the right-hand side. This is consistent with (\ref{SUSY relation chemical potentials}) and its counterpart on the second branch. Note that except in the special case $\Phi=\frac{1}{2}$, we cannot have $\Phi$ and $\beta$ real and also $\Omega$ imaginary in supersymmetric solutions.

In order to relate to the index, following the discussion of Section~\ref{part_to_index}, we can map the chemical potentials of our gravitational configurations to index parameters using \eqref{param_relation} (on both branches), namely%
\footnote{Recall that $\Phi_1 = \Phi_2 = \Phi_3 = \frac23 \Phi$, justifying the slightly different definition of $\Delta_\mg$ with respect to \eqref{param_relation}.}
\be
\label{def chemical potentials both branches}
\sigma_\mg = \tau_\mg = \frac{\beta \, (\Omega-1)}{2\pi i} \;, \qquad\qquad \Delta_\mg = \frac{\beta\, (\Phi-\frac{3}{2})}{3\pi i} \;.
\ee
Such a gravitational configuration contributes to the index with $\Delta_1 = \Delta_2 = \Delta_3 = \Delta_\mg$.
The index parameters on the first branch are given by
\be
\label{susyvalues}
\sigma_\mg = \tau_\mg = \frac{a \, (1-a)}{r_*^2 - 3ia r_+} \;,\qquad\qquad \Delta_\mg = - \frac{a \, (a-ir_+) }{ r_*^2 - 3iar_+} \;.
\ee
Note that they satisfy
\be \label{tau_delta_relation}
2\tau_\mg - 3\Delta_\mg = 1 \qquad\qquad\text{(on 1$^\text{st}$ branch)} \;.
\ee
Thus, one combination of $(a,r_+)$ controls $\tau_\mg,\Delta_\mg$ while another one controls $\beta$. On the second SUSY branch, the chemical potentials satisfy $2\tau_\mg - 3\Delta_\mg = - 1$. The on-shell action \eqref{on_shell_action} becomes
\be
\label{SUSY on-shell action}
I_\mathrm{SUGRA} = - \frac{i\pi^2}2 \, \frac{ a \, (a-ir_+)^3 }{ (1-a)^2(r_*^2 - 3ia r_+)} = \frac{i \pi^2}{2} \, \frac{\Delta_\mg^3}{\tau_\mg^2} \;.
\ee
The rightmost expression is valid on both branches. Notice that it does not depend on $\beta$.

\subsection{Comparison to field theory results} 

We can rewrite the on-shell action (\ref{SUSY on-shell action}) in terms of the field theory variables. We should reinstate dimensions using the relation (at leading order in $1/N$)
\be
\frac1{g^3 G_5} = \frac{8c}{\pi} = \frac{2N^2}\pi \;,
\ee
where $a=c$ is the field theory central charge, while $G_5$ is the five-dimensional Newton constant. Thus the logarithm of the classical contribution of these solutions to the partition function is (in the first branch)
\be
\label{on-shell action basic solution}
\log (Z) = - I_\mathrm{SUGRA} = - \frac1{g^3 G_5} \, \frac{i\pi^2}{2} \, \frac{\Delta_\mg^3}{\tau_\mg^2} = - \pi i N^2 \, \frac{\Delta_\mg^3}{\tau_\mg^2} = - \frac{\pi i N^2}{27} \, \frac{(2\tau_\mg-1)^3}{\tau_\mg^2} \;.
\ee
This is exactly the contribution \eqref{large N index special case final} of the basic Bethe Ansatz solution $\{1,N,0\}$ that falls in the first case. Note that, indeed, \eqref{tau_delta_relation} ensures that our parameters always satisfy the field theory conditions for this first case, namely $\Delta_\mg=[\Delta_\mg]_{\tau_\mg}$. 

Similarly, the value of $\Delta_\mg$ that we obtain on the second branch of solutions, which is $\Delta_\mg = (2\tau_\mg + 1) / 3$, always satisfies the condition for the second case in our analysis of the basic Bethe ansatz solution, $\Delta_\mg = [\Delta_\mg]'_{\tau_\mg}$, and also the action of our solution is consistent with this case,
\be
\log (Z) = - I_\mathrm{SUGRA} =  - \pi i N^2 \, \frac{\Delta_\mg^3}{\tau_\mg^2} = - \frac{\pi i N^2}{27} \, \frac{(2\tau_\mg+1)^3}{\tau_\mg^2}\,,
\ee
which coincides with the contribution \eqref{large N index special case final} of the solution $\{1,N,2\}$ (if we make the identification $\tau_\mg = \check\tau = \tau+2$).

So, the gravity results agree nicely with the basic Bethe ansatz solution. Moreover, recall from our discussion above that shifts of $\Omega$ and $\Phi$, which correspond to shifts of $\tau_\mg$ and $\Delta_\mg$ by integers, give new solutions which also contribute to the same index. However, we need to still satisfy the constraint \eqref{tau_delta_relation}. So, if we start (for instance) from some value of $\tau_\mg$ and $\Delta_\mg$ which satisfies \eqref{tau_delta_relation} on the first branch, we can shift $\tau_\mg \to \tau_\mg + 3n$ and $\Delta_\mg \to \Delta_\mg + 2n$ and obtain another solution contributing to the same index; it is easy to see that the contribution of this is precisely that of the shifted Bethe Ansatz solution $\{1,N,3n\}$ --- which is in the ``first case''.%
\footnote{Note that $n$ here can be positive or negative. In our analysis of shifted Bethe Ansatz solutions we took $r=0,\dots,N-1$ and assumed that $r$ does not scale with $N$. However, our analysis there works equally well for negative $r$ which does not scale with $N$, by identifying it with $(N+r)$.}
Moreover, starting from the same values, if we shift $\tau_\mg \to \tau_\mg + 3n-1$, $\Delta_\mg \to \Delta_\mg + 2n$ we obtain a SUSY solution on the second branch which also contributes to the same index, and which precisely reproduces the contribution of the shifted Bethe Ansatz solution $\{1, N, 3n - 1\}$ --- which is in the ``second case''. So, the sum over gravitational solutions contributing to the same index precisely reproduces the sum over two thirds of the Bethe Ansatz solutions with $m=1$. The leftover solutions are problematic since they lie on Stokes lines, as discussed in Section~\ref{universal}, but as in that section, this problem does not arise for generic (unequal) chemical potentials.
 
 However, before we declare success in our matching with the Bethe Ansatz solutions with $m=1$ (the solutions with $m>1$ will be discussed later in Section~\ref{sec: Orbifolds}), we should be more careful. So far in this subsection we only considered shifts of the solutions which retain equality of the angular momenta and of the $U(1)$ charges. However, the most general solution has three different chemical potentials --- parameterized by $\Delta_{\mg,a} = \frac{\beta}{2\pi i} ( \Phi_a - 1 )$ --- and two different angular chemical potentials $\sigma_\mg$, $\tau_\mg$. Each of these can be shifted independently by integers without affecting the boundary conditions, so we should sum over all these additional solutions as well, if they are supersymmetric. The action of the more general solutions is%
 \footnote{This is shown for various cases in Appendices \ref{app: Two angular momenta} and \ref{app: Three charges}, and we assume here that the general form holds.}
\be \label{general SUGRA action}
I_\mathrm{SUGRA} \bigl( \sigma_\mg, \tau_\mg, \Delta_{\mg,1}, \Delta_{\mg,2}, \Delta_{\mg,3} \bigr) = \pi i N^2 \, \frac{\Delta_{\mg,1} \, \Delta_{\mg,2} \, \Delta_{\mg,3} }{ \sigma_\mg \, \tau_\mg } \;,
\ee
and they are supersymmetric whenever
\be
\label{eq: general constituent relation}
\sigma_\mg + \tau_\mg - \Delta_{\mg,1} - \Delta_{\mg,2} - \Delta_{\mg,3} = \pm 1 \,, \qquad\qquad\bigl( 1^\text{st}/2^\text{nd} \text{ branch} \bigr) \;.
\ee
So we have two four-parameter families of shifted solutions, that all seem to contribute to the index (even when we happen to evaluate it for $\sigma = \tau$ and $\Delta_1=\Delta_2 = \Delta_3$).
Namely, it appears that in the supersymmetric case we should have contributions from supergravity of the form
\begin{multline}
Z_{S^1\times S^3}(\sigma, \tau, \Delta_1, \Delta_2) \;\stackrel{?}{=} \; \\
\sum_{\substack{ n_1, n_2, m_1, m_2 \,\in\, \bZ \\[.2em] s \,\in\, \{+1, -1\} }} e^{-I_\mathrm{SUGRA} \bigl( \sigma+n_1,\; \tau+ n_2,\; \Delta_1 + m_1,\; \Delta_2 + m_2,\; \sigma + \tau - \Delta_1 - \Delta_2 - s + n_1 + n_2 - m_1 - m_2 \bigr)} \;.
\end{multline}
However, the only shifts that show up in the Bethe ansatz computation of the field theory index are the specific ones discussed above, related to the $\{1,N,r\}$ family.

One may wonder whether these new shifts correspond to as yet undiscovered BA solutions, or if they might cancel after appropriate resummation. This does not seem to be the case. Indeed, the contribution of some shifted backgrounds to the partition function actually diverges exponentially with the size of the shift, so they had better not contribute for some reason. Specifically, consider starting from the equal chemical potential case, and shift to $\Delta_{\mg,1} = \Delta_\mg + n$, $\Delta_{\mg,2} = \Delta_\mg + n$, $\Delta_{\mg,3} = \Delta_\mg - 2n$, consistently with \eqref{eq: general constituent relation}. We then have for large $|n|$:
\be
\re \bigl( I_\mathrm{SUGRA} \bigr) = \pi N^2 \, \im \left(\frac{2n^3}{\tau_\mg^2}\right)  + \cO(n^2) \;.
\ee
So, the contribution $e^{-I_\mathrm{SUGRA}}$ from these backgrounds to the partition function would diverge, either for very positive or for very negative $n$.

In the next sections we will suggest a resolution to the problem: that most of the shifted solutions (which are all complex valued) should not be included in the sum over solutions, because they are unstable towards the condensation of D3-branes, and that when evaluating the gravitational partition function, only stable contributions should be considered. Presumably this criterion can be justified by a careful analysis of the analytic continuation to complex-valued solutions --- in particular through the study of Lefschetz thimbles --- but we will not attempt to do this here. We will show below that with that criterion, the acceptable shifted solutions (for arbitrary chemical potentials) precisely match the $m=1$ shifted solutions that contribute to the index.

\section{Wrapped D3-branes}
\label{sec: Branes}

Up to now we considered the classical on-shell action of our Euclidean solutions. In general, the action receives quantum corrections coming from loops of the gravity fields in these backgrounds. However, we can also have additional non-perturbative corrections coming from wrapped D-branes. It turns out (see Appendix~\ref{app: D3-branes}) that wrapped Euclidean D3-branes can be added to our backgrounds while still preserving the same supersymmetry, and without changing the boundary conditions, so that they give non-perturbative corrections to the contributions discussed in the previous section. In this section we compute those corrections for a specific class of wrapped D3-branes, and analyze their consequences.

\subsection{Uplift to 10 dimensions}
\label{sec: uplift to 10d}

In order to analyze configurations with D-branes, we must first find the 10d solution that corresponds to our 5d black holes. Luckily, \cite{Cvetic:1999xp} discusses the embedding of 5d supergravity with $U(1)^3$ gauge symmetry --- the so-called STU model --- into 10d type IIB supergravity, and the uplift of solutions of the former into the latter. In our context we view the action \eqref{5d action} as the $U(1)^3$ action with all three gauge fields taken to be equal.%
\footnote{See also (2.11) of \cite{Cvetic:1999xp}. Comparing to \eqref{5d action} we see that $A_\text{there} = \frac{2}{\sqrt{3}} A_\text{here}$, and $A^{i}_\text{there} = \frac{2}{3} A_\text{here}$ ($i=1,2,3$). The scalar fields in the $U(1)^3$ supergravity are all trivial when the charges are equal.}
The uplift of solutions with unequal angular momenta or charges is described in Appendices~\ref{app: Two angular momenta} and \ref{app: Three charges}.

The 10d metric for the equal-charge solutions is given by \cite{Cvetic:1999xp}
\be \label{BH metric 10}
ds_{10}^2 = ds_5^2 + \sum_{a=1}^3 \biggl( d\mu_a^2 + \mu_a^2 \, \Bigl( d\phi_a + \frac{2}{3} A\Bigr)^2 \biggr) \;,
\ee
where $ds_5^2$ is the 5d solution \eqref{BH metric 1}, $A$ is the gauge field in \eqref{BH metric 1}, and we write the $S^5$ in a way that is natural when it is embedded in $\bC^3$ with coordinates $z_a = \mu_a \, e^{i \phi_a}$, obeying the constraint $\mu_1^2 + \mu_2^2 + \mu_3^2 = 1$, such that for $A=0$ we have in \eqref{BH metric 10} the round metric on $S^5$. Note that despite the mixing in the last term, the determinant of this metric is the product of the determinant of $ds_5^2$ times that of the $S^5$ metric.

The only other 10d field turned on is the self-dual 5-form flux, which is given by 
\be
F_{(5)} = G_{(5)} + *\, G_{(5)} \;,
\ee
with
\be \label{5form}
G_{(5)} = - 4 \epsilon_{(5)} + \frac{1}{3} \sum_{a=1}^3 d \bigl( \mu_a^2 \bigr) \wedge d\phi_a \wedge *_5 F \;,
\ee
where $F = dA$, $\epsilon_{(5)}$ is the volume form of $ds_5^2$, and $*_5$ is the 5d Hodge dual with respect to that metric. From \eqref{BH metric 1} we see that
\be
F = -\frac{3 q r}{\rho^4 (1-a^2)} \, dr \wedge (dt - \nu) - \frac{3q}{2 \rho^2 (1-a^2)} \, d\nu \;.
\ee

In order to evaluate the D3-brane action we need to write $F_{(5)} = d C_{(4)}$, where the potential $C_{(4)}$ is defined up to gauge transformations. In this case, the exact form of $C_{(4)}$ can be chosen to be
\bea
\label{cfour}
C_{(4)} &= \frac{2r\rho^{2}\cos(2\theta)}{\Xi_a^2} \, dt\wedge dr\wedge d\phi\wedge d\psi + \frac{1}{3}\sum_{a=1}^{3} \mu_a^2 \, d\phi_{a} \wedge \biggl( {*_5} F - \frac23 A \wedge F \biggr) \\
&\quad - \frac{1}{2} \biggl( \mu_{1}^{2} \, d \bigl( \mu_{2}^{2} \bigr) - \mu_{2}^{2} \, d \bigl( \mu_{1}^{2} \bigr) \biggr) \wedge \left( d\phi_{1}+\frac{2}{3}A \right)\wedge \left( d\phi_{2}+\frac{2}{3}A \right) \wedge \left( d\phi_{3}+\frac{2}{3}A \right) \\
&\quad + \frac{1}{3} d\bigl( \mu_{2}^{2} \bigr) \wedge d\phi_{2} \wedge d\phi_{3} \wedge A + \frac{1}{3} d \bigl( \mu_{1}^{2} \bigr) \wedge d\phi_{1}\wedge d\phi_{3}\wedge A \;.
\eea
Locally, we can write
\be
{*_5} F - \frac{2}{3}A\wedge F = d \alpha_{(2)}
\ee
since the left-hand side is closed thanks to the equation of motion of $A$, in terms of the following 2-form on 5d spacetime:
\be
\label{alpha_2 equal ang momenta}
\alpha_{(2)} = \frac{(3+2\alpha)q}{2\Xi_a \rho^2} \, dt \wedge \nu - \frac{3q\cos(2\theta)}{4\Xi_a^{2}} \Bigl( a \, dt \wedge (d\phi - d\psi) + d\phi \wedge d\psi \Bigr) \;.
\ee
The expression in (\ref{cfour}) is not smooth at the horizon, so one might need to add suitable total-derivative terms in order to fix that, without modifying the integrations over D3-brane worldvolumes that are discussed below. An alternative gauge-equivalent choice for $C_{(4)}$ is presented in (\ref{general 5flux 4potential}).

Note that in the 10d solution the 5d gauge fields become geometrical, and in the asymptotic behavior of the solutions \eqref{BH metric 10}, given the boundary condition \eqref{boundary_gauge}, the chemical potentials are realized using the off-diagonal components of the metric, while the angular coordinates $\phi_a$ still have the standard identifications. Thus, these chemical potentials (unlike the ones for the angular momenta) are realized in our solutions using the ``metric realization''. If desired, we can think of these coordinates as ``tilded coordinates'' as in \eqref{tildephi}, and define new coordinates
\be \label{hatphi}
{\hat \phi}_a = \phi_a - \frac{2i}3  \Phi  t_\mrE
\ee
which have the original metric appearing in their 10d boundary condition, but obey twisted identifications under $t_\mrE \to t_\mrE + \beta$.

\subsection{The brane action}
\label{sec: brane action}

Consider now a Euclidean D3-brane in the background \eqref{BH metric 10}, whose worldvolume wraps a maximal $S^3$ inside the $S^5$ given (say) by $\mu_1 = 0$, and a maximal $S^1$ inside the $S^3$ on the horizon in AdS$_5$ given (say) by $r=r_+$, $\theta = \pi/2$ (where the $\psi$ direction shrinks) such that the brane wraps the $\phi$ direction. We can consider such an embedding either in the Lorentzian or in the Euclidean solution; in the Euclidean solution some combination of the time circle and the $\phi$ circle shrinks at the horizon, but we can still take our brane to sit at a fixed time and wrap the $\phi$ direction. This configuration is supersymmetric, as we show in Appendix~\ref{app: D3-branes}.

The action of a D3-brane is given in general by
\be
\label{D3 action}
S_\mathrm{D3} = -\frac{1}{g_s (2\pi)^3 (\alpha')^2} \int \biggl( d^4x \, \sqrt{-\det (g_\mathrm{D3}) } \mp P[C_{(4)}] \biggr) \;,
\ee
where $g_\mathrm{D3}$ is the induced metric on the D3-brane worldvolume, $P[C_{(4)}]$ is the pull-back of the 4-form $C_{(4)}$ to the worldvolume, and the $\mp$ sign refers to a brane/anti-brane, depending on conventions. Note that in our case the AdS radius (which we previously denoted by $1/g$ and set to one) is given by 
\be
\frac{1}{g^4} = 4\pi g_s N (\alpha')^2 \;,
\ee
so the prefactor in \eqref{D3 action} is given by $N/2 \pi^2$.

Let us compute each of the two terms in \eqref{D3 action}, for generic (not necessarily supersymmetric) solutions. The second term is quite simple, as only the $\frac{1}{3} d\bigl( \mu_{2}^{2} \bigr) \wedge d\phi_{2}\wedge d\phi_{3}\wedge A$ term in \eqref{cfour} contributes in this configuration. It factorizes into the integral of $\frac{1}{3}A$ on the $S^1$ in AdS$_5$, and the integral of $d \bigl( \mu_{2}^{2} \bigr) \wedge d\phi_{2}\wedge d\phi_{3}$ on the $S^3$ in $S^5$. The first integral gives
\be \label{intofa}
\frac1{3} \int_{S^1} A = -\frac{q}{2(r_+^2+a^2) (1-a^2)} \int_{S^1} \nu = -\frac{\pi q a}{(r_+^2+a^2)(1-a^2)} \;,
\ee
while the second integral just gives $4\pi^2$. 

Even though the metric induced on the D3-brane from \eqref{BH metric 10} is not diagonal between the $S^1$ and the $S^3$ in the D3-brane worldvolume, its form implies that the determinant of the induced metric on the D3-brane is the product of the determinant of the $S^3$ metric (coming from the $S^5$ coordinates) and of $g_{\phi \phi}$ in \eqref{BH metric 1} (not including the extra contribution to this from the second term in \eqref{BH metric 10}), as we discuss around equation \eqref{eq: detg 3 charges}. The integral of the first determinant is just ${\rm Vol(S^3)} = 2\pi^2$. On the other hand, the component $g_{\phi\phi}$ on the D3-brane evaluates to
\be
g_{\phi\phi} = \biggl[ \frac{ (r_+^2 + a^2)^2 + a^2q }{ r_+ (r_+^2 + a^2)(1-a^2) } \biggr]^2 \;.
\ee
Therefore, choosing a specific sign for the square root,
\be
\label{DBI part of D3 action}
\int d^4x\, \sqrt{-\det(g_\mathrm{D3})} = - 4\pi ^3 i \, \frac{ (r_+^2 + a^2)^2 + a^2q }{ r_+ (r_+^2 + a^2) (1-a^2) } \;.
\ee
The sign is chosen in such a way that in the background of a real and causally-well-behaved Lorentzian black hole, for which $a,q$ are real (with $a^2 <1$) and $r_+$ is sufficiently large, the contribution from (\ref{DBI part of D3 action}) to the path-integral measure $\exp(iS_\text{D3})$ in bounded in absolute value. Notice that the Wick rotation $t \to -it_\mathrm{E}$ of the black hole metric has no effect on the Euclidean D3-brane action, because such a brane does not wrap the time direction. The full D3-brane action is
\be
\label{D3 on-shell action general}
S_\text{D3} = - \frac{N}{2\pi^2} \biggl[ - 4\pi^3 i \, \frac{ (r_+^2 + a^2)^2 + a^2q }{ r_+ (r_+^2 + a^2) (1-a^2) } \pm 4\pi^3 \, \frac{qa}{(r_+^2 + a^2)(1-a^2)} \biggr]
\ee
where, as before, the $\pm$ sign refers to a brane/anti-brane. 

The D3-brane action considerably simplifies when the black hole background is supersymmetric; as we show in Appendix~\ref{app: D3-branes}, this is because the branes preserve all supersymmetries of the background. On the first branch where the SUSY condition (\ref{first SUSY branch}) is obeyed, and choosing the upper plus sign in (\ref{D3 on-shell action general}), we obtain
\be
\label{d3action}
S_\mathrm{D3} = 2\pi N \, \frac{a-ir_+}{a-1} = 2\pi N \, \frac{\Delta_\mg}{\tau_\mg} \;,
\ee
where in the last step we used the expressions \eqref{susyvalues} for the chemical potentials.

We have computed the on-shell action of similar branes in the case of two different angular momenta, as reported in Appendix~\ref{app: Two angular momenta}, and of three different charges, as reported in Appendix~\ref{app: Three charges}. The result leads us to assume that in the general case of chemical potentials $(\sigma_\mg, \tau_\mg, \Delta_{\mg,a})$, there are supersymmetric D3-branes that wrap either the $\phi$ or the $\psi$ circle in AdS$_5$ (and sit at $\theta=\frac\pi2$ or $\theta = 0$, respectively) and an $S^3$ inside $S^5$ for which $\mu_a = 0$ ($a=1,2,3$), and that their actions are
\be 
\label{eq: general brane action}
S_\mathrm{D3}^\phi = 2\pi N \, \frac{\Delta_{\mg,a}}{\sigma_\mg} \;, \qquad\qquad S_\mathrm{D3}^\psi = 2\pi N \, \frac{\Delta_{\mg,a}}{\tau_\mg} \;, \qquad\qquad {\rm (\first{} \ branch)} \;.
\ee 

On the second branch of solutions, the parameters of the background obey \eqref{first SUSY branch} but with $i \to -i$. It turns out that the on-shell action (\ref{D3 on-shell action general}) simplifies when choosing the lower minus sign. This is because the branes that are supersymmetric in the two branches of solutions must have opposite charge (corresponding to swapping branes with anti-branes). This also follows from the analysis of supersymmetry in Appendix~\ref{app: D3-branes}. Repeating the same computation as above we then find
\be 
\label{eq: general brane action2}
S_\mathrm{D3}^\phi = - 2\pi N \, \frac{\Delta_{\mg,a}}{\sigma_\mg} \;,\qquad\quad S_\mathrm{D3}^\psi = -2\pi N \, \frac{\Delta_{\mg,a}}{\tau_\mg} \;, \qquad\quad {\rm (\second{} \ branch)} \;.
\ee

The contribution of the Euclidean D3-brane configurations to the partition function is \mbox{$\exp\bigl( - I_\mathrm{SUGRA} + i S_\mathrm{D3} \bigr)$} where $I_\text{SUGRA}$ is the on-shell action of the background black hole solution. Note the relative factor of $i$ that appears in our Euclidean solutions since the branes do not extend in the time direction, and therefore they do not get a factor of $i$ from the Wick rotation to Euclidean time, as noticed earlier. Similarly, we can consider branes wrapping any positive integer linear combinations of these cycles, and we would have in the exponent the sum of the corresponding D3-brane actions.

The analysis above makes sense when the D3-branes decrease the contribution of the solution to the partition function, since otherwise considering an arbitrary number of D3-branes will increase their contribution without a bound.
We interpret the latter cases as unstable, perhaps to condensation of these D3-branes that would take us to some other background.
Thus, we consider a gravitational background to be stable only if \emph{every} D3-brane satisfies%
\footnote{Here we are referring to the D3-branes described above, that wrap an $S^3$ in $S^5$ and an $S^1$ in the AdS$_5$ coordinates. We have found additional SUSY D3-branes that wrap an $S^1$ in $S^5$ and the whole $S^3$ at the horizon in the AdS$_5$ coordinates, and such branes can induce additional instabilities (see (\ref{new D3-branes action general}) for their on-shell action). Their interpretation will be discussed elsewhere \cite{ABMMinprogress}.}
\be 
\label{eq: stability condition}
\im (S_\mathrm{D3}) > 0 \;,
\ee
and we suggest that only these backgrounds should be included in the analysis. Note that the equal-charge backgrounds happen to always satisfy this condition for all their D3-branes, so they are always stable, but generic solutions, including solutions related to the equal-charge case by shifts, do not. We will show in the next section that for these stable solutions the D3-brane contributions precisely match with the non-perturbative corrections to the Bethe Ansatz solutions computed in Section~\ref{non_pert}, and we will analyze the precise implications of the stability condition \eqref{eq: stability condition}.

Note that for real supersymmetric solutions we expect the action to be bounded from below, so that finding D3-branes that decrease the action would be impossible. However, once we continue to complex solutions this is no longer the case, which is why we have to impose the condition \eqref{eq: stability condition} for every D3-brane. A similar condition on complex configurations was recently suggested in a similar context in \cite{Copetti:2020dil}.

\section{Stable gravity solutions and the index}
\label{comparison}

Let us now find the stable gravitational backgrounds that contribute to the partition function with boundary conditions given by chemical potentials $(\sigma = \tau,\, \Delta_{1,2,3})$, so that we can compare them to the field theory analysis of Section~\ref{sec: Field theory analysis}. The gravitational solutions have chemical potentials $(\sigma_\mg, \tau_\mg, \Delta_{\mg,a})$ ($a=1,2,3$) which could be any integer shift of the ones labeling the index, as long as they satisfy the SUSY constraint \eqref{eq: general constituent relation}. Recall that the index is parameterized by $\sigma, \tau, \Delta_1, \Delta_2$, and that we defined in Section~\ref{sec: Field theory analysis} an auxiliary chemical potential $\Delta_3$ such that $\Delta_1+\Delta_2+\Delta_3-\sigma-\tau$ is an integer; from the gravity point of view, $\Delta_{\mg,3}$ (defined from the third chemical potential) always satisfies this condition (on both branches of supersymmetric solutions) thanks to \eqref{eq: general constituent relation}, so we can identify also $\Delta_{\mg,3}$ on the gravity side with the index parameter defined in Section~\ref{sec: Field theory analysis} (up to some integer shift).

We shall start by considering gravitational backgrounds in the first branch. Consider the stability condition \eqref{eq: stability condition} arising from having the union of three branes, one at $\mu_1 = 0$, one at $\mu_2 = 0$, and one at $\mu_3 = 0$, all wrapping the $\phi$ circle, and from a similar configuration which wraps the $\psi$ circle instead. The action is the sum of the individual brane actions \eqref{eq: general brane action}, so using \eqref{eq: general constituent relation} on the first branch, the two conditions are
\be
\label{stab_cond}
\im \left(\frac{\tau_\mg - 1}{\sigma_\mg}\right) > 0 \;, \qquad\qquad  \im \left(\frac{\sigma_\mg - 1}{\tau_\mg}\right) > 0 \;.
\ee
Now, remember that the gravitational chemical potentials are just integer shifts of the ones in the CFT, so they share the same imaginary part. Using this we can rewrite \eqref{stab_cond} as
\be \label{stab_cond2}
\im (\tau) > \im (\tau) \, \re \bigl(\tau_g - \sigma_g \bigr) > - \im (\tau) \;.
\ee
Recalling that $\tau_g-\sigma_g$ is an integer, this can be satisfied only if $\tau_g=\sigma_g$.
Thus, the only stable gravitational backgrounds have $\tau_g = \sigma_g$ in the bulk as well.%
\footnote{For $\tau_g-\sigma_g = \pm 1$ it seems from \eqref{stab_cond2} that we are on the boundary of the region of stability. However, recalling that this condition comes from the sum of the conditions for stability of three different D3-branes, for generic values of the chemical potentials at least one of these three D3-branes would lead to an instability.}

Next, consider the stability conditions arising from a single brane wrapping the $\psi$ cycle at $\mu_a = 0$, and from the union of two branes wrapping the other two cycles, $\mu_b = 0$ and $\mu_c = 0$ with $a,b,c$ all different. Using \eqref{eq: general constituent relation}, the two conditions are
\be
\im \left(\frac{\Delta_{\mg,a}}{\tau_\mg}\right) > 0 \;, \qquad\qquad \im \left(\frac{-1 - \Delta_{\mg,a}}{\tau_\mg}\right) > 0 \;,
\ee
and hence
\be 
\im \left(-\frac{1}{\tau_\mg}\right) > \im \left(\frac{\Delta_{\mg,a}}{\tau_\mg}\right) > 0 \;.
\ee
We see that we only have a stable solution when $\Delta_{\mg,a} = [\Delta_{\mg,a}]_{\tau_\mg}$ ($a=1,2,3$), and not for any other shift of the $\Delta_a$. Recall from the analysis of Section~\ref{sec: Field theory analysis} that this condition can only be satisfied if the values of the chemical potentials are in the ``first case''. Thus, for any specific value of $\tau_\mg$, if the chemical potentials satisfy the conditions of the ``first case'' then we have a single stable solution among all possible shifts of the electric chemical potentials, and this solution obeys $\Delta_{\mg,a} = [\Delta_{\mg,a}]_{\tau_\mg}$ such that its action \eqref{general SUGRA action} precisely reproduces the contribution of the basic Bethe Ansatz solution for $\tau = \tau_\mg$. On the other hand, if the chemical potentials satisfy the ``second case'', we do not find any shifted solution within the first branch that is stable for this $\tau_\mg$.

If we now consider gravitational solutions on the second branch of \eqref{eq: general constituent relation}, then an analogous stability analysis implies that the only stable bulk solutions have $\tau_\mg = \sigma_\mg$ and $\Delta_{\mg,a} = [\Delta_{\mg,a}]'_{\tau_\mg}$. So if the chemical potentials satisfy the ``second case'' condition we find here a single shifted solution (with this $\tau_\mg$), and we can check that the action of this solution \eqref{general SUGRA action} precisely reproduces 
the contribution of the basic Bethe Ansatz solution for the second case with $\tau=\tau_\mg$. On the other hand, if the chemical potentials satisfy the ``first case'' condition we do not find any stable solutions within the second branch.

The bottom line is that for each value of $\tau_\mg$, and of $\Delta_a$ (up to integer shifts), we find precisely one stable gravitational solution (on either the first branch or the second branch), and the gravitational action precisely reproduces the value of the basic Bethe Ansatz solution for $\tau=\tau_\mg$ (which satisfies the ``first case'' or the ``second case'' condition, respectively). Now, recalling that $\tau_\mg$ can take any value of the form $\tau_\mg = \tau + r$ for any integer $r$, we recognize that the contribution of that $\tau_\mg$ is exactly that of the $\{1,N,r\}$ solution to the Bethe Ansatz equations with chemical potential $\tau$. Thus, we have a one-to-one correspondence between the stable shifted solutions on the gravity side, and the $m=1$ shifted solutions to the Bethe Ansatz equations, with an exact match of the leading action between the two sides.

Moreover, given the values of the $\Delta_{\mg,a}$'s that we found, the form of the exponentials in the non-perturbative corrections that we found on the gravity side --- of the form $e^{i S_\mathrm{D3}}$ with $S_\mathrm{D3}$ given by \eqref{eq: general brane action} and \eqref{eq: general brane action2} --- precisely match with the exponentials that appeared in the corrections to the corresponding Bethe Ansatz solutions in Section~\ref{non_pert} (in each of the two cases). So our matching of the gravity side to the index extends also to these non-perturbative corrections. It would be interesting to match also the coefficients in front of the exponentials in the various non-perturbative corrections between the two sides, but this lies beyond the scope of this paper.

\section{Orbifolds and \matht{m>1} solutions}
\label{sec: Orbifolds}

So far we have found a precise match between stable gravitational black hole solutions and the $\{m,n,r\}$ Hong-Liu solutions to the Bethe Ansatz equations with $m=1$. It is interesting to ask whether we can find gravitational solutions that will agree with the $m>1$ BA solutions. In this section we show that we can construct orbifold configurations that precisely agree with these solutions (when $m$ remains finite in the large $N$ limit).

A clue to finding these solutions is the simple relation between the leading-order action of the $m>1$ solutions and that of the basic $\{1,N,0\}$ solution; they are related by taking $\tau \to \check\tau \equiv m\tau + r$, $\Delta_a \to m \Delta_a$, and dividing the action by $m$. 

Consider the Euclidean black hole solutions with equal charges and angular momenta, described in Section~\ref{sec: black holes 5d}. In 5d, the Euclidean solution with inverse temperature $\tilde\beta$ and chemical potentials $\wt\Omega$ and $\wt\Phi$ involves an identification of the coordinates by
\be
\label{twisted identifications2}
(t_\mathrm{E}, \phi, \psi) \,\cong\, \bigl( t_\mathrm{E} + \tilde\beta,\; \phi - i \wt\Omega \tilde\beta,\; \psi - i \wt\Omega \tilde\beta) \;,
\ee
and a gauge field given near the boundary by $A_{\rm bdry} = i \wt\Phi \,dt_\mathrm{E}$, such that the holonomy around the cycle \eqref{twisted identifications2} at infinity is $i \wt\Phi \tilde\beta$. From the ten-dimensional point of view, as discussed above, in our conventions the identifications of the $S^5$ coordinates are not shifted, so we have
\be
\label{twisted identifications3}
(t_\mathrm{E}, \phi, \psi, \phi_1, \phi_2, \phi_3) \,\cong\, \Bigl( t_\mathrm{E} + \tilde\beta,\; \phi - i \wt\Omega \tilde\beta,\; \psi - i \wt\Omega \tilde\beta,\; \phi_1,\; \phi_2,\; \phi_3 \Bigr) \;,
\ee
with off-diagonal metric components proportional at infinity to $\frac{2}{3} i \wt\Phi$.
These can alternatively be described using the hatted coordinates $\hat\phi_a$ in \eqref{hatphi} with an unmodified metric at infinity, but with identifications where $\hat\phi_a$ is shifted by $-\frac{2}{3} i \wt\Phi \tilde\beta$.
The generalization to different charges and/or angular momenta is straightforward, and the resulting identification on the hatted coordinates is:
\be
\label{twisted identifications4}
(t_\mathrm{E}, \phi, \psi, \hat\phi_1, \hat\phi_2, \hat\phi_3) \,\cong\, \Bigl( t_\mathrm{E} + \tilde\beta,\; \phi - i \wt\Omega_1 \tilde\beta,\; \psi - i \wt\Omega_2 \tilde\beta,\; \hat\phi_1 - i \wt\Phi_1 \tilde\beta,\; \hat\phi_2- i \wt\Phi_2 \tilde\beta,\; \hat\phi_3- i \wt\Phi_3\tilde\beta\Bigr) \;.
\ee
On the other hand, in the $(\tilde\phi,\tilde\psi,\phi_1,\phi_2,\phi_3)$ coordinates --- see (\ref{tildephi}) --- there is no shift under the identification \eqref{twisted identifications4} (such that these coordinates are real even when the chemical potentials are complex), and the chemical potentials are all realized through off-diagonal terms in the behavior of the metric near the boundary.

Now, let us take this 10d background (starting from the case of equal charges and angular momenta, for simplicity) and perform on it a $\bZ_m$ orbifold by identifying points related by
\begin{multline}
\label{twisted identificationsm}
(t_\mathrm{E}, \phi, \psi, \hat\phi_1, \hat\phi_2, \hat\phi_3) \,\cong\, 
\biggl( t_\mathrm{E} + \frac{\tilde\beta}{m},\; \phi - i \wt\Omega \frac{\tilde\beta}{m} - \frac{2\pi r}{m},\; \psi - i \wt\Omega \frac{\tilde\beta}{m} - \frac{2\pi r}{m}, \\
\hat\phi_1 - \frac{2i}{3} \wt\Phi \frac{\tilde\beta}{m} - \frac{2\pi s}{m},\; \hat\phi_2 - \frac{2i}{3} \wt\Phi \frac{\tilde\beta}{m} - \frac{2\pi s}{m},\; \hat\phi_3 - \frac{2i}{3} \wt\Phi \frac{\tilde\beta}{m} - \frac{2\pi s}{m} \biggr) \;,
\end{multline}
for some integers $r=0,\dots,m-1$ and $s=0,\dots,2m-1$ (note that we allowed a larger range for $s$ here, since shifting the three angles $\hat\phi_a$ by $2\pi$ acts as the identity on bosons but gives a minus sign to fermions, so it is a non-trivial operation). Performing this identification $m$ times brings us back to the identification \eqref{twisted identifications4}, so (given that our solutions are all independent of $t_\mrE$ and of the angular coordinates) this is indeed a $\bZ_m$ isometry of our original backgrounds, that we can orbifold by in string theory. While in the hatted coordinates it is not obvious that the shifts in \eqref{twisted identificationsm} are consistent with the ranges of the coordinates when the chemical potentials are complex, if we use the real coordinates $(\tilde\phi,\tilde\psi,\phi_1,\phi_2,\phi_3)$ we find that only the $\frac{2\pi r}{m}$ and $\frac{2\pi s}{m}$ terms in the shifts remain, consistent with these coordinates being real:
\be
\label{twisted identificationsmm}
(t_\mathrm{E}, \tilde\phi, \tilde\psi, \phi_1, \phi_2, \phi_3) \,\cong\, 
\biggl( t_\mathrm{E} + \frac{\tilde\beta}{m},\; \tilde\phi - \frac{2\pi r}{m},\; \tilde\psi - \frac{2\pi r}{m},\;
\phi_1 - \frac{2\pi s}{m},\; \phi_2 - \frac{2\pi s}{m},\; \phi_3 - \frac{2\pi s}{m} \biggr) \;.
\ee
Thus, \eqref{twisted identificationsm} is a consistent orbifold in string theory (we will discuss its fixed points below).

If we consider the behavior of the orbifolds near the boundary, we see that the identifications \eqref{twisted identificationsm} that we perform in the orbifold background take exactly the same form as our original identifications on some other black hole solutions \eqref{twisted identifications3} (with tilded quantities substituted by untilded ones, and in terms of hatted coordinates $\hat\phi_a$ as in \eqref{twisted identifications4}), if we identify
\be
\beta = \frac{\tilde\beta}{m} \;, \qquad\qquad \Omega = \wt\Omega - \frac{2 \pi i r}{\tilde \beta} \;, \qquad\qquad \Phi = \wt\Phi - \frac{3\pi i s}{\tilde\beta} \;.
\ee
Thus, the orbifold background contributes to the sphere partition function with these values of $\beta$, $\Omega$ and $\Phi$.
Note that the new background has
\bea
\tau_\mg &= \frac{\beta \, (\Omega-1)}{2\pi i} = \frac{\tilde\beta \, \bigl( \wt\Omega - \frac{2\pi i r}{\tilde\beta} - 1 \bigr)}{2\pi i m} = \frac{1}{m} \, {\tilde\tau}_\mg - \frac{r}{m} \;,\\[.5em]
\Delta_\mg &= \frac{\beta \, \bigl( \Phi - \frac{3}{2} \bigr)}{3 \pi i} = \frac{\tilde\beta \, \bigl( \wt\Phi - \frac{3\pi i s}{\tilde\beta} - \frac{3}{2} \bigr)}{3\pi i m} = \frac{1}{m} \, {\wt\Delta}_\mg - \frac{s}{m} \;,
\eea
which is equivalent to
\be
{\tilde\tau}_\mg = m \tau_\mg + r \qquad\qquad\text{and}\qquad\qquad {\wt\Delta}_\mg = m \Delta_\mg + s \;.
\ee

General orbifolds of this type will not preserve supersymmetry even if the original background does; the condition for preserving supersymmetry is that the Killing spinor should be anti-periodic under the new identification \eqref{twisted identificationsm} (note that the same Killing spinor can be used as in the original background). This happens precisely when $2\tau_\mg - 3 \Delta_\mg$ is an odd integer, \ie, when
\be
2r - 3s = m + (2\tilde\tau_\mg - 3 \wt\Delta_\mg)\qquad\qquad ({\rm mod}\ 2m) \;.
\ee
For any value of $r$ we can find a unique $s$ such that this is satisfied%
\footnote{More precisely, this is true when $m$ is not a multiple of $3$. In the more general case of unequal chemical potentials discussed below, this restriction does not arise.}
(moreover, recall that the ``parent'' black hole solution satisfies $2\tilde\tau_\mg - 3 \wt\Delta_\mg = \pm 1$).
When this condition is satisfied, the $\bZ_m$ orbifold of the black hole background with parameters $\bigl( \tilde\beta, \tilde\tau_\mg, \wt\Delta_\mg \bigr)$ contributes to the index with these parameters $(\beta, \tau_\mg, \Delta_\mg)$, since it has the same boundary conditions as the other backgrounds contributing to this index. 

The integral of the supergravity action density over the orbifolded background is $1/m$ times the action of the original black hole background, due to the $\bZ_m$ identification. Thus, we conclude that in the classical gravity approximation, the contributions of these orbifold backgrounds to the logarithm of the index with parameters $(\beta, \tau_\mg, \Delta_\mg)$ are precisely $1/m$ times those of a black hole with parameters $\bigl( m\beta,\, m \tau_\mg + r,\, m \Delta_\mg + s \bigr)$. This is precisely the same as what we found for the Hong-Liu BA solutions with parameters $\{m, N/m, r\}$ (recall that the leading-order contribution from the BA solutions is invariant under shifting $(m \Delta_g)$ by an integer). Thus, we have an exact match also for these solutions.
Note that the quantization condition for the 5-form flux on the orbifolded background is satisfied only when $N/m$ is an integer, so that exactly the same values of $m$ are allowed in the field theory and on the gravity side.

Consider next the non-perturbative contributions from D3-branes wrapped on the orbifolded background. Recall that our D3-branes sit at the horizon and wrap, say, the $\phi, \phi_2, \phi_3$ directions, together with an extra coordinate on the $S^5$. The position of the brane is such that the orbifold acts as a $\bZ_m$ shift within the worldvolume of the D3-brane, and thus its action is $1/m$ times the action of the D3-brane on the original background. On the first branch this was given by $2\pi N {\wt\Delta}_\mg / {\tilde\tau}_\mg$, so we find that on the orbifolded background the D3-brane action is given by $S_\text{D3} = 2\pi N (m \Delta_\mg + s) / m (m \tau_\mg + r)$. This agrees with the non-perturbative corrections found in Section~\ref{non_pert} if $m \Delta_\mg + s = [m \Delta_\mg]_{m\tau_\mg + r}$.

Of course, in the orbifolded backgrounds we still have the freedom of shifting the various chemical potentials by integers, and we need to perform a similar stability analysis to that of the previous section. And moreover, the generalization of the analysis above to different chemical potentials is straightforward, but we then get separate parameters $r_1$, $r_2$ for the shifts of $\phi$ and $\psi$ in \eqref{twisted identificationsm}, and separate parameters $s_1$ and $s_2$ for the shifts of $\phi_1$ and $\phi_2$ there (all in the range $0,\dots,m-1$), while the shift $s_3$ of $\phi_3$ (in the range $0,\dots,2m-1$) is uniquely determined by the supersymmetry condition
\be
\label{SUSY_cond}
r_1 + r_2 - s_1 - s_2 - s_3 = m + (2\tilde\tau_\mg - 3 \wt\Delta_\mg)\qquad\qquad ({\rm mod}\ 2m) \;.
\ee
So we need to generalize the analysis of Section~\ref{comparison} to allow for all these shifts.
The generalization is straightforward and it implies, for instance, that on the first branch the only stable solutions are those obeying $m \Delta_{\mg,a} + s_a = [m \Delta_{\mg,a}]_{m\tau_\mg + r}$ for all $a=1,2,3$, and that they obey $\tau_\mg = \sigma_\mg$. As in the $m=1$ case, this leads to a precise agreement between the stable solutions on the gravity side and the BA solutions, both for the leading-order contribution to the logarithm of the index, and for the form of the non-perturbative corrections discussed in the previous paragraph.

Finally, let us describe the fixed points of these general orbifolds, in which the identifications may be written as
\be
\label{twisted identificationsmmm}
(t_\mathrm{E}, \tilde\phi, \tilde\psi, \phi_1, \phi_2, \phi_3) \,\cong\, 
\biggl( t_\mathrm{E} + \frac{\tilde\beta}{m},\; \tilde\phi - \frac{2\pi r_1}{m},\; \tilde\psi - \frac{2\pi r_2}{m},\;
\phi_1 - \frac{2\pi s_1}{m},\; \phi_2 - \frac{2\pi s_2}{m},\; \phi_3 - \frac{2\pi s_3}{m} \biggr) \;.
\ee
It is clear that there are no fixed points away from the horizon. At the horizon we have one coordinate shrinking to zero size, and the remaining coordinates $\tilde\phi$ and $\tilde\psi$ form (together with $\theta$) a round $S^3$, while the $S^5$ coordinates $\phi_a$ and $\mu_a$ form a round $S^5$ (recall that our gauge fields in 5d supergravity vanish at the horizon). Generically all the angular coordinates are shifted by the orbifold. However, if (say) $r_1$ and one of the $s_a$ vanish, then the manifold where $r = r_+$, $\theta=\pi/2$ (such that the $\tilde\psi$ circle shrinks) and where $\mu_a=1$ (such that the two circles of the coordinates $\phi_b$ with $b\neq a$, that are shifted, shrink as well) is fixed under the orbifold action \eqref{twisted identificationsmmm}. This is a two dimensional manifold (parameterized by $\tilde\phi$ and $\phi_a$), and near this fixed manifold the orbifold acts on the transverse space as a supersymmetric $\bC^4/\bZ_m$ orbifold.

Similarly, if three of the $(r_1,r_2,s_1,s_2,s_3)$ vanish (including at least one of the $r_i$) we have a four-dimensional space of fixed points (with a transverse $\bC^3/\bZ_m$ orbifold action), and if four of them vanish we have a six-dimensional space of fixed points (with a transverse $\bC^2/\bZ_m$ orbifold action).%
\footnote{Note that in these cases it is possible that the $\bZ_m$ symmetry does not act within the worldvolume of the wrapped D3-branes. For instance, if $r_1=s_1=s_2=0$, it does not act in the worldvolume of a D3-brane wrapping the $\phi$, $\phi_1$ and $\phi_2$ directions, so its worldvolume action is not divided by $m$. However, in such cases the wrapped D3-brane sits at a fixed point of the $\bZ_m$ orbifold, and then one can construct also ``fractional D3-branes'' wrapping the same manifold whose action is smaller by a factor of $m$, such that the action of the minimal D3-brane still matches what we found from the Bethe Ansatz analysis.}
Note that due to \eqref{SUSY_cond}, it is not possible for all five of these numbers to vanish, consistent with the fact that a $\bC/\bZ_m$ orbifold cannot be supersymmetric.

When there are fixed points, the orbifolds have light twisted sector states living there, while otherwise all twisted sector states are heavy. Supersymmetry ensures that these twisted sector states do not lead to tachyonic instabilities. In principle the loops of the twisted sector states (whether they are light or heavy) contribute perturbative corrections to the supergravity action, and it would be interesting to verify that their contributions are consistent with the order $\cO(1)$ contributions that we found from the Bethe Ansatz approach.

Note that all of our orbifolds involve a shift action on the Euclidean time circle. Thus, they are not related to any AdS$_5\times S^5/\bZ_m$ backgrounds, and their Lorentzian interpretation is unclear.

\section*{Acknowledgements}

We thank C.~Closset, Z.~Komargodski, A.~Sharon and E.~Urbach for useful discussions. We thank the organizers and participants of the SCGP seminar series on ``Supersymmetric black holes, holography and microstate counting'' for interesting discussions and for inviting us to present our work there.
The work of O.A. and O.M. was supported in part  by an Israel Science Foundation center for excellence grant (grant number 2289/18), by grant no. 2018068 from the United States-Israel Binational Science Foundation (BSF), and by the Minerva foundation with funding from the Federal German Ministry for Education and Research. O.A. is the Samuel Sebba Professorial Chair of Pure and Applied Physics. F.B. is partially supported by the MIUR-PRIN contract 2015 MP2CX4, and by the INFN ``Iniziativa Specifica ST\&FI''. F.B. is also supported by the MIUR-SIR grant RBSI1471GJ, and by the ERC-COG grant NP-QFT no.~864583. E.M. is supported by the Israel Science Foundation under grant No. 2289/18, and by the I-CORE Program of the Planning and Budgeting Committee.


\appendix


\section{Special functions}
\label{app: functions}

We will use fugacities and chemical potentials related by
\be
z = e^{2\pi i u} \;,\qquad\qquad p = e^{2\pi i \sigma} \;,\qquad\qquad q = e^{2\pi i \tau} \;.
\ee

\paragraph{\matht{q}-Pochhammer symbol.} We use the standard notations
\be
(z; q)_n = \prod_{k=0}^{n-1} (1-zq^k) , \qquad\qquad (z; q)_\infty = \prod_{k=0}^\infty (1-zq^k) \qquad\qquad\text{for } |q|<1 \;.
\ee
The expression $(z;q)_\infty$ admits the series expansion and the plethystic representation
\be
(z;q)_\infty = \sum_{n=0}^\infty \frac{(-1)^n \, q^{\frac{n(n-1)}2} }{ (q;q)_n} \, z^n
= \exp\left[ - \sum_{k=1}^\infty \frac1k \, \frac{z^k}{1-q^k} \right] \;,
\ee
respectively, where the first one converges on the whole domain $|q|<1$, while the second one converges for $|z|, |q|<1$. Noticing the relation 
\be
\eta(\tau) = e^{\frac{\pi i \tau}{12}} (q;q)_\infty
\ee
with the Dedekind eta function, one obtains the modular transformation properties of the $q$-Pochhammer symbol:
\be
\label{q-Pochhammer modularity}
(\tilde q; \tilde q)_\infty = \sqrt{-i\tau} \, e^{\frac{\pi i}{12} (\tau + 1/\tau)} \, (q;q)_\infty
\ee
where $\tilde q = e^{-2\pi i/\tau}$. The square root is taken with the principal determination, recalling that $\im(\tau)>0$. Finally, we have the asymptotic behaviors
\be
(z;q)_\infty \sim (1-z) \qquad\text{for } q \to 0 \;;\qquad\qquad \log\bigl[ (z;q)_\infty \bigr] \sim - \frac z{1-q} \qquad\text{for } z \to 0 \;,
\ee
where $f \sim g$ means that $\lim f/g = 1$.

\paragraph{Function \matht{\theta_0}.} The elliptic theta function is defined as
\be
\theta_0(u;\tau) = (z;q)_\infty (q/z; q)_\infty = \prod_{k=0}^\infty (1-zq^k)(1-z^{-1}q^{k+1}) \;.
\ee
This gives an analytic function on $|q|<1$ with simple zeros at $z = q^k$ for $k\in \bZ$ and no singularities. The infinite product is convergent on the whole domain. We can also give a plethystic definition
\be
\label{plethystic theta0}
\theta_0(u; \tau) = \exp \left[ - \sum_{k=1}^\infty \frac1k \, \frac{z^k + (q/z)^k}{1-q^k} \right]
\ee
which converges for $|q|<|z|<1$.

The periodicity relations are
\bea
\label{theta0 periodicities}
& \theta_0 \bigl(u+n+m\tau; \tau \bigr) = (-1)^m \, e^{-2\pi i m u - \pi i m(m-1)\tau} \, \theta_0(u; \tau) \qquad\qquad\text{for } n,m\in \bZ \\
& \theta_0(u; \tau) = \theta_0(\tau - u; \tau) = - e^{2\pi i u} \, \theta_0(-u; \tau) \;.
\eea
The modular properties are
\be
\label{theta0 modularity}
\theta_0(u; \tau+1) = \theta_0(u; \tau) \;,\qquad \theta_0 \left( \frac u\tau; - \frac1\tau \right) = - i \, e^{\tfrac{\pi i}\tau \left( u^2 + u + \frac16 \right) - \pi i u + \frac{\pi i \tau}{6}} \, \theta_0(u; \tau) \;.
\ee

\paragraph{Function \matht{\psi}.} Define, for $\im (t) < 0$, the function
\be
\label{function psi definition}
\psi(t) = \exp \left[ t \log \bigl( 1-e^{-2\pi i t} \bigr) - \frac1{2\pi i} \text{Li}_2 (e^{-2\pi i t}) \right] 
= \exp \left[ - \sum_{\ell=1}^\infty \left( \frac t\ell + \frac1{2\pi i \, \ell^2} \right) e^{-2\pi i t \, \ell} \right] \;.
\ee
The branch of the logarithm is determined by its series expansion $\log(1-z) = - \sum_{\ell=1}^\infty z^\ell/ \ell$, whereas $\text{Li}_2(z) = \sum_{\ell=1}^\infty z^\ell / \ell^2$ is the dilogarithm. One can show that the branch cut discontinuities of the logarithm and the dilogarithm cancel in the definition of $\psi(t)$, therefore the latter extends to a meromorphic function on the whole complex plane. Some useful properties of $\psi(t)$ are:
\be
\psi(t) \, \psi(-t) = e^{-\pi i (t^2 - 1/6)} \;,\qquad\qquad \psi(t+n) = (1-e^{-2\pi i t})^n \, \psi(t) \qquad\text{for}\quad n \in \bZ \;.
\ee
In particular, from (\ref{function psi definition}), $\psi(0) = e^{\pi i / 12}$.

\paragraph{Function \matht{\wt\Gamma}.} The elliptic gamma function is defined, in terms of chemical potentials, as
\be
\wt\Gamma(u; \sigma, \tau) = \prod_{m,n=0}^\infty \frac{1- p^{m+1} q^{n+1}/z}{1-p^m q^n z} \;.
\ee
This definition gives a meromorphic single-valued function on $|p|, |q|<1$, with simple zeros at $z = p^{m+1}q^{n+1}$ and simple poles at $z = p^{-m} q^{-n}$ for $m,n \geq 0$. The infinite product is convergent on the whole domain. We can also give a plethystic definition
\be
\wt\Gamma(u; \sigma, \tau) = \exp\left[ \sum_{k=1}^\infty \frac1k \, \frac{z^k - (pq/z)^k}{(1-p^k)(1-q^k)} \right]
\ee
which converges for $|pq| < |z|< 1$.

The function satisfies the following periodicity relations:
\bea
\label{Gamma periodicities}
& \wt\Gamma(u; \sigma, \tau) = \wt\Gamma(u; \tau, \sigma) \\
& \wt\Gamma(u+1; \sigma, \tau) = \wt\Gamma(u; \sigma+1, \tau) = \wt\Gamma(u; \sigma, \tau+1) = \wt\Gamma(u; \sigma, \tau) \\
& \wt\Gamma(u + \sigma; \sigma, \tau) = \theta_0(u; \tau) \, \wt\Gamma(u; \sigma, \tau) \;,\qquad \wt\Gamma(u + \tau; \sigma, \tau) = \theta_0(u; \sigma) \, \wt\Gamma(u; \sigma, \tau) \;.
\eea
Moreover
\be
\label{Gamma product}
\wt\Gamma(u; \sigma, \tau) \, \wt\Gamma(\sigma + \tau - u; \sigma, \tau) = 1 \;.
\ee
The elliptic gamma function has $SL(3,\bZ)$ modular properties.
For $\sigma$, $\tau$, $\sigma/\tau$, $\sigma + \tau \in \bC\setminus\bR$ there is a ``modular formula'' \cite{Felder:1999}:
\be
\wt\Gamma(u; \sigma, \tau) = e^{-\pi i \cQ(u; \sigma, \tau)} \frac{\wt\Gamma\bigl( \frac u\tau; \frac\sigma\tau, - \frac1\tau \bigr) }{ \wt\Gamma\bigl( \frac{u-\tau}\sigma ; - \frac1\sigma, - \frac\tau\sigma \bigr) } = e^{-\pi i \cQ(u; \sigma, \tau)} \frac{\wt\Gamma\bigl( \frac u\sigma; - \frac1\sigma, \frac\tau\sigma \bigr) }{ \wt\Gamma\bigl( \frac{u-\sigma}\tau ; - \frac\sigma\tau, - \frac1\tau \bigr) }\,,
\ee
where $\cQ$ is the cubic polynomial
\be
\label{Q polynomial}
\cQ(u; \sigma, \tau) = \frac{u^3}{3\sigma\tau} - \frac{\sigma + \tau -1}{2\sigma\tau} u^2 + \frac{\sigma^2 + \tau^2 + 3\sigma\tau -3\sigma - 3\tau + 1}{6\sigma\tau} u + \frac{(\sigma + \tau -1)(\sigma + \tau - \sigma\tau)}{12\sigma\tau} \;.
\ee
In the degenerate case $\sigma = \tau$ the inversion formula above is not valid. For $u \in \bC \setminus (\bZ + \tau \bZ)$, however, there is a degenerate relation:
\be
\label{degenerate modular formula}
\wt\Gamma(u; \tau, \tau) = \frac{e^{-\pi i \cQ(u; \tau, \tau)} }{ \theta_0\bigl( \frac u\tau; - \frac1\tau \bigr) } \prod_{k=0}^\infty \frac{ \psi\bigl( \frac{k+1+u}\tau \bigr) }{ \psi\bigl( \frac{ k- u}\tau \bigr) } \;.
\ee
The polynomial $\cQ$ reduces to
\be
\label{Q polynomial degenerate}
\cQ (u; \tau,\tau) = \frac{(2u-2\tau+1) \bigl( 2u(u+1) - 2 \tau(2u+1) + \tau^2 \bigr) }{ 12\tau^2} \;.
\ee
The function $\psi$ is defined in (\ref{function psi definition}). Using
\be
\cQ(u+1; \tau,\tau) - \cQ(u; \tau,\tau) = \frac{(u+1)(u+1-2\tau)}{\tau^2} + \frac56 \;,
\ee
one can check that both sides of (\ref{degenerate modular formula}) are invariant under $u \to u+1$.

\paragraph{Function \matht{\cG}.} This function of $u, \Delta, \tau$ is defined as
\be
\cG(u, \Delta;\tau) = \frac1{2\pi i} \, \parfrac{}{u} \log \left(\frac{ \theta_0(\Delta-u; \tau) }{ \theta_0(\Delta + u; \tau)} \right) \;.
\ee
We can write the series expansion
\be
\label{G series expansion}
\cG(u,\Delta; \tau) = \sum_{\ell=1}^\infty \frac{ (z^\ell + z^{-\ell}) \bigl( y^\ell - (q/y)^\ell \bigr) }{ 1-q^\ell}
\ee
which converges for $|q|< |yz| < 1$ and $|q| < |y/z| < 1$, with $z = e^{2\pi i u}$ and $y = e^{2\pi i \Delta}$. Such a domain could be too restrictive; in that case, notice that $\cG$ can be written as the sum of two series, each one convergent in one of the two domains, respectively.

We have
\be
\cG(u, \Delta; \tau) = \cG(-u, \Delta; \tau) = - \cG(u, \tau - \Delta; \tau) \;.
\ee
We also have modular properties, which follow from the ones of $\theta_0$:
\be
\label{G modularity}
\cG(u,\Delta; \tau+1) = \cG(u,\Delta; \tau) \;,\qquad \cG \left( \frac u\tau, \frac\Delta\tau; - \frac1\tau \right) = \tau - 2\Delta -1 + \tau \, \cG(u,\Delta; \tau) \;.
\ee
The periodicity properties are
\bea
\cG(u, \Delta; \tau) &= \cG(u+1, \Delta; \tau) = \cG(u+\tau, \Delta; \tau) = \cG(u, \Delta+1; \tau) \\
\cG(u, \Delta+\tau; \tau) &= 2 + \cG(u, \Delta; \tau) \;.
\eea
In particular, $\cG$ is an elliptic function of $u$, and quasi-elliptic of $\Delta$.

\section{Contribution of Hong-Liu solutions}
\label{app: computation}

In this appendix we carry out the computations that lead to the contributions of the HL solutions to the superconformal index, reported in Section~\ref{sec: HL contrib}. We will first compute the quantity $\gamma_\Delta$, and then the Jacobian $H$.

We proceed to explicitly evaluate the sum in (\ref{def gamma Delta}) that defines the function $\gamma_\Delta$. As explained in the main text, the strategy is to expand various functions in a common domain of convergence, manipulate the expansions and then obtain exact expressions. By analytic continuation, the latter will be valid everywhere.

\subsection[The case \texorpdfstring{$\Delta \neq 0$}{\unichar{"0394} \unichar{"2260} 0}]{The case $\boldsymbol{\Delta \neq 0}$}
\label{app: case Delta neq 0}

Let us start with the generic case that $\Delta \neq 0$. For $|q|^2 < |y| < 1$, we use the plethystic expansion to evaluate the second summation in the second line of (\ref{def gamma Delta}):
\begin{multline}
n \sum_{j_1 \neq j_2}^{m-1} \log \left( \wt\Gamma\bigl( v_{j_1j_2} + \Delta; \tau, \tau \bigr) \right) = n \sum_{j_1 \neq j_2}^{m-1} \sum_{\ell=1}^\infty \frac1\ell \, \frac{ \left( \frac{\xi_{j_1}}{\xi_{j_2}} \right)^\ell y^\ell - \left( \frac{\xi_{j_2}}{\xi_{j_1}} \right)^\ell y^{-\ell} q^{2\ell} }{ (1-q^\ell)^2 } \\
= n \sum_{\ell=1}^\infty \frac1\ell \, \frac{y^\ell - y^{-\ell} q^{2\ell}}{ (1-q^\ell)^2} A_{\ell,m} = N \log \left( \frac{ \wt\Gamma( m\Delta; m\tau, m \tau ) }{ \wt\Gamma(\Delta; \tau, \tau) } \right) \;.
\end{multline}
Here we introduced
\be
A_{\ell,m} \equiv \sum_{j_1 \neq j_2}^{m-1} e^{2\pi i (j_1 - j_2)\ell/m} = \begin{cases} m^2 - m &\text{for } \ell = 0 \mod m \\ -m &\text{for } \ell \neq 0 \mod m \;. \end{cases}
\ee
We similarly expand the first summation:
\be
\label{expansion first term in a computation}
\sum_{j_1, j_2=0}^{m-1} \sum_{k_1 \neq k_2}^{n-1} \log \Bigl( \wt\Gamma \bigl( v_{j_1 j_2} + w_{k_1k_2} + \Delta; \tau, \tau\bigr) \Bigr) =
\sum_{j_1, j_2=0}^{m-1} \sum_{k_1 \neq k_2}^{n-1} \sum_{\ell=1}^\infty \frac{ \left( \frac{\xi_{j_1} \zeta_{k_1} }{ \xi_{j_2} \zeta_{k_2} } \right)^\ell \! y^\ell - \left( \frac{\xi_{j_2} \zeta_{k_2} }{ \xi_{j_1} \zeta_{k_1} } \right)^\ell \! y^{-\ell} q^{2\ell} }{ \ell \; (1-q^\ell)^2 } \;.
\ee
This expansion converges if
\be
0 < \im \bigl( v_{j_1j_2} + w_{k_1k_2} + \Delta \bigr) = \frac{k_1 - k_2}n \im(\tau) + \im(\Delta) < 2\im (\tau) \qquad\qquad\forall\; k_1 \neq k_2 \;.
\ee
Taking into account the ranges of $k_{1,2}$, this is the case if
\be
\frac{n-1}n \im(\tau) < \im(\Delta) < \frac{n+1}n \im (\tau) \;.
\ee
For any value of $n$ there exists a (small) domain of convergence that we can use to perform our manipulations.%
\footnote{Actually, since we are resumming over $j_1, j_2$ at fixed $k_1, k_2$, the domain of convergence is even larger.}
We find
\be
(\mathrm{\ref{expansion first term in a computation}}) = \sum_{k_1 \neq k_2}^{n-1} \sum_{\ell=1}^\infty \frac{ \left(\frac{ \zeta_{k_1} \, y}{ \zeta_{k_2} } \right)^\ell \! - \Bigl(\frac{ \zeta_{k_2} \, q^2}{ \zeta_{k_1} \, y} \Bigr)^\ell }{ \ell \; (1-q^\ell)^2 } \bigl( A_{\ell, m} + m \bigr)
= m \! \sum_{k_1 \neq k_2}^{n-1} \log\Bigl( \wt\Gamma\bigl( m w_{k_1k_2} + m\Delta; m\tau, m\tau \bigr) \Bigr) .
\ee

Putting the two terms together we obtain
\be
\label{intermediate expression}
\gamma_\Delta = m \sum_{k_1 \neq k_2}^{n-1} \log\left( \wt\Gamma\bigl( m w_{k_1k_2} + m\Delta; m\tau, m\tau \bigr) \right) + N \log \left( \frac{ \wt\Gamma( m\Delta; m\tau, m \tau ) }{ \wt\Gamma(\Delta; \tau, \tau) } \right) \;,
\ee
which, by analyticity, extends to the whole domain of definition of the functions. The sum in the first term on the right-hand side can be evaluated using the ``modular formula'' (\ref{degenerate modular formula}). Recall that $\wt\Gamma(u; \tau,\sigma)$ is invariant under integer shifts of $\tau,\sigma$, and thus we can shift $m\tau \to m\tau + r$. It is convenient to define $\check\tau \equiv m\tau + r$. We obtain
\begin{align}
& m \sum_{k_1 \neq k_2}^{n-1} \log\biggl[ \wt\Gamma \biggl( \frac{k_1 - k_2}n \check\tau + m\Delta; m\tau, m\tau \biggr) \biggr] = - \pi i m \sum_{k_1 \neq k_2}^{n-1} \cQ \biggl( \frac{k_1 - k_2}n \check\tau + m\Delta; \check\tau, \check\tau \biggr) \\
& \qquad - m \sum_{k_1 \neq k_2}^{n-1} \log \biggl[ \theta_0 \biggl( \frac{k_1 - k_2}n + \frac{m\Delta}{\check\tau}; - \frac1{\check\tau} \biggr) \biggr] + m \sum_{k_1 \neq k_2}^{n-1} \sum_{k=0}^\infty \log \left( \frac{ \psi \bigl( \frac{k+1 + m\Delta}{\check\tau} + \frac{k_1 - k_2}n \bigr) }{ \psi \bigl( \frac{ k-m\Delta}{\check\tau} + \frac{k_2 - k_1}n \bigr) } \right) \,. \nn
\end{align}
Here $\cQ$ is the cubic polynomial (\ref{Q polynomial degenerate}). Let us evaluate the sums in the second line. For the first sum we use the plethystic expansion (\ref{plethystic theta0}) of $\theta_0$, valid for $0 < \im \bigl( m\Delta/ \check\tau \bigr) < \im \bigl( - 1/\check\tau \bigr)$. It is convenient to define the variables
\be
\tilde\zeta_k = e^{2\pi i m w_k/\check\tau} = e^{2\pi i k/n} \;,\qquad\qquad \tilde y = e^{2\pi i m\Delta/ \check\tau} \;,\qquad\qquad \check q = e^{-2\pi i / \check\tau} \;.
\ee
Then
\begin{multline}
- m \sum_{k_1 \neq k_2}^{n-1} \log\biggl[ \theta_0 \biggl( \frac{k_1 - k_2}n + \frac{m\Delta}{\check\tau}; - \frac1{\check\tau} \biggr) \biggr]  = m \sum_{k_1 \neq k_2}^{n-1} \sum_{\ell =1}^\infty \frac1\ell \, \frac{ \Bigl( \frac{\tilde\zeta_{k_1}}{\tilde\zeta_{k_2}} \Bigr)^\ell \tilde y^\ell + \Bigl( \frac{ \tilde\zeta_{k_2} }{ \tilde\zeta_{k_1}} \Bigr)^\ell \tilde y^{-\ell} \check q^\ell }{ 1-\check q^\ell } \\
{} = m \sum_{\ell=1}^\infty \frac1\ell \, \frac{\tilde y^\ell + \tilde y^{-\ell} \check q^\ell }{ 1-\check q^\ell} \, A_{\ell, n} = N \log \left( \frac{ \theta_0 \bigl( \frac{m \Delta}{\check\tau}; - \frac1{\check\tau} \bigr) }{ \theta_0 \bigl( \frac{N\Delta}{\check\tau}; - \frac n{\check\tau} \bigr) } \right) \;.
\end{multline}
For the second sum, we use the expansion (\ref{function psi definition}) valid in the same range:
\begin{multline}
\label{piece 2}
m \sum_{k_1 \neq k_2}^{n-1} \sum_{k=0}^\infty \log \Biggl( \frac{ \psi \bigl( \frac{ k+1 + m\Delta}{\check\tau} + \frac{k_1 - k_2}n \bigr) }{ \psi \bigl( \frac{ k-m\Delta}{\check\tau} + \frac{k_2 - k_1}n \bigr) } \Biggr)
= m \sum_{k_1 \neq k_2}^{n-1} \sum_{k=0}^\infty \sum_{\ell=1}^\infty \left[ \frac1\ell \left( \frac{k - m\Delta}{\check\tau} \tilde y^\ell \right. \right. \\
\left. \left. - \frac{ k+1+m\Delta}{\check\tau} \Bigl( \frac{\check q}{\tilde y} \Bigr)^\ell \right)
+ \frac1\ell \left( \frac{k_2 - k_1}n + \frac1{2\pi i\, \ell} \right) \bigl( \tilde y^\ell - (\check q/\tilde y)^\ell \bigr) \right] \biggl( \frac{\tilde\zeta_{k_1}}{ \tilde\zeta_{k_2} } \biggr)^\ell \check q^{k\ell} \;.
\end{multline}
Then we use that the following sum vanishes:
\be
B_{\ell,n} = \sum_{k_1 \neq k_2}^{n-1} (k_1 - k_2) \, e^{2\pi i \ell(k_1 - k_2)/n} = 0 \;.
\ee
Therefore
\begin{multline}
(\mathrm{\ref{piece 2}}) = m \sum_{k=0}^\infty \sum_{\ell=1}^\infty \left[ \frac1\ell \left( \frac{k-m\Delta}{\check\tau} \tilde y^\ell - \frac{k+1+m\Delta}{\check\tau} (\check q/ \tilde y)^\ell \right) + \frac1{2\pi i \, \ell^2} \bigl( \tilde y^\ell - (\check q/ \tilde y)^\ell \bigr) \right] \check q^{k\ell} A_{\ell, n} \\
= \sum_{k=0}^\infty \left[ m \log \left( \frac{ \psi \bigl( \frac{ n (k+1+m\Delta)}{\check\tau} \bigr) }{ \psi \bigl( \frac{ n(k-m\Delta) }{ \check\tau} \bigr) } \right) - N \log \left( \frac{ \psi\bigl( \frac{k +1 + m\Delta}{\check\tau} \bigr) }{ \psi \bigl( \frac{ k- m\Delta}{\check\tau} \bigr) } \right) \right] \;.
\end{multline}

Finally we put all terms together. The expression simplifies using
\bea
N \log \left( \wt\Gamma(m\Delta; m\tau, m\tau) \right) &= - \pi i N \cQ\bigl( m\Delta; \check\tau, \check\tau \bigr) \\
&\quad - N \log \biggl[ \theta_0 \biggl( \frac{m\Delta}{\check\tau}; - \frac1{\check\tau} \biggr) \biggr] + N \sum_{k=0}^\infty \log \left( \frac{ \psi \bigl( \frac{k+1+m\Delta}{\check\tau} \bigr) }{ \psi\bigl( \frac{k - m\Delta}{\check\tau} \bigr) } \right) \;.
\eea
We obtain:
\bea
\gamma_\Delta &= -\pi i m \sum_{k_1, k_2 = 0}^{n-1} \cQ \left( \frac{k_1 - k_2}n \,\check\tau + m\Delta; \check\tau, \check\tau \right) - N \log \left( \wt\Gamma(\Delta; \tau, \tau) \right) \\
&\quad - N \log \biggl[ \theta_0 \biggl( \frac{N\Delta}{\check\tau}; - \frac{n}{\check\tau} \biggr) \biggr] + m \sum_{k=0}^\infty \log \left( \frac{ \psi\bigl( \frac{k+1+m\Delta}{\check\tau/n} \bigr) }{ \psi \bigl( \frac{ k-m\Delta}{\check\tau/n} \bigr) } \right) \;.
\eea
The sum over $k_1,k_2$ can be performed exactly:
\begin{multline}
-\pi i m \sum_{k_1, k_2=0}^{n-1} \cQ \left( \frac{ k_1 - k_2}n \, \check\tau + m \Delta; \check\tau, \check\tau \right) = - \frac{\pi i m}6 \bigl( \check\tau - m\Delta - \tfrac12 \bigr) \\
{} + \pi i \, \frac{N^2}m \, \frac{\bigl( \check\tau - m\Delta \bigr) \bigl( \check\tau - m\Delta - \frac12 \bigr) \bigl( \check\tau -m\Delta - 1 \bigr) }{ 3 \check\tau^2 } \;.
\end{multline}
We thus obtain the exact expression in (\ref{gamma_Delta resummed}).

\subsection[The case \texorpdfstring{$\Delta = 0$}{\unichar{"0394} = 0}]{The case $\boldsymbol{\Delta=0}$}
\label{app: case Delta = 0}

The case $\Delta=0$ requires a separate treatment. Using (\ref{theta0 periodicities}), (\ref{Gamma periodicities}) and (\ref{Gamma product}), we find
\begin{align}
\gamma_0 &= - \sum_{i\neq j}^N \log \bigl( \theta_0(u_{ij}; \tau) \bigr) \Big|_{(\ref{HL solution u_ij})} \\
&= - \sum_{j_1, j_2=0}^{m-1} \sum_{k_1 \neq k_2}^{n-1} \log \Bigl( \theta_0 \bigl( v_{j_1j_2} + w_{k_1k_2}; \tau \bigr) \Bigr) - n \sum_{j_1 \neq j_2}^{m-1} \log \Bigl( \theta_0( v_{j_1j_2}; \tau) \Bigr) \;. \nn
\end{align}
The sum in the second term on the second line is computed as follows:
\begin{align}
\label{piece 3}
& - n \sum_{j_1 \neq j_2}^{m-1} \log\Bigl( \theta_0( v_{j_1j_2}; \tau)\Bigr) = -n \sum_{j_1 \neq j_2}^{m-1} \left[ \log\left( 1 - \frac{\xi_{j_1}}{\xi_{j_2}} \right) + 2 \sum_{k=1}^\infty \log \left( 1 - \frac{\xi_{j_1}}{\xi_{j_2}} q^k \right) \right] \\
&\qquad\qquad = - N \log (m) + 2n \sum_{k=1}^\infty \sum_{\ell=1}^\infty \frac1\ell \, A_{\ell, m} q^{k\ell} = - N \log (m) + 2N \log \biggl[ \frac{ (q;q)_\infty }{ (q^m; q^m)_\infty }\biggr] \;. \nn
\end{align}
We used that $\sum_{j\neq k}^m \log \left( 1 - e^{2\pi i (j-k)/m} \right) = m \log (m)$. The sum in the first term can be computed using the plethystic expansion (\ref{plethystic theta0}) which converges for $0 < \frac{k_1-k_2}n \im(\tau) < \im(\tau)$. This is not satisfied for all $k_1,k_2$, therefore we split the sum using (\ref{theta0 periodicities}) as follows:
\begin{align}
-\sum_{j_1,j_2=0}^{m-1} & \left[ \sum_{k_1 > k_2}^{n-1} \log\Bigl( \theta_0 \bigl( v_{j_1j_2} + w_{k_1k_2}; \tau \bigr)\Bigr) + \sum_{k_1<k_2}^{n-1} \log \Bigl( - e^{2\pi i (v_{j_1j_2} + w_{k_1k_2})} \, \theta_0 \bigl( -v_{j_1j_2} - w_{k_1k_2}; \tau \bigr) \Bigr) \right] \nn \\
&\quad = - 2 \sum_{j_1,j_2=0}^{m-1} \sum_{k_1 > k_2}^{n-1} \log \Bigl( \theta_0 \bigl( v_{j_1j_2} + w_{k_1k_2}; \tau \bigr) \Bigr) - m^2 \sum_{k_1<k_2}^{n-1} \log \Bigl( - e^{2\pi i w_{k_1k_2}} \Bigr) \nn \\
&\quad = - 2 \sum_{k_1 > k_2}^{n-1} \frac1\ell \, \frac{ \left( \frac{\zeta_{k_1}}{\zeta_{k_2}} \right)^\ell + \left( \frac{\zeta_{k_2}}{\zeta_{k_1}} \right)^\ell q^\ell }{ 1-q^\ell} \bigl( A_{\ell,m} + m \bigr) - \sum_{k_1 < k_2}^{n-1} \log \Bigl( (-1)^{m^2} e^{2\pi i m^2 w_{k_1k_2}} \Bigr) \nn \\
&\quad = - 2m \sum_{k_1 > k_2}^{n-1} \log \Bigl( \theta_0\bigl( mw_{k_1k_2}; m\tau \bigr) \Bigr) - \sum_{k_1 < k_2}^{n-1} \log \Bigl( (-1)^{m^2} e^{2\pi i m^2 w_{k_1k_2}} \Bigr) \nn \\
&\quad = - m \sum_{k_1 \neq k_2}^{n-1} \log\Bigl( \theta_0 \bigl( mw_{k_1k_2}; m\tau \bigr) \Bigr) \;.
\end{align}
In the last equality we used $(-1)^{m^2 -m} = 1$. The sum can be computed using the modular properties (\ref{theta0 modularity}):
\begin{align}
-m \sum_{k_1 \neq k_2}^{n-1} & \log \Bigl( \theta_0 \bigl( mw_{k_1k_2}; m\tau + r \bigr) \Bigr) \\
& = -\pi i m \sum_{k_1 \neq k_2}^{n-1} \left[ \frac12 - \frac{(k_1 -k_2)^2}{n^2} \check\tau - \frac1{6\check\tau} - \frac{\check\tau}6 \right] - m \sum_{k_1 \neq k_2}^{n-1} \log \biggl[ \theta_0 \left( \frac{k_1 - k_2}n; - \frac1{\check\tau} \right) \biggr] \nn \\
& = \pi i N (n-1) \frac{\check\tau \bigl( \check\tau - \frac12 \bigr) (\check\tau -1) }{ 3 \check\tau^2} + \pi i m (n-1) \frac{\check\tau}6 - N \log (n) + 2N \log \biggl[ \frac{ (\check q; \check q)_\infty }{ (\check q^n; \check q^n)_\infty } \biggr] \;. \nn
\end{align}
The computation of the sum in the second line is as in (\ref{piece 3}).
Finally, we use the modular properties (\ref{q-Pochhammer modularity}) of the $q$-Pochhammer symbol to obtain
\be
2N \log \left[ \frac{ (q;q)_\infty }{ (q^m; q^m)_\infty} \, \frac{ (\check q; \check q)_\infty }{ (\check q^n; \check q^n)_\infty } \right] = 2 N \log \biggl[ \frac{ (q;q)_\infty }{ (\check q^n; \check q^n)_\infty } \biggr] + N \log (-i\check\tau) + \frac{\pi i N}6 \left(\check\tau + \frac1{\check\tau} \right) \;.
\ee
Recall that $\im (\check\tau)>0$, therefore $\log(-i\check\tau) = \log(\check\tau) - i\pi/2$ with the logarithm in its principal determination.
Putting the various pieces together we find (\ref{gamma_0 resummed}).

\subsection[The Jacobian $H$]{The Jacobian \matht{H}}
\label{app: Jacobian}

Given the BA operators $Q_i$ written as in (\ref{BA operators alternative}), we compute
\be
\frac1{2\pi i} \parfrac{\log (Q_i)}{u_j} = \sum_{k=1}^N \parfrac{u_{ik}}{u_j} \left[ -1 + \sum_{a=1}^3 \cG(u_{ik}, \Delta_a) \right] \;,
\ee
where we defined the function (see also Appendix~\ref{app: functions})
\be
\cG(u, \Delta; \tau) = \frac1{2\pi i} \parfrac{}{u} \log \left( \frac{\theta_0(-u+\Delta; \tau)}{\theta_0(u+\Delta; \tau)} \right) =  - \frac1{2\pi i} \left[ \frac{\theta_0'(u+\Delta; \tau)}{\theta_0(u+\Delta; \tau)} + \frac{\theta_0'(-u+\Delta; \tau)}{\theta_0(-u+\Delta; \tau)} \right]
\ee
with implicit dependence on $\tau$ (unless specified). This is an even function of $u$, namely \mbox{$\cG(-u,\Delta) = \cG(u, \Delta)$}. Since we treat $u_1,\ldots, u_{N-1}$ as the independent variables while $u_N$ is fixed by the $SU(N)$ constraint, we have $\partial_{u_j} u_{ik}= \delta_{ij} - \delta_{kj} - \delta_{iN} + \delta_{kN}$. Therefore
\be
\frac1{2\pi i} \parfrac{\log (Q_i)}{u_j} = (\delta_{ij} - \delta_{iN}) \left( - N + \sum_{k=1}^N \sum_{a=1}^3 \cG(u_{ik}, \Delta_a) \right) + \sum_{a=1}^3 \biggl[ \cG(u_{iN}, \Delta_a) - \cG(u_{ij}, \Delta_a) \biggr] \;.
\ee
This leads to the following expression for the Jacobian matrix:
\begin{multline}
\cA_{ij} = - N(1+\delta_{ij}) + \sum_{k=1}^N \sum_{a=1}^3 \biggl[ \delta_{ij} \, \cG(u_{ik}, \Delta_a) + \cG(u_{kN}, \Delta_a) \biggr] \\
{} + \sum_{a=1}^3 \biggl[ \cG(u_{iN}, \Delta_a) + \cG(u_{jN}, \Delta_a) - \cG(u_{ij}, \Delta_a) - \cG(0, \Delta_a) \biggr] \;,
\end{multline}
where $i,j = 1, \dots, N-1$.
Let us compute the following quantity:
\be
\label{def Upsilon}
\Upsilon \,\equiv\, \sum_{l=1}^N \cG\bigl( u_{i_1 l}, \Delta_a; \tau \bigr) = \sum_{j=0}^{m-1} \sum_{k=0}^{n-1} \cG \left( \frac{j_1 - j}m + \frac{k_1 - k}n \, \frac{\check\tau}m, \Delta_a; \tau \right) \;.
\ee
The sum over $j$ is computed easily, using the series expansion (\ref{G series expansion}). One has to be careful about the domain of convergence of the series. We sum over $j$ with $k$ fixed, however there is no domain of $\Delta_a, \tau$ such that the full series is convergent. One can instead break the series of $\cG$ in two, each one convergent in a different domain, and recombine the two pieces at the end. Exploiting
\be
\label{sum of phases}
\sum_{j=0}^{m-1} e^{2\pi i (j_1-j)\ell/m} = \begin{cases} m \qquad &\text{for } \ell = 0 \mod m \\ 0 &\text{for } \ell \neq 0 \mod m \;, \end{cases}
\ee
we obtain
\be
\Upsilon = m \sum_{k=0}^{n-1} \cG \left( \frac{k_1 - k}n \, \check\tau, m\Delta_a; \check\tau \right) \;.
\ee
Then we use the modular property (\ref{G modularity}) and perform the second sum. We obtain
\be
\label{Upsilon evaluation}
\Upsilon = N \, \frac{2m\Delta_a +1-\check\tau}{\check\tau} + \frac N{\check\tau} \, \cG\left( 0, \frac{N\Delta_a}{\check\tau}; - \frac n{\check\tau} \right) \;.
\ee
Notice in particular that $\Upsilon$ does not depend on $i_1$ or equivalently on $(j_1, k_1)$. Moreover, $\Upsilon$ is invariant under $m\Delta_a \to m\Delta_a + 1$. Therefore,
\begin{multline}
\label{matrix Aij}
\cA_{ij} = N(1+\delta_{ij}) \left[ -4 + \frac1{\check\tau} \sum_{a=1}^3 \left( 2m\Delta_a + 1 + \cG\Bigl( 0, \frac{N\Delta_a}{\check\tau}; - \frac n{\check\tau} \Bigr) \right) \right] \\
{} + \sum_{a=1}^3 \biggl[ \cG(u_{iN}, \Delta_a) + \cG(u_{jN}, \Delta_a) - \cG(u_{ij}, \Delta_a) - \cG(0, \Delta_a) \biggr] \;.
\end{multline}
Here we recognize the expression in (\ref{Aij resummed}).

\subsection{Perturbative corrections}
\label{app: pert corrections}

We give some evidence that the term $\log\bigl[ \det \bigl( \unit + \cB^{-1} \cC / b_N N \bigr) \bigr]$ in (\ref{large N result case 1}) and (\ref{large N result case 2}) only leads to a contribution of order $\cO(1)$ and to non-perturbative corrections in the large $N$ limit, but no $1/N$ perturbative corrections. In order to do that, we expand
\be
- \log \biggl[ \det \biggl( \unit + \frac{\cB^{-1}\cC}{b_N N} \biggr) \bigg] = \sum_{\ell=1}^\infty \frac{(-1)^{\ell}}{\ell} \Tr \biggl[ \biggl( \frac{\cB^{-1}\cC}{b_N N} \biggr)^{\!\!\ell\;} \biggr]
\ee
and compute the first few terms.

\paragraph{First order.} We compute $\Tr\bigl( \cB^{-1} \cC\bigr)$. The two matrices $\cB$ and $\cC$ are defined in (\ref{elements of Jacobian}) and have size $(N-1)\times (N-1)$. Using the fact that $\cG$ is an even function, the trace can be recast as
\be
\Tr\bigl( \cB^{-1}\cC \bigr) = \sum_{i=1}^{N-1} \cC_{ii} - \frac1N \sum_{i,j=1}^{N-1} \cC_{ij} = \sum_{a=1}^3 \left[ \frac1N \sum_{i,j=1}^N \cG(u_{ij}, \Delta_a; 
\tau) - N \cG(0, \Delta_a; \tau) \right] \;.
\ee
The sum inside square brackets was already computed in (\ref{def Upsilon})--(\ref{Upsilon evaluation}) and it equals $\Upsilon$, therefore
\bea
\Tr\bigl( \cB^{-1}\cC \bigr) &= N \sum_{a=1}^3 \Biggl[ \frac{ 2[m\Delta_a]_{\check\tau} - \check\tau + 1}{\check\tau} + \frac1{\check\tau} \, \cG \biggl( 0, \frac{[m\Delta_a]_{\check\tau} }{ \check\tau/n}; - \frac n{\check\tau} \biggr) - \cG\bigl( 0, \Delta_a; \tau \bigr) \Biggr] \\
&= N\, \frac{\check\tau \pm 1}{\check\tau} + N \sum_{a=1}^3 \Biggl[ \frac1{\check\tau} \, \cG \biggl( 0, \frac{ [m\Delta_a]_{\check\tau}}{\check\tau/n} ; - \frac{n}{\check\tau} \biggr) - \cG\bigl( 0, \Delta_a; \tau \bigr) \Biggr] \;.
\eea
In the first line we used that $\Upsilon$ is invariant under integer shifts of $m\Delta_a$, while the $\pm$ signs in the second line refer to the \first{} and \second{} case of parameter space, respectively. Then, using the large $N$ behavior of $\cG$ in (\ref{large N suppressions}) and
\be
b_N\bigl( [m\Delta]_{\check\tau} \bigr) = \pm \frac1{\check\tau} + \cO(e^{-N}) \;,
\ee
we obtain the large $N$ result
\be
\frac{\Tr\bigl( \cB^{-1} \cC \bigr) }{ b_N N} = 1 \pm \check\tau \Biggl( 1 - \sum_{a=1}^3 \cG\bigl( 0, \Delta_a; \tau \bigr) \Biggr) + \cO(e^{-N})
\ee
where, once again, the $\pm$ signs refer to the \first{} and \second{} case, respectively. We see that this quantity only receives non-perturbative corrections at large $N$.

\paragraph{Second order.} We compute $\Tr\bigl( \cB^{-1}\cC \cB^{-1} \cC \bigr)$. Using that $\cC$ is a symmetric matrix, with some lengthy algebra we obtain
\bea
\Tr\bigl( \cB^{-1}\cC \cB^{-1} \cC \bigr) &= \sum_{i,j=1}^{N-1} (\cC_{ij})^2 - \frac2N \sum_{i,j,k=1}^{N-1} \cC_{ik} \cC_{jk} + \frac1{N^2} \left( \sum_{i,j=1}^{N-1} \cC_{ij} \right)^2 \\
&= \sum_{a,b=1}^3 \left[ \sum_{i,j=1}^N \cG(u_{ij}, \Delta_a; \tau) \, \cG(u_{ij}, \Delta_b; \tau) - \Upsilon(\Delta_a; \tau) \, \Upsilon(\Delta_b, \tau) \right] \;.
\eea
This expression is valid for any HL solution $\{m,n,r\}$. However, evaluating the summation in the second line by brute force is quite complicated and here we will content ourselves with the T-transformed solutions, \ie, with the case $m=1$. 

We define the following quantity:
\be
\cU \,\equiv\, \sum_{a,b=1}^3 \sum_{i,j=1}^N \cG(u_{ij}, \Delta_a; \tau) \, \cG(u_{ij}, \Delta_b; \tau)
\ee
where $u_{ij}$ is as in (\ref{HL solution u_ij}) but with $m=1$. Using the fact that $\cG$ is invariant under integer shifts of each of its arguments (see Appendix~\ref{app: functions}), the modular transformation formula (\ref{G modularity}) and the constraint among $[m\Delta_a]_{\check\tau}$, we obtain
\bea
\cU &= \sum_{a,b=1}^3 \sum_{i,j=1}^N \cG\bigl( u_{ij}, [\Delta_a]_{\check\tau}; \check\tau \bigr) \, \cG\bigl( u_{ij}, [\Delta_b]_{\check\tau}; \check\tau \bigr) \\
&= N^2 \, \frac{(\check\tau \pm 1)^2}{\check\tau^2} + \frac{2(\check\tau \pm 1)}{\check\tau^2} \sum_{a=1}^3 \sum_{i,j=1}^N \cG \biggl( \frac{u_{ij}}{\check\tau}, \frac{[\Delta_a]_{\check\tau}}{\check\tau}; - \frac1{\check\tau} \biggr) \\
&\qquad + \frac1{\check\tau^2} \sum_{a,b=1}^3 \sum_{i,j=1}^N \cG\biggl( \frac{u_{ij}}{\check\tau}, \frac{[\Delta_a]_{\check\tau}}{\check\tau}; - \frac1{\check\tau} \biggr) \, \cG\biggl( \frac{u_{ij}}{\check\tau}, \frac{[\Delta_b]_{\check\tau}}{\check\tau}; - \frac1{\check\tau} \biggr)
\eea
where the $\pm$ signs refer to the \first{} and \second{} case, respectively. Using the series expansion of $\cG$ in (\ref{G series expansion}) and the same method we used to compute $\Upsilon$ after (\ref{def Upsilon}), the summation on the second line gives
\be
\sum_{i,j=1}^N \cG\biggl( \frac{u_{ij}}{\check\tau}, \frac{ [\Delta_a]_{\check\tau}}{\check\tau}; - \frac1{\check\tau} \biggr) = N^2 \, \cG\biggl( 0, \frac{N[\Delta_a]_{\check\tau}}{ \check\tau}; - \frac{N}{\check\tau} \biggr) \;.
\ee
The summation on the third line involves the object
\be
\cW_{ab} \equiv \sum_{i,j=1}^N \cG\biggl( \frac{u_{ij}}{\check\tau}, \frac{[\Delta_a]_{\check\tau}}{\check\tau}; - \frac1{\check\tau} \biggr) \, \cG\biggl( \frac{u_{ij}}{\check\tau}, \frac{[\Delta_b]_{\check\tau}}{\check\tau}; - \frac1{\check\tau} \biggr)
= 2 \sum_{k,\ell=1}^\infty c_a^k c_b^\ell \sum_{i,j=1}^N \biggl[ \biggl( \frac{\tilde z_i}{\tilde z_j} \biggr)^{\! k+\ell} \! + \biggl( \frac{\tilde z_i}{\tilde z_j} \biggr)^{\! k - \ell} \biggr]
\ee
where the coefficients $c_a^\ell$ are given by
\be
\label{coefficients c}
c_a^\ell = \frac{\tilde y_a^\ell - (\check q/\tilde y_a)^\ell}{1-\check q^\ell} = \sum_{r=0}^\infty \Bigl[ \bigl( \check q^r \tilde y_a \bigr)^\ell - \bigr( \check q^{r+1}/ \tilde y_a \bigr)^\ell \Bigr] \;,
\ee
and $\tilde z_i = e^{2\pi i u_i/\check\tau}$, $\tilde y_a = e^{2\pi i [\Delta_a]_{\check\tau}/ \check\tau}$, $\check q = e^{-2\pi i /\check\tau}$. The sum over $i,j$ can be performed exploiting (\ref{sum of phases}), and we obtain
\be
\cW_{ab} = 4N^2 \sum_{\alpha,\beta=0}^\infty c_a^{N(\alpha+1)} c_b^{N(\beta+1)} + 2N^2 \sum_{\alpha,\beta=0}^\infty \sum_{\gamma=1}^{N-1} c_a^{N\alpha+\gamma} \Bigl( c_b^{N\beta + \gamma} + c_b^{N(\beta+1)-\gamma} \Bigr) \;.
\ee
The first sum can be performed exactly using
\be
\sum_{\alpha=1}^\infty c_a^{N\alpha} = \frac12 \, \cG\biggl( 0, \frac{N [\Delta_a]_{\check\tau}}{\check\tau}; - \frac{N}{\check\tau} \biggr) \;.
\ee
Substituting back in $\cW_{ab}$, in $\cU$ and finally in the trace we obtain the expansion
\be
\Tr\bigl( \cB^{-1}\cC \cB^{-1} \cC \bigr) = \frac{2N^2}{\check\tau^2} \sum_{a,b=1}^3 \sum_{\alpha,\beta=0}^\infty \sum_{\gamma=1}^{N-1} c_a^{N\alpha+\gamma} \Bigl( c_b^{N\beta + \gamma} + c_b^{N(\beta+1)-\gamma} \Bigr) \;.
\ee

Let us study, in the large $N$ limit, the summations containing the two terms in parenthesis separately. The second term involves
\begin{align}
\chi^{(2)}_{ab} &\,\equiv\, \sum_{\alpha,\beta=0}^\infty \sum_{\gamma=1}^{N-1} c_a^{N\alpha+\gamma} c_b^{N(\beta+1)-\gamma} \\
&\,=\, \sum_{\alpha,\beta=0}^\infty \sum_{\gamma=1}^{N-1} \sum_{r,s=0}^\infty \biggl[ \bigl( \check q^r \tilde y_a \bigr)^{N\alpha+\gamma} - \bigl( \check q^{r+1} / \tilde y_a \bigr)^{N\alpha+\gamma} \biggr]\biggl[ \bigl( \check q^s \tilde y_b \bigr)^{N\beta + N - \gamma} - \bigl( \check q^{s+1} / \tilde y_b \bigr)^{N\beta + N - \gamma} \biggr] \;. \nn
\end{align}
The fundamental property of $[\Delta_a]_{\check\tau}$ in (\ref{properties square bracket function}) guarantees that $| \check q| < | \tilde y_a | < 1$, and similarly $|\check q| < |\tilde y_b| < 1$. Therefore the defining sum of $\chi^{(2)}_{ab}$ involves terms whose absolute value is smaller than a number smaller than 1, elevated to the powers $N(\alpha+\beta)+N$. This implies that $\chi^{(2)}_{ab}$ is of order $\cO(e^{-N})$, and thus it only produces non-perturbative corrections. Next we consider
\be
\chi^{(1)}_{ab} \,\equiv\, \sum_{\alpha,\beta=0}^\infty \sum_{\gamma=1}^{N-1} c_a^{N\alpha+\gamma} c_b^{N\beta + \gamma} \;.
\ee
By expanding out the coefficients $c_a^\ell$ as in (\ref{coefficients c}), one obtains four sums. One of them is
\begin{align}
& \sum_{\alpha,\beta=1}^\infty \sum_{\gamma=1}^{N-1} \sum_{r,s=0}^\infty \bigl( \check q^r \tilde y_a \bigr)^{N\alpha+\gamma} \bigl( \check q^s \tilde y_b \bigr)^{N\beta + \gamma} = {} \\
&\quad = \sum_{\alpha,\beta,r,s=0}^\infty \bigl( \check q^r \tilde y_a \bigr)^{N\alpha} \bigl( \check q^s \tilde y_b \bigr)^{N\beta} \, \frac{ \check q^{r+s} \tilde y_a \tilde y_b - \bigl( \check q^{r+s} \tilde y_a \tilde y_b \bigr)^N }{ 1-\check q^{r+s} \tilde y_a \tilde y_b}
= \sum_{\ell = 0}^\infty \frac{ (\ell +1) \, \check q^\ell \, \tilde y_a \, \tilde y_b }{ 1 - \check q^\ell \, \tilde y_a \, \tilde y_b } + \cO(e^{-N}) \;. \nn
\end{align}
In the first equality we summed the finite geometric series in $\gamma$, while in the second equality we only kept the terms that are finite in the large $N$ limit. The other three sums that constitute $\chi^{(1)}_{ab}$ can be treated similarly. The result, up to non-perturbative corrections, can be nicely expressed in terms of the function
\be
\wt\gamma(u;\tau) \,\equiv\, \frac1{2\pi i} \, \partial_u \log\Bigl( \wt\Gamma(u;\tau, \tau) \Bigr)
\ee
that can be expanded as follows:
\be
\label{expansion gamma tilde}
\wt\gamma(u; \tau) = \sum_{k=0}^\infty (k+1)\biggl( \frac{ q^k z}{1-q^k z} + \frac{ q^{k+2}/z}{ 1-q^{k+2}/z} \biggr) \;.
\ee
One obtains
\be
\chi^{(1)}_{ab} = \wt\gamma \biggl( \frac{[\Delta_a]_{\check\tau} + [\Delta_b]_{\check\tau}}{ \check\tau}; - \frac1{\check\tau} \biggr) - \wt\gamma\biggl( \frac{ [\Delta_a]_{\check\tau} - [\Delta_b]_{\check\tau} - 1}{ \check\tau}; - \frac1{\check\tau} \biggr) + \cO(e^{-N}) \;.
\ee

We can finally substitute into the trace, and obtain the second-order contribution
\be
\frac{ \Tr\bigl( \cB^{-1}\cC \cB^{-1} \cC \bigr) }{ b_N^2 N^2 } = \frac12 \sum_{a,b=1}^3 \Biggl[ \wt\gamma \biggl( \frac{[\Delta_a]_{\check\tau} + [\Delta_b]_{\check\tau}}{ \check\tau}; - \frac1{\check\tau} \biggr) - \wt\gamma\biggl( \frac{ [\Delta_a]_{\check\tau} - [\Delta_b]'_{\check\tau}}{ \check\tau}; - \frac1{\check\tau} \biggr) \Biggr] + \cO(e^{-N}) \;.
\ee
Also at this order we see that there are only non-perturbative corrections at large $N$.

\section{Solutions with two angular momenta}
\label{app: Two angular momenta}

In this appendix we generalize our discussion in the main body of the paper to black hole solutions with two different angular momenta (and correspondingly two different chemical potentials for them), but still with three equal $U(1)$ charges (and chemical potentials), so that the solutions can be constructed using 5d minimal gauged supergravity. We also discuss the supersymmetry of those solutions, and the uplift to 10d type IIB supergravity. Finally, we study the supersymmetric embeddings of Euclidean D3-branes.

In Appendix~\ref{app: giant gravitons}, as an aside, we will present supersymmetric embeddings of Lorentzian D3-branes that generalize the giant graviton and dual giant graviton solutions in empty AdS$_5$ \cite{McGreevy:2000cw, Grisaru:2000zn, Hashimoto:2000zp} to Lorentzian black hole backgrounds.

\subsection{5d minimal gauged supergravity}

The bosonic Lagrangian of 5d minimal gauged supergravity is (\ref{5d action}), that we repeat here:%
\footnote{Given a $p$-form $\omega_{(p)} = \frac1{p!} \, \omega_{\mu_1 \dots \mu_p} dx^{\mu_1} \dots dx^{\mu_p}$ in $d$ dimensions, we define its Hodge dual in Lorentzian sig\-na\-ture as $*\omega_{(p)} = \frac{\sqrt{-g}}{p! \, (d-p)!} \, \omega^{\nu_1 \dots \nu_p} \epsilon_{\nu_1 \dots \nu_p \mu_1 \dots \mu_{d-p}} dx^{\mu_1} \dots dx^{\mu_{d-p}}$ where $\epsilon_{0\dots (d-1)}=1$.
	Then $*^2 = (-1)^{p(d-p)+1}$, as well as $*1 = \dvol_d$ and $\omega_{(p)} \wedge *\omega_{(p)} = \frac1{p!} \omega_{\mu_1 \dots \mu_p} \omega^{\mu_1 \dots \mu_p} \dvol_d$.
\label{foo: notation}}
\be
\label{5d minimal gauged sugra action}
\cL = \bigl( R + 12g^2 \bigr) *1 - \frac2{3} \, F \wedge *F + \frac8{27} \, F \wedge F \wedge A \;,
\ee
where $A$ is the graviphoton potential, $F=dA$ is its field strength, and we followed the notation of \cite{Cabo-Bizet:2018ehj}. As in the main text, we set the dimensionful coupling $g$ to $1$.

The authors of \cite{Chong:2005hr}, generalizing previous work of \cite{Gutowski:2004ez, Cvetic:2004hs, Madden:2004ym}, constructed a four-parameter family of charged and rotating black hole solutions:
\bea
\label{CCLP solution I}
ds^2 &= - \frac{\Delta_\theta \bigl[ (1+ r^2) \, \rho^2 \, dt + 2q\,\nu \bigr] dt}{\Xi_a \, \Xi_b \, \rho^2} + \frac{2q\,\nu\,w}{\rho^2} + \frac{f_t}{\rho^4} \left( \frac{\Delta_\theta \, dt}{\Xi_a \, \Xi_b} - w \right)^2 + \frac{ \rho^2 \, dr^2}{\Delta_r} \\
&\quad\; + \frac{\rho^2 \, d\theta^2}{\Delta_\theta} + \frac{r^2 + a^2}{\Xi_a} \sin^2(\theta)\, d\phi^2 + \frac{r^2 + b^2}{\Xi_b} \cos^2(\theta)\, d\psi^2\,, \\[.5em]
A &= \frac{3 q}{2\rho^2} \left( \frac{\Delta_\theta \, dt}{\Xi_a \, \Xi_b} - w \right) - \alpha \, dt \;,
\eea
where
\bea
\label{CCLP solution II}
\nu &= b\sin^2(\theta)\, d\phi + a \cos^2(\theta)\, d\psi \,, \qquad&
w &= a \sin^2(\theta)\, \frac{d\phi}{\Xi_a} + b \cos^2(\theta)\, \frac{d\psi}{\Xi_b} \,, \\
\Xi_a &= 1-a^2 \;,\qquad \Xi_b = 1-b^2 \;, \qquad&
f_t &= 2(m+abq)\rho^2 - q^2 \,, \\[.5em]
\Delta_\theta &= 1-a^2 \cos^2(\theta) - b^2 \sin^2(\theta) \,, \qquad&
\rho^2 &= r^2 + a^2\cos^2(\theta) + b^2\sin^2(\theta) \,, \\[.5em]
\Delta_r &= \frac{ (r^2 + a^2)(r^2 + b^2)(1+ r^2) + q^2 + 2ab q}{r^2} - 2m \;. \hspace{-1cm}
\eea
The field strength can be written as
\begin{align}
F &= \frac{3qr}{\rho^4} \, dr \wedge w - \frac{3q}{\Xi_a \Xi_b \rho^4} \left( r \Delta_\theta \, dr - \frac{a^2 - b^2}2 \, (1+r^2) \sin(2\theta)\, d\theta \right) \wedge dt \nn\\
&\quad\;\; - \frac{3q \sin(2\theta)}{2\rho^4} \, d\theta \wedge \left( a\, \frac{r^2 + a^2}{\Xi_a} \, d\phi - b \, \frac{r^2 + b^2}{\Xi_b} \, d\psi \right) \;,
\end{align}
while $*_5F - \frac23 A \wedge F = d\alpha_{(2)}$ with
\be
\label{alpha_2 generic ang mom}
\alpha_{(2)} = \frac{3q}{2\rho^2} \biggl( \frac{\Delta_\theta\, dt}{\Xi_a \Xi_b} - w \biggr) \wedge \biggl( \nu + \frac23 \, \alpha\, dt \biggr) - \frac{3q \cos(2\theta)}{ 4 \Xi_a \Xi_b} \Bigl( dt \wedge (b \, d\phi - a \, d\psi) + d\phi \wedge d\psi \Bigr) \;.
\ee
The coordinates are $(t,r,\theta, \phi, \psi)$, where $\theta \in [0, \pi/2]$ while $\phi,\psi$ have period $2\pi$. The constant $\alpha$ is arbitrary in Lorentzian space (it can be shifted by gauge transformations). The four parameters are $(a,b,q,m)$, and correspondingly there are four independent conserved charges: the energy $E$ (associated to the Killing vector $\parfrac{}{t}$), two angular momenta $J_{1,2}$ (associated to the Killing vectors $\parfrac{}{\phi}$ and $\parfrac{}{\psi}$, respectively), and the electric charge $Q$. Setting $a=b$ one recovers the solution in (\ref{BH metric 1}) which has $J_1 = J_2 \equiv J$.

The outer horizon is the largest positive root of $\Delta_r(r_+) = 0$, and we denote it by $r_+$. This is a Killing horizon generated by the Killing vector field
\be
\label{Killing vector of horizon}
V = \parfrac{}{t} + \Omega_1 \, \parfrac{}{\phi} + \Omega_2 \, \parfrac{}{\psi} \;.
\ee
Here $\Omega_{1,2}$ are the angular velocities at the horizon, measured in a non-rotating frame at infinity. Evaluating the surface gravity one determines the Hawking temperature \mbox{$T \equiv 1/\beta$}. Besides, one defines the electrostatic potential at the horizon, $\Phi = \imath_V A\big|_{r_+} - \imath_V A\big|_\infty$, and the entropy $S$, equal to a quarter of the horizon area. Explicit expressions for the charges and chemical potentials can be found in \cite{Cabo-Bizet:2018ehj}. Those quantities satisfy the first law of thermodynamics,
\be
dE = T\, dS + \Omega_1 \, dJ_1 + \Omega_2 \, dJ_2 + \Phi \, dQ \;.
\ee
It turns out to be convenient to trade the parameter $m$ for $r_+$, and thus use $(a,b,q,r_+)$ as the independent parameters. Since $r_+$ is a root of $\Delta_r$, the relation
\be
\label{from m to rplus}
m = \frac{ (r_+^2 + a^2)(r_+^2 + b^2)(1+ r_+^2) + q^2 + 2ab q}{2r_+^2}
\ee
can be used to eliminate $m$ in favor of $r_+$.

The analytic continuation to Euclidean signature, studied in detail in \cite{Cabo-Bizet:2018ehj, Cassani:2019mms}, is obtained by rotating $t \to - i t_\mrE$, $a \to i \tilde a$, and $b \to i \tilde b$. This yields a real Euclidean metric, though the gauge field becomes imaginary. Regularity of such a metric around $r = r_+$ requires to compactify the Euclidean time $t_\mrE$ with a period equal to $\beta$, and more precisely it requires to identify
\be
(t_\mrE, \phi, \psi) \cong \bigl(t_\mrE + \beta,\; \phi - i \Omega_1 \beta,\; \psi - i \Omega_2 \beta \bigr) \cong (t_\mrE, \phi + 2\pi , \psi) \cong (t_\mrE, \phi, \psi+2\pi) \;.
\ee
Notice that the Killing vector (\ref{Killing vector of horizon}) precisely generates rotations of the Euclidean circle that shrinks at $r_+$. Regularity of the gauge field at $r_+$, in a gauge which is regular at the horizon, requires $\imath_V A \big|_{r_+} = 0$, which implies
\be
\label{gauge field regularity cond}
\alpha = \Phi \;.
\ee
The on-shell action can then be computed for regular Euclidean metrics, regularized using background subtraction (which assigns vanishing action to thermal AdS$_5$). The family of solutions can then be extended to generic complex values of the parameters $(a,b,q,r_+)$, and the on-shell action $I_\text{SUGRA}$ is extended analytically. It satisfies the so-called ``quantum statistical relation''
\be
I_\mathrm{SUGRA} = \beta \bigr( E - T S - \Omega_1 J_1 - \Omega_2 J_2 - \Phi Q \bigl) \;.
\ee

\subsubsection{Supersymmetry}

The solutions (\ref{CCLP solution I}) become supersymmetric for
\be
\label{SUSY condition q}
m = (1+a+b) \, q \;.
\ee
This gives a three-parameter family of complex (and, generically, non-extremal) solutions in terms of $(a,b,q)$. The charges take the values
\bea
E &= \frac{ \pi q \bigl( 3 + ab - (1+a)b^2 - (1+b)a^2 \bigr) }{ 4 (1-a)(1-a^2)(1-b)(1-b^2) } \;,\qquad&
Q &= \frac{\pi q }{ 2(1-a^2)(1-b^2) }\,, \\
J_1 &= \frac{\pi q (2a+b+ab) }{ 4(1-a)(1-a^2)(1-b^2) } \;,\qquad&
J_2 &= \frac{\pi q (2b+a+ab) }{ 4(1-a^2)(1-b)(1-b^2) }\,,
\eea
and satisfy the linear supersymmetric relation $E = J_1 + J_2 + \frac32 Q$. In the following it will be convenient to use $(a,b,r_+)$ as the independent parameters, and combining the change of variables (\ref{from m to rplus}) with the SUSY condition (\ref{SUSY condition q}) one finds
\be
q = -ab + (1+a+b)r_+^2 \mp i r_+ \bigl( r_+^2 - r_*^2 \bigr) \qquad\text{with}\qquad r_*^2 = a+b+ab \;.
\ee
Here $r_*$ is the position of the outer horizon for real supersymmetric and extremal solutions, \ie, for the Euclidean rotation of Lorentzian BPS black hole solutions. The $\mp$ signs come from the branch cut of a square root. For the upper sign, the supersymmetry condition becomes
\be
\label{q on first branch generic}
q = - (a-ir_+)(b-i r_+) (1-ir_+) \qquad\qquad\qquad \text{(\first{} branch)} \;.
\ee
We refer to the corresponding family of solutions as the ``\first{} branch''.
The ``\second{} branch'', that follows from the lower sign, is obtained by sending $i \to -i$. In the first branch, the chemical potentials take the values
\bea
\Phi &= \frac{ 3ir_+ (1-ir_+) }{ 2(r_*^2 + i r_+) } \;,\qquad\qquad\quad&
\beta &= \frac{2\pi (a-i r_+) (b-i r_+) (r_*^2 + i r_+) }{ (r_+^2 - r_*^2) \bigl( 2 (1+a+b)r_+ + i (r_*^2 - 3r_+^2) \bigr) } \;, \\
\Omega_1 &= \frac{ (r_*^2 + i a r_+)(1-ir_+) }{ (r_*^2 + i r_+)(a - i r_+) } \;,\quad&
\Omega_2 &= \frac{ (r_*^2 + i b r_+)(1-ir_+) }{ (r_*^2 + i r_+)(b - i r_+) } \;.
\eea
They satisfy the relation
\be
\label{SUSY relation chemical pot}
\beta \, (1 + \Omega_1 + \Omega_2 - 2 \Phi) = \pm 2\pi i, \qquad\qquad (\text{\first/\second{} branch})
\ee
where the $\pm$ sign depends on the branch.
Following \cite{Cabo-Bizet:2018ehj}%
\footnote{The potentials $\omega_1, \omega_2, \varphi$ defined in \cite{Cabo-Bizet:2018ehj} differ from ours by a simple rescaling.}
and the discussion in Section~\ref{part_to_index},
we define new chemical potentials $\sigma_\mrg$, $\tau_\mrg$, $\Delta_\mrg$ as
\be
\sigma_\mrg = \frac{\beta}{2\pi i} \bigl( \Omega_1 - 1 \bigr) \;,\qquad \tau_\mrg = \frac{\beta}{2\pi i} \bigl( \Omega_2 -1 \bigr) \;,\qquad \Delta_\mrg = \frac{\beta}{2\pi i} \left( \frac23\, \Phi - 1 \right) \;.
\ee
In the \first{} branch they are given by
\bea
\sigma_\mrg &= \frac{ (a-1)(b-ir_+) }{ 2i(1+a+b)r_+ - r_*^2 + 3r_+^2} \;,\qquad\qquad  \Delta_\mrg = \frac{ (a-ir_+)(b-ir_+) }{ 2i(1+a+b)r_+ - r_*^2 + 3r_+^2 } \;,  \\
\tau_\mrg &= \frac{ (b-1)(a -ir_+) }{ 2i(1+a+b)r_+ - r_*^2 + 3r_+^2} \;,
\eea
(while the values in the \second{} branch are obtained sending $i \to -i$ and changing the overall sign). These potentials satisfy
\be
\sigma_\mrg + \tau_\mrg - 3 \Delta_\mrg = \pm 1 \qquad\qquad (\text{\first/\second{} branch}) \;.
\ee
In terms of them, the on-shell action takes the very simple form
\be
I_\mathrm{SUGRA} = \frac{i\pi^2}{2g^3 G_5} \, \frac{\Delta_\mrg^3}{\sigma_\mrg \, \tau_\mrg}
\ee
in both branches. Notice in particular that it does not depend on $\beta$. Here, for clarity, we reinstated the coupling $g$ and the 5d Newton constant $G_5$ that were previously set to $1$. This expression generalizes (\ref{SUSY on-shell action}).

\subsubsection{Killing spinor}
\label{sec: Killing spinor}

The solutions (\ref{CCLP solution I}) with the SUSY condition (\ref{SUSY condition q}) admit a Killing spinor $\epsilon$. The Killing vector arising as a bilinear of the Killing spinor is $K = \parfrac{}{t} + \parfrac{}{\phi} + \parfrac{}{\psi}$.
The spinor satisfies
\be
\cL_{\parfrac{}{t}} \epsilon = \frac i2 \, (1- 2\alpha) \, \epsilon \;,\qquad\qquad\qquad \cL_{\parfrac{}{\phi}} \epsilon = \cL_{\parfrac{}{\psi}} \epsilon = \frac i2 \, \epsilon \;,
\ee
where $\cL$ is the Lie derivative. The second equation implies that $e^{2\pi \cL_{\partial/\partial\phi}} \epsilon = e^{2\pi \cL_{\partial/\partial\psi}} \epsilon = - \epsilon$, namely, that $\epsilon$ is anti-periodic along the two circles parametrized by $\phi$ and $\psi$, which is a necessary condition for a spinor to be well-defined since the two circles shrink somewhere inside $S^3$. The first equation, combined with the gauge-field regularity condition (\ref{gauge field regularity cond}) and the SUSY relation (\ref{SUSY relation chemical pot}) among the chemical potentials, implies $\cL_V \epsilon = \mp \frac\pi\beta \epsilon$, where $V$ is the Killing vector (\ref{Killing vector of horizon}) and $\mp$ correspond to the \first/\second{} branch. In turn this guarantees that $e^{-i\beta \cL_V} \epsilon = - \epsilon$, and thus that $\epsilon$ is anti-periodic along the Euclidean time circle that shrinks at the horizon.

In order to construct the Killing spinor $\epsilon$, it is convenient to use orthotoric coordinates $(t, \xi, \eta, \Phi, \Psi)$ \cite{Cassani:2015upa}. The coordinate change is given by
\bea
\label{change to orthotoric coordinates}
r^2 &= r_*^2 + (a+b) \wt m + \frac{(a+b)^2}2 \, \wt m + \frac{a^2 - b^2}2 \, \wt m\, \xi \;,\qquad\qquad
\theta = \frac12 \arccos (\eta) \;, \\[.5em]
\phi &= t - \frac{4 (1-a^2)}{(a^2-b^2) \wt m} \, (\Phi - \Psi) \;,\qquad\qquad\qquad
\psi = t - \frac{4 (1-b^2)}{(a^2 - b^2) \wt m} \, (\Phi + \Psi) \;,
\eea
where we defined the new mass parameter%
\footnote{The extremality condition for supersymmetric solutions, namely $r_+ = r_*$, simply reads $\wt m = 0$ \cite{Cabo-Bizet:2018ehj}.}
\be
\wt m = \frac{m}{(a+b)(1+a)(1+b)(1+a + b)} - 1 \;.
\ee
In these coordinates, supersymmetric black hole metrics are described by the simple vielbein
\bea
\label{vielbein orthotoric}
E^0 &= f (dt - \omega) \;, \\
E^1 &= \frac1{f^{1/2}} \, \sqrt{ \frac{\eta - \xi}{\cF(\xi)}} \, d\xi \;,\qquad&
E^2 &= \frac1{f^{1/2}} \, \sqrt{ \frac{\cF(\xi)}{(\eta- \xi)}} \, \bigl( d\Phi + \eta\, d\Psi \bigr) \;, \\
E^3 &= - \frac1{f^{1/2}} \, \sqrt{ \frac{\eta - \xi}{\cG(\eta)}} \, d\eta \;,\qquad&
E^4 &= \frac1{f^{1/2}} \, \sqrt{ \frac{\cG(\eta)}{(\eta- \xi)}} \, \bigl( d\Phi + \xi \, d\Psi \bigr) \;,
\eea
where $\cF(\xi)$ and $\cG(\eta)$ are the cubic polynomials
\bea
\cG(\eta) &= - \frac{4 \, (1-\eta^2) }{(a^2 - b^2) \, \wt m} \, \Bigl[ (1-a^2 ) (1+\eta) + (1-b^2)(1-\eta) \Bigr]\,, \\
\cF(\xi) &= - \cG(\xi) - 4 \, \frac{1+\wt m}{\wt m} \left( \frac{2+a + b}{a-b} + \xi \right)^3 \;,
\eea
while
\bea
f &= \frac{24(\eta - \xi)}{\cF''(\xi) + \cG''(\eta)} \,, \\
\omega &= - \frac{\cF''' + \cG'''}{48 \, (\eta - \xi)^2} \left\{ \left[ \cF + (\eta - \xi) \left( \frac{\cF'}2 - \frac14 \left( \frac{2+a +b }{ a-b} + \xi \right)^2 \cF''' \right) \right] (d\Phi + \eta\, d\Psi) \right. \\
&\hspace{3cm} + \cG \, (d\Phi + \xi\, d\Psi) \Biggr\} + \frac{\cF''' \, \cG'''}{288} \Bigl[ (\eta + \xi) \, d\Phi + \eta \, \xi\, d\Psi \Bigr] + \frac2{\wt m} \, d\Psi \;.
\eea
Explicitly, the function $f$ takes the form
\be
f = - \frac{\wt m \, (a-b) \, (\eta - \xi) }{ (2+a+b)(1+\wt m) + (a-b)(\eta + \wt m \xi) } \;.
\ee
One can also verify that at the location $\xi_+$ of the outer horizon $r=r_+$, we have
\be
\cF(\xi_+) = 0 \;.
\ee
The metric is then $ds^2 = - (E^0)^2 + (E^1)^2 + (E^2)^2 + (E^3)^2 + (E^4)^2$. The gauge potential is
\begin{align}
\label{gauge potential orthotoric}
A &= \frac{3(1-f)}2 dt - \alpha\, dt \\
&\quad\; - \frac{6(1+a)(1+b)(1+\wt m) f}{ (a-b)^2 (a+b) \wt m^2 (\eta - \xi) } \, \biggl[ \Bigl( a+b-(a-b)\eta \Bigr) d\Phi + \Bigl( (a+b) \eta - (a-b) \Bigr) d\Psi \biggr] \;. \nn
\end{align}
Using vielbein indices, the non-vanishing components of the field strength are $F_{01}$, $F_{03}$, $F_{12}$, $F_{14}$, $F_{23}$, $F_{34}$ and antisymmetrizations. The metric determinant is $-\det(g) = (\eta-\xi)^2 / f^2$.

The Killing spinor equation of 5d minimal gauged supergravity reads
\be
\label{KSE in 5d minimal sugra}
\left[ \nabla_\mu - i A_\mu - \frac12 \, \gamma_\mu - \frac{i}{12} \Bigl( \gamma\du{\mu}{\nu\rho} - 4 \, \delta_\mu^\nu \, \gamma^\rho \Bigr) F_{\nu\rho} \right] \epsilon = 0 \;.
\ee
We use the orientation $\gamma^{01234} = -i$. Then we impose the relation $\frac i2 \bigl( \gamma^{12} -\gamma^{34} \bigr) \epsilon = \epsilon$, which implies the following projectors:
\be
\label{spinor projections}
i\, \epsilon = - \gamma^0 \epsilon = - \gamma^{12} \epsilon = \gamma^{34} \epsilon \;.
\ee
One verifies that the following spinor solves (\ref{KSE in 5d minimal sugra}):
\be
\label{Killing spinor BH general}
\epsilon = \exp \left\{ \frac{i}{2} \left[ (3 - 2\alpha)t - \frac{4(1-a^2)}{(a^2 - b^2) \wt m} (\Phi - \Psi) - \frac{4 (1-b^2)}{ (a^2 - b^2) \wt m} (\Phi + \Psi) \right] \right\} \, \sqrt{f} \; \epsilon_0 \;,
\ee
where $\epsilon_0$ is a constant spinor satisfying (\ref{spinor projections}). When checking the Killing spinor equation, one encounters a series of square roots. We started from a domain in which all radicands in (\ref{vielbein orthotoric}) are positive, and then analytically continued from there.

\subsubsection{Killing spinor for equal angular momenta}
\label{sec: Killing spinor a=b}

Black hole solutions with equal angular momenta $J_1 = J_2$ correspond to $a=b$ and have been discussed in the main text. However, the orthotoric coordinates in (\ref{change to orthotoric coordinates}) are singular for $a=b$, and thus the $a=b$ limit must be taken carefully. Following Section~4 of \cite{Cassani:2015upa}, we set%
\footnote{In this case, the extremality condition is $m= 2a (1+a)^2 (1+2a)$, or more easily $\wt m =0$.}
\bea
b &= a + \frac{4(1-a^2)}{a \, \wt m} \, \lambda \qquad\qquad\text{with}\qquad \wt m = \frac{m}{2a(1+a)^2(1+2a)} -1 \,, \\[.4em]
\Phi &= \lambda \, \wt\Phi \;,\qquad\qquad\qquad
\Psi = \lambda \, \wt\Psi \;,\qquad\qquad\qquad
\xi = - \frac1\lambda\, \wt\xi\,,
\eea
and take $\lambda \to 0$. The change of coordinates reduces to
\bea
\label{change to orthotoric coords a=b}
r^2 &= r_*^2 + 2a(1+a)\wt m + 4 (1-a^2) \, \wt\xi \;,\qquad\qquad& \theta &= \frac12 \arccos(\eta) \;, \\
\phi &= t + \frac{\wt\Phi - \wt\Psi}2 \;,\qquad\qquad& \psi &= t + \frac{\wt\Phi + \wt\Psi}2 \;,
\eea
with $r_*^2 = 2a+a^2$. The functions $\cF$ and $\cG$ behave as
\be
\cG(\eta) = \frac{\wt\cG(\eta)}\lambda + \cO(\lambda^0) \;,\qquad\qquad \cF(\xi) = \frac{\wt\cF(\wt\xi)}{\lambda^3} + \cO(\lambda^{-2}) \;,
\ee
with
\be
\label{orthotoric a=b functions 1}
\wt\cG(\eta) = 1- \eta^2 \;,\qquad\qquad \wt\cF(\wt\xi) = \biggl( 1 - \frac{4\wt\xi}{\wt m} \biggr) \, \wt\xi^2 + 4 \, \frac{1+\wt m}{\wt m} \left( \frac{a \, \wt m}{2(1-a)} + \wt\xi \right)^3 \;.
\ee
The function $f$ and the 1-form $\omega$ reduce to
\bea
\label{orthotoric a=b functions 2}
f &= \frac{2(1-a)\wt\xi}{ a(1+\wt m) + 2(1-a) \wt\xi } \;,\qquad
\omega = \left[ \frac1{3f} \biggl( \frac{\wt\cF'(\wt\xi)}{2\wt\xi} -1 \biggr) - \frac{(1+\wt m) a^2 }{ 4 (1-a)^2 \wt\xi } \right] \bigl( d\wt\Phi + \eta\, d\wt\Psi \bigr) \;.
\eea
In these coordinates, the outer horizon $r=r_+$ is located at
\be
\label{horizon location xi tilde +}
\wt\xi = \wt\xi_+ = \frac{(a-ir_+)(r_*^2 - r_+^2) }{ 4(1-a)(1+a)^2}
\ee
on the first branch (while the value on the second branch is obtained by sending $i \to -i$), and one verifies that
\be
\wt\cF \bigl( \wt\xi_+ \bigr) = 0 \;.
\ee
On the other hand, $f$ vanishes at the horizon only in the extremal BPS limit $r_+ \to r_*$.

The vielbein (\ref{vielbein orthotoric}) reduces to
\begin{align}
E^0 &= f (dt - \omega) \;,\quad&
E^1 &= -\frac1{f^{1/2}} \sqrt{ \frac{\wt\xi}{\wt\cF(\wt\xi)}} \, d\wt\xi \;,\quad&
E^2 &= \frac1{f^{1/2}} \sqrt{ \frac{\wt\cF(\wt\xi)}{\wt\xi}} \, \bigl( d\wt\Phi + \eta\, d\wt\Psi \bigr) \;, \nn \\[.4em]
&&
E^3 &= - \frac1{f^{1/2}} \sqrt{ \frac{\wt\xi}{\wt\cG(\eta)} }\, d\eta \;,\quad&
E^4 &= - \frac1{f^{1/2}} \sqrt{ \wt\cG(\eta) \, \wt\xi }\, d\wt\Psi \;.
\label{vielbein orthotoric a=b}
\end{align}
In these coordinates the black hole solution reads
\begin{align}
ds^2  
&= - f^2 \bigl( dt - \omega \bigr)^2 + \frac1f \left[ \frac{\wt\xi}{\wt\cF} \, d\wt\xi^2 + \frac{\wt\cF}{\wt\xi} \bigl( d\wt\Phi + \eta\, d\wt\Psi \bigr)^2 + \wt\xi \left( \frac{d\eta^2}{1-\eta^2} + (1-\eta^2) \, d\wt\Psi^2 \right) \right]\,, \nn\\[.4em]
A &= \biggl( \frac{3}2(1-f) - \alpha \biggr) dt - \frac{3 a^2 (1+\wt m) f }{8 (1-a)^2 \wt\xi } \bigl( d\wt\Phi + \eta\, d\wt\Psi\bigr) \;.
\label{BH metric orthotoric a=b}
\end{align}
In this case, using vielbein indices, the non-vanishing components of the field strength are $F_{01}$, $F_{12}$, $F_{34}$ and their antisymmetrizations.

One can check that the following spinor solves the Killing spinor equation (\ref{KSE in 5d minimal sugra}):
\be
\epsilon = \exp \left\{ \frac i2 \left[ (3-2\alpha)\, t \, + \, \wt\Phi \right] \right\} \, \sqrt{f} \; \epsilon_0 \;,
\ee
where $\epsilon_0$ is a constant spinor that satisfies the same projections as in (\ref{spinor projections}).

\subsection{10d type IIB supergravity}

In order to discuss D3-brane embeddings, we need to uplift the black hole solutions to 10d type IIB supergravity on AdS$_5 \times S^5$ \cite{Chamblin:1999tk, Cvetic:1999xp}. It turns out that 5d minimal gauged supergravity can be consistently embedded into AdS$_5 \times \mathrm{SE}_5$ for any Sasaki-Einstein 5-manifold $\mathrm{SE}_5$ \cite{Buchel:2006gb}. This general point of view is particularly useful when discussing supersymmetry, so let us review it here.

The bosonic action of 10d type IIB supergravity restricted to the metric $G_{MN}$ and the 5-form flux $F_{(5)}$ (this is a consistent truncation) is
\be
S_\text{IIB} = \frac1{2\kappa_{10}^2} \int d^{10}x\, \sqrt{-G} \, \left[ R_\text{10d} - \frac1{480} F_{M_1 \dots M_5} F^{M_1 \dots M_5} \right] \;,
\ee
supplemented by the self-duality condition $F_{(5)} = *F_{(5)}$. We reduce it on a Sasaki-Einstein manifold $\mathrm{SE}_5$, which in general can be locally written as a $U(1)$ fibration over a K\"ahler Einstein base $B$. We consider the ansatz:
\bea
ds_{10}^2 &= ds_5^2 + ds^2(B) + (e^9)^2 \;, \qquad\qquad& e^9 &= \frac13 \Bigl( d\psi_s + \cA + 2 A \Bigr) \\
F_{(5)} &= (1+*) \, G_{(5)} \;,\qquad\qquad& G_{(5)} &= - 4\epsilon_{(5)} + \frac23 J \wedge *_5 F \;.
\eea
Here $B$ is a K\"ahler-Einstein 4-manifold, normalized such that $R_\text{KE} = 24$ is its scalar curvature, with K\"ahler form $J$ (so that $\frac12 J\wedge J = \dvol_B$ is the volume form), and $\cA$ is a specific $U(1)$ connection on $B$ with $d\cA = 6J$. The resulting Sasaki-Einstein space has $R_\text{SE} = 20$. Then $ds_5^2$ is the 5d spacetime metric, $\epsilon_{(5)}$ is its volume form, $*_5$ is the 5d Hodge dual operator, $A$ is the 5d graviphoton field, and $F =dA$ is its field strength. Using
\be
*G_{(5)} = 2 \, J \wedge J \wedge e^9 - \frac23 F \wedge J \wedge e^9 \;,
\ee
the Bianchi identity for $F_{(5)}$ gives the 5d equation of motion:
\be
\label{Maxwell's equation}
d \, {*_5 F} = \frac23 F \wedge F \;.
\ee
The full expression for $F_{(5)}$, and a possible choice of potential such that $F_{(5)} = dC_{(4)}$, are
\bea
\label{general 5flux 4potential}
F_{(5)} &= -4 \epsilon_{(5)} + 2 \, J \wedge J \wedge e^9 - \frac23 F \wedge J \wedge e^9 + \frac23 J \wedge *_5 F \\
C_{(4)} &= - 4 \beta_{(4)} + \frac19 \wt\cA \wedge \Bigl[ (3J-F) \wedge e^9 + *_5 F \Bigr] \;.
\eea
Here $\beta_{(4)}$ is a 4-form such that $\epsilon_{(5)} = d\beta_{(4)}$, while $\wt\cA$ is any connection on the K\"ahler-Einstein base $B$ such that $d\wt\cA = 6J$ (we could for instance choose $\cA$, but not necessarily). Einstein's equations give the 5d equation of motion
\be
\label{Einstein's equation}
R_{\mu\nu} = - 4 g_{\mu\nu} + \frac23 F_{\mu\rho} F\du{\nu}{\rho} - \frac19 \, g_{\mu\nu} \, F_{\rho\lambda} F^{\rho\lambda} \;.
\ee
In turn, equations (\ref{Maxwell's equation}) and (\ref{Einstein's equation}) follow from the (bosonic part of the) 5d supergravity action (\ref{5d minimal gauged sugra action}).

One can similarly reduce the supersymmetry condition from 10d to 5d \cite{Buchel:2006gb}. Restricting to the metric and the 5-form flux, the 10d dilatino variation vanishes automatically while the gravitino variation reads
\be
\label{10d gravitino variation}
\delta_\varepsilon \psi_M = \nabla_M \varepsilon + \frac{i}{16\cdot 5!} \, F_{P_1 \dots P_5} \Gamma^{P_1 \dots P_5} \Gamma_M \varepsilon \;,
\ee
where $\varepsilon$ is the 10d spinor.
We decompose the 10d gamma matrices as
\be
\Gamma^M = \Bigl\{ \gamma^\mu \otimes \unit \otimes \sigma_1 \,,\; \unit \otimes \hat\gamma^{\underline a} \otimes \sigma_2 \Bigr\}
\ee
where the index $\underline{a}=5,6,7,8,9$ runs over the directions of the Sasaki-Einstein 5-manifold, and in particular $\underline{a}= 5,6,7,8$ runs over the directions of the K\"ahler-Einstein base $B$, while $\sigma_{1,2}$ are Pauli matrices. We take
\be
\gamma^{01234} = -i \;,\qquad\qquad
\hat\gamma^{56789} = 1 \;,\qquad\qquad
\Gamma_{11} \equiv \Gamma^0 \cdots \Gamma^9 = \unit \otimes \unit \otimes \sigma_3 \;.
\ee
The 10d chiral spinor $\varepsilon$ of type IIB supergravity satisfies $\Gamma_{11} \varepsilon = - \varepsilon$ (compatibly with our definition of $*$), in other words $\sigma_3 \varepsilon = - \varepsilon$. We thus decompose
\be
\varepsilon = \epsilon \otimes \chi \otimes \biggl( \!\! \begin{array}{c} 0 \\ 1 \end{array} \!\! \biggr) \;.
\ee
We assume we are using a vielbein that ``diagonalizes'' the K\"ahler form of $B$, namely, such that $J = e^{56} + e^{78}$. This leads to the relations
\bea
F_{M_1 \dots M_5} \Gamma^{M_1 \dots M_5} \Gamma_\mu \varepsilon &= 40i \Bigl[ 24 + 2 \bigl( \hat\gamma^{56} + \hat\gamma^{78} \bigr) F_{\rho\lambda} \gamma^{\rho\lambda} \Bigr] \gamma_\mu \varepsilon \\
F_{M_1 \dots M_5} \Gamma^{M_1 \dots M_5} \Gamma_{\underline a} \varepsilon &= 40 \Bigl[ 24 + 2 \bigl( \hat\gamma^{56} + \hat\gamma^{78} \bigr) F_{\rho\lambda} \gamma^{\rho\lambda} \Bigr] \hat\gamma_{\underline a} \varepsilon \;.
\eea
Sasaki-Einstein manifolds admit a Killing spinor $\chi$,
\be
\label{Killing spinor SE5}
\left[ \hat\nabla_{\underline{a}} + \frac i2 \, \hat\gamma_{\underline{a}} \right] \chi = 0 \;,
\ee
where $\hat\nabla$ is the covariant derivative on $\mathrm{SE}_5$. With our choice of vielbein, the spinor satisfies the projectors
\be
\label{projectors SE5}
\hat\gamma^{56} \chi = \hat\gamma^{78} \chi = i \chi \;,\qquad\qquad \hat\gamma^9 \chi = - \chi \;,
\ee
which imply $\partial\chi / \partial\psi_s = \frac i2 \chi$. Substituting into the 10d gravitino variation (\ref{10d gravitino variation}) one obtains the gravitino variation (\ref{KSE in 5d minimal sugra}) of 5d minimal gauged supergravity.

\paragraph{The case of \matht{S^5}.}
Let us specialize the discussion to the case of the Sasaki-Einstein manifold $S^5$, which is a globally-defined $U(1)$ fibration over $\bC\bP^2$. We use coordinates $(\rho_s, \theta_s, \varphi_s, \zeta_s, \psi_s)$ and take the following vielbein:
\bea
\label{vielbein S5}
e^5 &= d\rho_s \;,\qquad&
e^6 &= \frac14 \sin(2\rho_s) \, \bigl( d\zeta_s - \cos(\theta_s)\, d\varphi_s \bigr) \;,\quad&
\cA &= 3 \tan(\rho_s) \, e^6 - d\zeta_s \\
e^7 &= \frac12 \sin(\rho_s) \, d\theta_s \;,\quad&
e^8 &= \frac12 \sin(\rho_s) \, \sin(\theta_s) \, d\varphi_s \;,\qquad&
e^9 &= \frac13 \bigl( d\psi_s + \cA  + 2A \bigr) \;.
\eea
The ranges of coordinates are $\rho_s \in \bigl[ 0, \frac\pi2 \bigr]$, $\theta_s \in [0, \pi]$, $\varphi_s \in [0,2\pi)$, $\zeta_s \in [0, 4\pi)$, $\psi_s \in [0,6\pi)$ with the identifications
\be
\label{identifications S5}
\mat{\psi_s \\ \zeta_s \\ \varphi_s} \,\simeq\, \mat{ \psi_s + 6\pi \\ \zeta_s \\ \varphi_s } \,\simeq\, \mat{ \psi_s - 2\pi \\ \zeta_s + 4\pi \\ \varphi_s } \,\simeq\, \mat{ \psi_s - 2\pi \\ \zeta_s - 2\pi \\ \varphi_s + 2\pi } \;.
\ee
The coordinates $(\rho_s, \theta_s, \varphi_s, \zeta_s, \psi_s)$ are related to the coordinates $(\mu_a, \phi_a)$ (with $a=1,2,3$) of Section~\ref{sec: uplift to 10d} by%
\footnote{We can also introduce complex coordinates on $\bC\bP^2$ given by
\be
w_1 = \frac{\mu_1}{\mu_3} \, e^{i(\phi_1 - \phi_3)} = \tan(\rho_s) \, e^{i (\zeta_s - \varphi_s)/2} \cos(\theta_s/2) \;,\qquad
w_2 = \frac{\mu_2}{\mu_3} \, e^{i(\phi_2 - \phi_3)} = \tan(\rho_s) \, e^{i (\zeta_s + \varphi_s)/2} \sin(\theta_s/2) \;,
\ee
then the K\"ahler potential is $K = \log\bigl( 1 + |w_1|^2 + |w_2|^2 \bigr)$ while the K\"ahler form is $J = \frac i2 \partial \bar\partial K$.}
\bea
\mu_1 &= \sin(\rho_s) \cos(\theta_s/2) \;,\qquad\qquad& \phi_1 &= \bigl( 2 \psi_s + \zeta_s - 3\varphi_s \bigr)/6 \,, \\
\mu_2 &= \sin(\rho_s) \sin(\theta_s/2) \;,\qquad\qquad& \phi_2 &= \bigl( 2 \psi_s +  \zeta_s + 3\varphi_s \bigr) / 6 \,, \\
\mu_3 &= \cos(\rho_s) \;,\qquad\qquad& \phi_3 &= \bigl( \psi_s - \zeta_s \bigr) / 3 \;.
\eea
One can explicitly verify that the spinor
\be
\chi = e^{\frac i2 \psi_s} \, \chi_0 \qquad\text{with}\qquad i \hat\gamma^{56} \chi_0 = i \hat\gamma^{78}\chi_0 = \hat\gamma^9 \chi_0 = - \chi_0 \;,
\ee
where $\chi_0$ is a constant spinor, solves the Killing spinor equation (\ref{Killing spinor SE5}).

Notice that $\cA$ is singular along the loci $\theta_s = \pi$ (which is $\mu_1=0$), $\theta_s = 0$ (which is $\mu_2 = 0$), $\rho_s = \frac\pi2$ (which is $\mu_3=0$), as well as $\rho_s = 0$. Therefore it will not be well-suited to describe the 4-form potential through (\ref{general 5flux 4potential}) when discussing the embedding of Euclidean D3-branes and their on-shell action. Instead, we will take
\be
\wt\cA = 3 \tan(\rho_s) \, e^6 \;.
\ee
This is well-defined for all $\rho_s < \pi/2$, and will lead to smooth limits as we take $\rho_s \to \pi/2$.

Combining the five-dimensional analysis in spacetime with the one in the internal manifold, we conclude that for the 10d uplift of complex supersymmetric (but not necessarily extremal) black hole solutions with two angular momenta, using the frames (\ref{vielbein orthotoric}) and (\ref{vielbein S5}) with the gauge field (\ref{gauge potential orthotoric}), the Killing spinor $\varepsilon$ takes the form
\be
\label{10d Killing spinor BH general}
\varepsilon = \exp \left\{ \frac{i}{2} \left[ (3 - 2\alpha)t - \frac{4(1-a^2)}{(a^2 - b^2) \wt m} (\Phi - \Psi) - \frac{4 (1-b^2)}{ (a^2 - b^2) \wt m} (\Phi + \Psi) + \psi_s\right] \right\} \, \sqrt{f} \; \varepsilon_0
\ee
where $\varepsilon_0$ is a constant spinor, moreover
\be
\label{10d spinor projectors}
\Gamma^{09} \varepsilon = \varepsilon \;,\qquad \Gamma^{12} \varepsilon = -i \varepsilon \;,\qquad \Gamma^{34} \varepsilon = \Gamma^{56} \varepsilon = \Gamma^{78} \varepsilon = i \varepsilon \;,
\ee
and obviously the same projections hold for $\varepsilon_0$. In the case of black hole solutions with two equal angular momenta, using the frames (\ref{vielbein orthotoric a=b}) and (\ref{vielbein S5}) with gauge field (\ref{BH metric orthotoric a=b}), the 10d Killing spinor takes the form
\be
\label{10d Killing spinor BH a=b}
\varepsilon = \exp \left\{ \frac i2 \left[ (3-2\alpha) t + \wt\Phi + \psi_s \right] \right\} \, \sqrt{f} \; \varepsilon_0 \;,
\ee
with exactly the same projections as above.

\subsection{D3-brane embeddings}
\label{app: D3-branes}

Let us discuss various types of supersymmetric embeddings of D3-branes in the black hole geometries, and their on-shell actions. Let $\sigma^{0,1,2,3}$ be worldvolume coordinates on the D3-brane, and $X^\mu(\sigma)$ the embedding. The supersymmetry (or $\kappa$-symmetry) condition, in the absence of worldvolume flux on the D3-branes, is \cite{Cederwall:1996pv, Cederwall:1996ri, Bergshoeff:1996tu, Bergshoeff:1997kr}
\be
\label{kappa symm condition}
\Theta \, \varepsilon = \mp i \, \varepsilon
\ee
where
\be
\label{def Theta}
\Theta = \frac1{4!} \, \frac{\epsilon^{\alpha_1 \dots \alpha_4}}{\sqrt{-h}} \, \parfrac{X^{\mu_1}}{\sigma^{\alpha_1}} \dots \parfrac{X^{\mu_4}}{\sigma^{\alpha_4}} \, e\ud{M_1}{\mu_1} \dots e\ud{M_4}{\mu_4} \, \Gamma_{M_1 \dots M_4} \;.
\ee
Here $\alpha=0,\dots,3$ are worldvolume spacetime indices on the brane, $\mu = \{t, \dots, \psi_s\}$ are spacetime indices in 10d, while $M = 0, \dots, 9$ are vielbein indices in 10d. We take $\epsilon^{0123} = 1$, then $\mp$ in (\ref{kappa symm condition}) correspond to brane/anti-brane depending on conventions and worldvolume orientation. Finally
\be
h_{\alpha_1 \alpha_2} = \parfrac{X^{\mu_1}}{\sigma^{\alpha_1}} \, \parfrac{X^{\mu_2}}{\sigma^{\alpha_2}} \, G_{\mu_1 \mu_2}
\ee
is the induced metric on the D3-brane, while $h = \det (h_{\alpha\beta})$.

Notice that the supersymmetry condition (\ref{kappa symm condition})-(\ref{def Theta}) is valid for standard Lorentzian embeddings, as well as for Euclidean embeddings upon analytic continuation. On the other hand, once we discuss Euclidean embeddings, the Wick rotation $t \to -i t_\mathrm{E}$ of the background metric does not affect the D3-branes and their supersymmetry because they do not wrap the time coordinate.

\subsubsection{Equal angular momenta}

Consider first the case, discussed in the main text, of complex supersymmetric black hole solutions with two equal angular momenta, namely, with $a=b$. Working with the orthotoric coordinates of Section~\ref{sec: Killing spinor a=b} in spacetime and the coordinates of (\ref{vielbein S5}) on $S^5$, we consider the embedding
\bea
t &= \text{const} \quad& \wt\xi &= \text{const} \quad& \eta &= -1 \quad& \wt\Phi &= \text{const} -\sigma^0 \quad& \wt\Psi &= \sigma^0 \\
\rho_s &= \sigma^1 \quad& \theta_s &= \text{const} \quad& \varphi_s &= \text{const} \quad& \zeta_s &= \sigma^2 \quad& \psi_s &= \sigma^3 \;.
\eea
This is a Euclidean D3-brane wrapping a spatial $S^1$ in AdS$_5$ (so far at generic radial position $\wt\xi$ and at $\theta = \pi/2$) as well as a maximal $S^3 \subset S^5$ located at $z_1 / z_2 = \text{constant}$ in complex coordinates.
We compute
\be
\sqrt{-h} = i \, \frac{\sin(2\rho_s)}6 \sqrt{ \frac{\wt\cF}{f \, \wt\xi} - f^2 \omega_{\wt\Phi}^2} \;\;,\qquad
\Theta = \frac{i}{\sqrt{ \frac{\wt\cF}{f \, \wt\xi} - f^2 \omega_{\wt\Phi}^2}} \left[ f \omega_{\wt\Phi} \, \Gamma^{0569} + \sqrt{ \frac{\wt\cF}{f \, \wt\xi}} \; \Gamma^{2569} \right] \;.
\ee
Here $\omega_{\wt\Phi}$ is the component of the 1-form $\omega$ in (\ref{orthotoric a=b functions 2}) along $d\wt\Phi$.
We see that, in general, this is not a supersymmetric embedding. However at the horizon, located at $\wt\xi = \wt\xi_+$, we have $\wt\cF\bigl(\wt\xi_+ \bigr) = 0$. Choosing the sign of the square root in such a way to match the analytic continuation of the Lorentzian expression (\ref{DBI part of D3 action}), we find
\be
\sqrt{-h} = \pm \, \frac{\sin(2\rho_s) \, f \, \omega_{\wt\Phi}}6 \qquad\text{and}\qquad \Theta = \mp \Gamma^{0569}
\ee
for an embedding at the horizon, where the upper/lower signs refer to the \first/\second{} branch of supersymmetric solutions. Using the projectors (\ref{10d spinor projectors}) we obtain $\Theta\varepsilon = \mp i\varepsilon$, thus the embedding is supersymmetric when placed at the horizon. We also see that (depending on conventions) one branch of solutions host a supersymmetric brane while the other one a supersymmetric anti-brane.

The embeddings we just described wrap a particular family of $S^3$'s inside $S^5$. However, recall that the uplift of 5d minimal gauged supergravity into 10d type IIB on $S^5$ breaks $SO(6)_R \cong SU(4)_R \to U(1)_R \times SU(3)$ and is invariant under $SU(3)$. Therefore, there is a more general class of embeddings in which the $S^3 \subset S^5$ is given by the linear complex equation $a \, z_1 + b \, z_2 + c \, z_3 = 0$ inside $|z_1|^2 + |z_2|^2 + |z_3|^2 = 1$ \cite{Mikhailov:2000ya}.

For instance, let us exhibit the embedding at $\mu_3 = 0$ (and at the horizon). This corresponds to
\bea
\label{embedding m3=0 for a= b}
t &= \text{const} \quad& \wt\xi &= \wt\xi_+ \quad& \eta &= -1 \quad& \wt\Phi &= \text{const} -\sigma^0 \quad& \wt\Psi &= \sigma^0 \\
\rho_s &= \frac\pi2 \quad& \theta_s &= \sigma^1\quad& \varphi_s &= \sigma^2 \quad& \zeta_s &= \sigma^3 \quad& \psi_s &= \sigma^3 \;.
\eea
With a little bit of algebra one finds
\be
\label{Euclidean emb induced metric}
\sqrt{-h} = \pm \, \frac{\sin(\theta_s) \, f \, \omega_{\wt\Phi} }4 \qquad\text{and}\qquad \Theta = \mp \Gamma^{0789} \;.
\ee
Using the projectors (\ref{10d spinor projectors}) we obtain $\Theta\varepsilon = \mp i\varepsilon$, thus the embedding is supersymmetric.

\paragraph{On-shell action.}
Let us determine the on-shell action for the Euclidean embedding (\ref{embedding m3=0 for a= b}). Using the change of coordinates (\ref{change to orthotoric coords a=b}) as well as the identifications (\ref{twisted identifications}) and (\ref{identifications S5}), we conclude that $\sigma^0$ has period $2\pi$, $\sigma^1$ has range $\pi$, $\sigma^2$ has (twisted) period $2\pi$ while $\sigma^3$ has period $4\pi$. The D3-brane action (\ref{D3 action}) in our conventions reads
\be
\label{general on-shell action}
S_\text{D3} = - \frac{N}{2\pi^2} \int \biggl( d^4\sigma \, \sqrt{-h} \mp P[ C_{(4)} ]\biggr) \;,
\ee
where $P[C_{(4)}]$ is the pull-back of $C_{(4)}$ to the brane worldvolume, while the $\mp$ sign refers to branes/anti-branes.

Using the horizon location $\wt\xi_+$ on the first branch, given in (\ref{horizon location xi tilde +}), one finds the values of the functions
\be
f \, \omega_{\wt\Phi} \Big|_{\xi = \wt\xi_+} = \frac{(a-ir_+)(r_*^2 + ir_+) }{ 2(1-a^2)(a + i r_+) } \;,
\ee
as well as
\be
L \,\equiv\, \frac{a^2 (1+\wt m) f}{ 2(1-a)^2\wt\xi } \bigg|_{\xi=\xi_+} = - \frac{ a (1-ir_+)(a-i r_+) }{ (1-a^2) (a+ir_+)}
\ee
which equals $- \frac43 A_{\wt\Phi}$ and appears in the connection (\ref{BH metric orthotoric a=b}).
Given the integration ranges, we obtain the integral of the metric part:
\be
\int d^4\sigma\, \sqrt{-h} = 16 \pi^3 \int \frac{\sin(\theta_s)}4 \, f\omega_{\wt\Phi} \Big|_{\wt\xi_+} d\theta_s = 4\pi^3 \, \frac{(2a+a^2 + ir_+)(a-ir_+) }{ (1-a^2)(a + i r_+) } \;.
\ee
Let us now move to the 4-form potential. Using (\ref{general 5flux 4potential}), the only term that contributes to the pull-back is $\frac29 \wt\cA \wedge J \wedge A$. We compute
\be
\int P[C_{(4)}] = - \frac L8 \int \sin(\sigma^1) \, d\sigma^0 \wedge d\sigma^1 \wedge d\sigma^2 \wedge d\sigma^3 = 4\pi^3 \, \frac{ a (1-ir_+)(a-i r_+) }{ (1-a^2) (a+ir_+)} \;.
\ee
Summing the two contributions using the upper minus sign in (\ref{general on-shell action}), we obtain
\be
S_\text{D3} = 2\pi N \, \frac{a-i r_+}{a-1} = 2\pi N \, \frac{\Delta_\mg}{\tau_\mg}
\ee
as in Section~\ref{sec: brane action}, where in the second equality we used the values (\ref{susyvalues}) of the chemical potentials. Recall that this result is obtained on the first branch.

To obtain the on-shell action on the second branch, we should send $i \to -i$ in the expression of $q$ and then of $\wt\xi_+$ and the formulas that follow. According to (\ref{Euclidean emb induced metric}), we should also choose the lower sign in the expression of $\sqrt{-h}$ in order to match the analytic continuation of the Lorentzian expression. Moreover, we should take the lower plus sign in (\ref{general on-shell action}) because anti-branes are supersymmetric on the second branch, as follows from the $\kappa$-symmetry operator in (\ref{Euclidean emb induced metric}). Taking all of this into account, we obtain
\be
S_\text{D3} = -2\pi N \, \frac{a+i r_+}{a-1} = - 2\pi N \, \frac{\Delta_\mg}{\tau_\mg}
\ee
on the second branch.

\subsubsection{Generic angular momenta}

The discussion in the general case is similar to the previous one and the one in Section~\ref{sec: brane action}. We consider a Euclidean D3-brane located at the black hole horizon $r=r_+$, wrapping the spatial $S^1$ in AdS$_5$ along $\phi$ at $\theta = \frac\pi2$, as well as the maximal $S^3 \subset S^5$ located at $\mu_3 = 0$ (that is $z_3 = 0$ in complex coordinates). By the same argument that lead to (\ref{DBI part of D3 action}), by evaluating the metric component $g_{\phi\phi}$ in generic black hole solutions (\ref{CCLP solution I}) one concludes that
\be
\label{DBI part general a neq b}
\int d^4x \, \sqrt{- \det(g_\text{D3})} = - 4\pi^3 i \, \frac{ (r_+^2 + a^2)(r_+^2 + b^2) + abq }{ r_+ (r_+^2 + b^2) (1-a^2) } \;.
\ee
The sign of the square root has been chosen in such a way that in the background of real and causally-well-behaved Lorentzian black holes, the contribution to $\exp(iS_\text{D3})$ is bounded in absolute value. We will use this expression to analytically continue the square root to complex metrics.

In order to discuss supersymmetry of the embedding, it is better to use the orthotoric coordinates of Section~\ref{sec: Killing spinor} in spacetime and the coordinates of (\ref{vielbein S5}) on $S^5$. We thus consider the embedding
\bea
t &= \text{const} \quad& \xi &= \xi_+ \quad& \eta &= -1 \quad& \Phi &= \text{const} -\sigma^0 \quad& \Psi &= \sigma^0 \\
\rho_s &= \frac\pi2 \quad& \theta_s &= \sigma^1 \quad& \varphi_s &= \sigma^2 \quad& \zeta_s &= \sigma^3 \quad& \psi_s &= \sigma^3 \;.
\eea
Given the value of $q$ in (\ref{q on first branch generic}), the location of the horizon in orthotoric coordinates is
\be
\xi_+ = \frac{ ir_+(a+b+2) - a - a^2 - b - b^2 }{ (a-b)(1+a+b - i r_+) }
\ee
on the first branch, while the value on the second branch is obtained by sending $i \to -i$. One then derives the following useful identities:
\be
\cF(\xi_+) = 0 \;,\qquad \cG(\pm1) = 0 \;,\qquad \omega_\Phi - \omega_\Psi \Big|_{\eta = 1} = 0 \;,\qquad \omega_\Phi + \omega_\Psi \Big|_{\eta = -1} = 0 \;.
\ee
With them, one computes
\be
\sqrt{-h} = \pm \, \frac{\sin(\theta_s) \, f \, (\omega_\Phi - \omega_\Psi) }8 \;,\qquad\qquad \Theta = \mp \Gamma^{0789} \;.
\ee
In the first formula, the sign of the square root has been chosen in such a way to match%
\footnote{Notice that $\sqrt{-\det(g_\text{D3})}$ and $\sqrt{-h}$ are computed in different coordinates. In order to compare them, one should include the factor $\wt m(b^2 - a^2) / 8(1-a^2)$ from the Jacobian of the coordinate change from $\sigma^0$ to $\phi$.}
the analytic continuation of (\ref{DBI part general a neq b}), leading to the $\pm$ sign that refers to the \first/\second{} branch of solutions. The second formula implies $\Theta\varepsilon = \mp i \varepsilon$, showing that the embedding is supersymmetric. It also implies that branes with opposite charge are supersymmetric on the two branches.

Computing the on-shell action is easier in the original coordinates of (\ref{CCLP solution I}) and for generic (not necessarily supersymmetric) black hole backgrounds. We already computed the Dirac-Born-Infeld part in (\ref{DBI part general a neq b}). The Wess-Zumino part immediately follows from $\frac13 \int_{S^1} A$. The full D3-brane on-shell action is
\be
S_\text{D3}^\phi = - \frac{N}{2\pi^2} \biggl[ - 4\pi^3 i \, \frac{ (r_+^2 + a^2)(r_+^2 + b^2) + abq }{ r_+ (r_+^2 + b^2)(1-a^2) } \pm 4\pi^3 \, \frac{qa}{ (r_+^2 + b^2)(1-a^2) } \biggr] \;,
\ee
where the $\pm$ sign refers to branes/anti-branes. Substituting the supersymmetric value of $q$ on the two branches, and considering branes ($+$ sign) on the first branch and anti-branes ($-$~sign) on the second branch, we obtain
\be
S^\phi_\text{D3} = 2\pi N \, \frac{a-i r_+}{a-1} 
\quad(\text{\first{} branch}) \;,\qquad\quad
S^\phi_\text{D3} = - 2\pi N \, \frac{a+i r_+}{a-1}
\quad(\text{\second{} branch}) \;.
\ee
These formulas reproduce the expressions in (\ref{eq: general brane action}) and (\ref{eq: general brane action2}), in the case that $\Delta_{\mg,a} \equiv \Delta_\mg$.

Next, consider a Euclidean D3-brane located at the black hole horizon $r=r_+$, wrapping the spatial $S^1$ in AdS$_5$ along $\psi$ at $\theta = 0$, as well as the $S^3$ at $\mu_3 = 0$. By evaluating the metric component $g_{\psi\psi}$ in generic black hole solutions (\ref{CCLP solution I}) one concludes that
\be
\int d^4x \, \sqrt{- \det(g_\text{D3})} = - 4\pi^3 i \, \frac{ (r_+^2 + a^2)(r_+^2 + b^2) + abq }{ r_+ (r_+^2 + a^2) (1-b^2) } \;.
\ee
We will use this expression to perform analytic continuations. In orthotoric coordinates, the embedding reads
\bea
t &= \text{const} \quad& \xi &= \xi_+ \quad& \eta &= 1 \quad& \Phi &= \text{const} + \sigma^0 \quad& \Psi &= \sigma^0 \\
\rho_s &= \frac\pi2 \quad& \theta_s &= \sigma^1 \quad& \varphi_s &= \sigma^2 \quad& \zeta_s &= \sigma^3 \quad& \psi_s &= \sigma^3 \;.
\eea
One computes%
\footnote{This time the Jacobian factor of the coordinate change from $\sigma^0$ to $\psi$ is $\wt m(a^2 - b^2) / 8(1-b^2)$.}
\be
\sqrt{-h} = \mp \, \frac{\sin(\theta_s) \, f \, (\omega_\Phi + \omega_\Psi) }8 \;,\qquad\qquad \Theta = \mp \Gamma^{0789} \;.
\ee
This implies $\Theta\varepsilon = \mp i \varepsilon$, showing that the embedding is supersymmetric.

Finally, we compute the D3-brane on-shell action on generic black hole backgrounds:
\be
S_\text{D3}^\psi = - \frac{N}{2\pi^2} \biggl[ - 4\pi^3 i \, \frac{ (r_+^2 + a^2)(r_+^2 + b^2) + abq }{ r_+ (r_+^2 + a^2)(1-b^2) } \pm 4\pi^3 \, \frac{qb}{ (r_+^2 + a^2)(1-b^2) } \biggr] \;.
\ee
Substituting the value (\ref{q on first branch generic}) of $q$ on the two branches, and choosing the upper/lower sign for branes/anti-branes on the \first/\second{} branch, we obtain
\be
S^\psi_\text{D3} = 2\pi N \, \frac{b-i r_+}{b-1} \quad(\text{\first{} branch}) \;,\qquad\quad
S^\psi_\text{D3} = - 2\pi N \, \frac{b+i r_+}{b-1} \quad(\text{\second{} branch}) \;,
\ee
that reproduce the expressions in (\ref{eq: general brane action}) and (\ref{eq: general brane action2}).

\subsubsection{More supersymmetric D3-branes}
\label{sec:more branes}

It turns out that there exists another class of Euclidean supersymmetric D3-branes, located at the black hole horizon, wrapping the full spatial $S^3$ in AdS$_5$ as well as a maximal $S^1 \subset S^5$ (say at $\mu_2 = \mu_3 = 0$ for definiteness). Considering the case of generic angular momenta (\ref{CCLP solution I}), one finds
\be
\label{volume form new D3-branes}
\int_\text{D3} d^4x\, \sqrt{ -\det\bigl( g_\text{D3}\bigr)} = - 4\pi^3 i \, \frac{ (r_+^2+a^2) (r_+^2 + b^2) + abq }{ r_+ (1-a^2)(1-b^2) } \;,
\ee
where the sign has been chosen so as to give a contribution bounded in absolute value to the path-integral measure $e^{iS_\text{D3}}$ on real Lorentzian  black-hole backgrounds.

In order to compute the contribution from the WZ term, we use the local expression (\ref{alpha_2 generic ang mom}) for $*_5 F - \frac23 A \wedge F = d\alpha_{(2)}$. Using (\ref{general 5flux 4potential}), the only term in $C_{(4)}$ that contributes to the WZ term is $\frac19 \wt \cA \wedge\bigl( *_5 F - \frac23 A \wedge F \bigr)$. The integral of the spacetime part is simply
\be
\int_{S^3} d\alpha_{(2)} = 4\pi^2 \, \frac{3q}{2\Xi_a \Xi_b} \;.
\ee
The integral of $\wt\cA$ along $S^1$ should be computed carefully, taking into account the possible gluing of patches. One obtains $\int \wt\cA = 6\pi$.

Combining the two contributions, we obtain
\be
S_\text{D3} = - \frac{N}{2\pi^2} \biggl[ - 4\pi^3 i \, \frac{(r_+^2 + a^2) (r_+^2 + b^2) + abq}{ r_+(1-a^2)(1-b^2) } \mp 4\pi^3 \, \frac{ q}{ (1-a^2)(1-b^2) } \biggr] \;.
\ee
On the first branch, where $q = - (a - ir_+)(b - ir_+)(1 - ir_+)$, and choosing the upper minus sign for anti-D3-branes, we obtain
\be
\label{action new D3-branes equal ang mom}
S_\text{D3} = - 2\pi N \, \frac{(a-ir_+)(b-ir_+)}{ (a-1)(b-1) } = - 2\pi N \, \frac{\Delta_\mg^2}{\sigma_\mg \tau_\mg} \qquad\qquad (\text{\first{} branch}) \;.
\ee
On the second branch, choosing the lower plus sign for D3-branes, we obtain
\be
S_\text{D3} = 2\pi N \, \frac{(a + ir_+)(b + ir_+)}{ (a-1)(b-1) } = 2\pi N \, \frac{\Delta_\mg^2}{\sigma_\mg \tau_\mg} \qquad\qquad (\text{\second{} branch}) \;.
\ee

Let us also check that the embeddings are supersymmetric. In terms of the orthotoric coordinates of Section~\ref{sec: Killing spinor}, the embedding reads
\bea
t &= \text{const} \qquad& \xi &= \xi_+ \qquad& \eta &= \sigma^0 \qquad& \Phi &= \sigma^1 \qquad& \Psi &= \sigma^2 \\
\rho_s &= \frac\pi2 \qquad& \theta_s &= 0 \qquad& \varphi_s &= -\sigma^3 \qquad& \zeta_s &= \sigma^3 \qquad& \psi_s &= \sigma^3 \;.
\eea
One computes
\be
\sqrt{h} = \pm (\xi_+ \omega_\Phi - \omega_\Psi) \;,\qquad\qquad \Theta = \pm \Gamma^{0349} \;,
\ee
where the sign of the square root has been chosen in such a way to match the analytic continuation of (\ref{volume form new D3-branes}) to the \first/\second{} branch of supersymmetric solutions. The second formula implies $\Theta \varepsilon = \pm i \varepsilon$, implying that Euclidean anti-D3-branes are supersymmetric on the \first{} branch while D3-branes are supersymmetric on the \second{} branch.

\section{Solutions with three electric charges}
\label{app: Three charges}

\newcommand{\HHH}{(H_1 H_2 H_3)}

In this appendix we generalize the solution described in the main text to the case of three different $U(1)$ charges, but with equal angular momenta.
There are known non-supersymmetric asymptotically locally AdS$_5$ black hole solutions with three different $U(1)$ charges and equal angular momenta \cite{Cvetic:2004ny} (see also \cite{Cvetic:2005zi, Cassani:2019mms}).%
\footnote{The general black hole solutions with six independent charges --- mass, two angular momenta, and three electric charges --- have been constructed in \cite{Wu:2011gq}.}
They are classical solutions to the $U(1)^3$ 5d gauged supergravity action (which is a consistent truncation of 10d type IIB supergravity on $S^5$)
\begin{multline}
	S_\text{5d} = \frac{1}{16\pi} \int \Biggl[ \biggl( R + 4\sum_{I=1}^{3}(X^{I})^{-1}-\frac{1}{2}\partial\vec{\chi}^{2} \biggr) * 1 -\frac{1}{2}\sum_{I=1}^{3} \bigl( X^{I} \bigr)^{-2} F^{I} \wedge * F^{I} \\
	{} - \frac{1}{6}|\epsilon_{IJK}|A^{I}\wedge F^{J}\wedge F^{K} \Biggr] \;,
\end{multline}
where $A^{I}$ (with $I=1,2,3$) are Abelian gauge fields, $F^{I}=dA^{I}$ are their field strengths, \mbox{$\vec{\chi}=(\chi_{1},\chi_{2})$} are real scalar fields, $X^{1} = \exp \left( -\frac{1}{\sqrt{6}} \, \chi_{1} - \frac{1}{\sqrt{2}} \, \chi_{2}\right)$, $X^{2} = \exp\left(-\frac{1}{\sqrt{6}} \, \chi_{1}+\frac{1}{\sqrt{2}} \, \chi_{2}\right)$, $X^{3}=\exp\left(\frac{2}{\sqrt{6}} \, \chi_{1}\right)$ such that $X^1X^2X^3=1$.

The solutions relevant to us are
\begin{align}
	ds_5^{2} &= \left(H_{1}H_{2}H_{3}\right)^{\frac{1}{3}} \biggr[ -\frac{r^{2}Y}{f_{1}}dt^{2} + \frac{r^{4}}{Y}dr^{2} + \frac{r^{2}}{4} \bigl( \sigma_{1}^{2}+\sigma_{2}^{2} \bigr) + \frac{f_{1}}{4r^{4}H_{1}H_{2}H_{3}} \left( \sigma_{3}-\frac{2f_{2}}{f_{1}}dt \right)^{2} \biggr] \nn\\
	X^{I}&=\frac{\left(H_{1}H_{2}H_{3}\right)^{\frac{1}{3}}}{H_{I}} \nn\\
	A^{I}&=\left(\frac{2m}{r^{2}H_{I}}s_{I}c_{I} - \alpha_{I}\right)dt+\frac{ma}{r^{2}H_{I}} \bigl( c_{I}s_{J}s_{K}-s_{I}c_{J}c_{K} \bigr) \sigma_{3}\equiv A_{t}^{I}dt+A_{\psi}^{I}\sigma_{3}
\end{align}
where the $\sigma_{i}$ are left-invariant one-forms on a three-sphere $S^{3}$ parameterized by $(\theta,\phi,\psi)$:%
\footnote{We choose angular coordinates that agree with the body of the paper. They can be related to those of \cite{Cassani:2019mms} by $\theta_\text{there} = 2\theta_\text{here}$ and $\phi_\text{there} = \psi_\text{here} - \phi_\text{here}$, $\psi_\text{there} = \psi_\text{here} + \phi_\text{here}$.}
\be
	\sigma_{1}+i\sigma_{2} = e^{-i(\phi+\psi)} \Bigl( 2 d\theta + i \sin(2\theta)\, (d\psi-d\phi) \Bigr) \;, \qquad \sigma_{3} = 2\Bigl(\cos^2(\theta) \, d\psi + \sin^2(\theta) \, d\phi \Bigr) \;,
\ee
while in the definition of $A^I$ we took $I,J,K$ all different.
The coefficients are radial functions:
\begin{equation}
	\begin{aligned}
		H_I &= 1+\frac{2ms_{I}^{2}}{r^{2}} \\
		f_1 &=  r^6 H_1 H_2 H_3 +2ma^{2}r^{2}+4m^{2}a^{2} \Bigl[ 2 \bigl( c_{1}c_{2}c_{3}-s_{1}s_{2}s_{3} \bigr) s_{1}s_{2}s_{3}-s_{1}^{2}s_{2}^{2}-s_{2}^{2}s_{3}^{2}-s_{3}^{2}s_{1}^{2} \Bigr] \\
		f_2 &= 2ma \bigl( c_{1}c_{2}c_{3}-s_{1}s_{2}s_{3} \bigr) r^{2}+4m^{2}as_{1}s_{2}s_{3} \\
		f_3 &= f_1 - r^6 H_1 H_2 H_3 + 2ma^2 \\
	\end{aligned}
\end{equation}
and 
\be
s_{I}=\sinh(\delta_{I}) \;, \qquad c_{I}=\cosh(\delta_{I}) \;,\qquad I=1,2,3 \;.
\ee
The solution therefore depends on five parameters:%
\footnote{Despite their similar names, they are not identical to those of the single-charge case of Section \ref{sec: black holes 5d} when the charges are equal.}
$m, a, \delta_1, \delta_2, \delta_3$. Once again, $r_+$ is the largest root of $Y$. A useful way to relate it to $m$ is by a change of coordinates to
\begin{equation}
	\label{eq:rToR_coord}
	R^2 = r^2 + 2ms_1^2 \,.
\end{equation} 
Now $Y$ becomes quadratic in $m$, and the relation between $m$ and $R_+$ can be easily solved to produce a lengthy expression for $m$ in terms of $R_+$ that will be omitted here. It requires a branch choice for a square root, and this choice determines whether we will be in the first or second branch. 

To the horizon at $r = r_+$ we can associate entropy $S$, inverse temperature $\beta$, angular velocity relative to a non-rotating frame at infinity $\Omega$, and electrostatic potentials $\Phi^I$: 
\begin{equation}
	\begin{aligned}
		S &= \frac{\pi^2}{2}\sqrt{f_1(r_+)} \;,\qquad & \beta &=4\pi r_+ \sqrt{f_1(r_+)}\left(\frac{dY}{dr} \Big\vert_{r=r_+} \right)^{-1} , \\
		\Omega &= \frac{f_2(r_+)}{f_1(r_+)} \;,  &\Phi^I &= \frac{m}{r_+^2 \, H_I(r_+)} \Bigl( 2s_Ic_I + 2a\Omega \bigl( c_I s_J s_K - s_I c_J c_K \bigr)\Bigr) \;.
	\end{aligned}
\end{equation}
The energy $E$, angular momentum $J$, and electric charges $Q_I$ are
\begin{equation}
	\begin{aligned}
		E &= E_0 + \frac{1}{4}m\pi \bigl( 3 + a^2 + 2s_1 + 2s_2 + 2s_3 \bigr) \;,\qquad\quad&
				Q_I &= \frac{1}{2}m\pi s_I c_I \;, \\
		J &= \frac{1}{2}m a \pi \bigl( c_1 c_2 c_3 - s_1 s_2 s_3 \bigr) \;,
	\end{aligned}
\end{equation}
where $E_0  = \frac{3\pi}{32}$ is the energy of empty AdS$_5$ using holographic renormalization \cite{Cassani:2019mms}.

The solution can be rotated to Euclidean signature where regularity determines the global identification 
\be
(t_\mathrm{E}, \psi, \phi)\, \cong \, (t_\mathrm{E} + \beta, \psi - i\Omega\beta, \phi - i\Omega\beta) \;,
\ee
and the gauge fixing to be $\alpha_{I} = \Phi_{I}$, and thus the gauge fields on the boundary are
\begin{equation}
	A^{I} \big|_{\rm bdry} = i\Phi_{I} dt_\mathrm{E} \;.
\end{equation}

The boundary metric is again that of $S^1 \times S^3$, and $\Omega$ appears only through the identification of coordinates. Again, we can shift $\Omega$ or $\Phi_I$ by $2\pi i / \beta$ to find new solutions with the same boundary conditions.

The solution is supersymmetric if 
\begin{equation}
	\label{eq:susy_cond_3Q}
	a = e^{-\delta_1 - \delta_2 - \delta_3} = \prod_{I=1,2,3} (c_I - s_I) \;,
\end{equation}
and in that case we can define chemical potentials as the deviation from their BPS values \cite{Cassani:2019mms}
\begin{equation}
	\tau_\mrg = \frac{\beta}{2\pi i} \bigl( \Omega - 1 \bigr) \;,\qquad\qquad\Delta_{\mrg,a} = \frac{\beta}{2\pi i} \bigl( \Phi_a - 1 \bigr) \;.
\end{equation}
One finds that they satisfy the constraint
\begin{equation}
	\label{eq:branches 3 charges def}
	2\tau_\mrg - \Delta_{\mrg,1} - \Delta_{\mrg,2} - \Delta_{\mrg,3} = \pm 1 \qquad\qquad\bigl( 1^\text{st}/2^\text{nd} \text{ branch} \bigr) \;.
\end{equation}

Let us embed this black hole into a solution of 10d type IIB supergravity, and ask what would be the contribution of wrapped D3-branes similar to the ones considered before. The uplift of this solution to 10d using \cite{Cvetic:1999xp} has metric
\begin{equation}
	ds_{10}^2 = \wt{\Lambda \mkern 0mu}^{1/2} ds_5^2 + \wt{\Lambda \mkern 0mu}^{-1/2} \sum_{I=1}^{3} \bigl( X^I \bigr)^{-1} \Bigl( d\mu_I^2 + \mu_I^2 \bigl( d\phi_I + A^I \bigr)^2 \Bigr) \;,
\end{equation}
where $\wt\Lambda = \sum_{I=1}^{3} X^I \mu_I^2$ and $\phi_{1,2,3}$, $\mu_{1,2,3}$ (with $\sum_{I=1}^3 \mu_I^2 = 1$) are a parameterization of $S^5$ as the phases and magnitudes of three complex numbers $\mu_I \, e^{i\phi_I}$ on the unit sphere in $\bC^3$. Its vielbeins are
\bea
		e^0 &= \wt{\Lambda \mkern 0mu}^{1/4}\sqrt{\frac{\HHH^{1/3} \, r^2 \, Y}{f_1}} \, dt \;, & e^1 &= \wt{\Lambda \mkern 0mu}^{1/4} \sqrt{\frac{\HHH^{1/3} \, r^4}{Y}} \, dr \;, \\ 
		e^2 &= \wt{\Lambda \mkern 0mu}^{1/4} \frac{\HHH^{1/6} \, r}{2} \, \sigma_1 \;, & e^3 &= \wt{\Lambda \mkern 0mu}^{1/4} \frac{\HHH^{1/6} \, r}2 \, \sigma_2 \;, \\
		e^4 &= \frac{\wt{\Lambda \mkern 0mu}^{1/4} \sqrt{f_1}}{2\HHH^{1/3} \, r^2}\left(\sigma_3 - \frac{2f_2}{f_1} \, dt\right) \;,\qquad & e^5 &= \frac{\mu_1}{\wt{\Lambda}^{1/4} \, \sqrt{X^1}} \bigl( d\phi_1 + A^1\bigr) \;, \\ 
		e^6 &=\frac{\mu_2}{\wt\Lambda^{1/4} \, \sqrt{X^2}} \bigl( d\phi_2 + A^2\bigr) \;, & e^7 &=\frac{\mu_3}{\wt\Lambda^{1/4} \, \sqrt{X^3}} \bigl( d\phi_3 + A^3 \bigr) \;,
	\end{aligned}
\end{equation}
\begin{equation}
	\begin{aligned}
		e^8 &= \wt{\Lambda \mkern 0mu}^{-1/4} \sqrt{\frac{X^3 \mu_3^2 + X^1 \mu_1^2}{\mu_3^2 X^1 X^3}}\left(d\mu_1 + \frac{\mu_1\mu_2 X^1}{X^3 \mu_3^2 + X^1 \mu_1^2} \, d\mu_2 \right) \;, \\	
		e^9 &= \wt{\Lambda \mkern 0mu}^{-1/4} \sqrt{\frac{\left(X^3\mu_3^2 + X^2\mu_2^2 \right) \left(X^3\mu_3^2 + X^1\mu_1^2\right) - \mu_1^2 \mu_2^2 X^1 X^2}{\mu_3^2 X^3 X^2 \left(X^3\mu_3^2 + X^1\mu_1^2\right)}} \, d\mu_2 \;. \hspace{5.4em}
	\end{aligned}
\end{equation}

When $\mu_1 = 0$, the last two simplify to 
\be
	e^8 = \frac{d\mu_1}{\wt\Lambda^{1/4} \, \sqrt{X^1}} \;,\qquad\qquad e^9 = \wt{\Lambda \mkern 0mu}^{-1/4} \sqrt{\frac{X^3\mu_3^2 + X^2 \mu_2^2}{\mu_3^2 X^3 X^2}} \, d\mu_2 \;.
\ee

There is also a self-dual 5-form flux $F_{(5)} = G_{(5)} + * \,G_{(5)}$ in the ten-dimensional solution, which is given by%
\footnote{There is a typo in the formula for $G_{(5)}$, eqn. (2.8), in \cite{Cvetic:1999xp}. See footnote 18 in \cite{Benini:2013cda}.}
\be
\label{eq:G5 def}
G_{(5)} = \sum_{I=1}^3 \left[ 2X^I \bigl( X^I \mu_I^2 - \wt\Lambda \bigr) \, \epsilon_{(5)} + \frac1{2 (X^I )^2} \, d\bigl( \mu_I^2 \bigr) \wedge \Bigl( \bigl( d\phi_I + A^I \bigr) \wedge *_5 F^I + X^I *_5 dX^I \Bigr) \right]
\ee
where $\epsilon_{(5)}$ the volume form of the original 5d metric, and hence
\bea
\label{eq:*G5 def}
* \, G_{(5)} &= \sum_{I=1}^3 \biggl[ 2X^I \bigl( X^I \mu_I^2 - \wt\Lambda \bigr) * \epsilon_{(5)} - \frac{\wt\Lambda^{-2}}{2X^I} \, dX^I \wedge \tilde *_5 \, d\bigl( \mu_I^2 \bigr) \\
&\qquad\qquad + \frac{\wt\Lambda^{-1/4}}{2(X^I)^2} \, \tilde *_5 \Bigl( d\bigl( \mu_I^2 \bigr) \wedge \bigl( d\phi_I + A^I \bigr) \Bigr) \wedge F^I \biggr] \;,
\eea
where $\tilde *_5$ is the Hodge dual with respect to the metric on $S^5$.
At $\mu_1 = 0$ the third term in $* \, G_{(5)}$ simplifies to
$d\bigl( \mu_2^2 \bigr) \wedge \left(d\phi_2 + A^2\right)\wedge \left(d\phi_3 + A^3\right) \wedge F^1$,
implying
\begin{equation}
	C_{(4)} = \frac{1}{2} \, d \bigl( \mu_2^2 \bigr) \wedge \left(d\phi_2 + A^2\right)\wedge \left(d\phi_3 + A^3\right) \wedge A^1 + \text{other terms} \,.
\end{equation}

Let us now consider the action of a D3-brane at $r=r_+$, $\theta = \frac{\pi}{2}$, $\mu_1 = 0$, along the $\phi$ direction.
We assume that the term discussed above is the only one contributing to the brane action via the term 
\begin{equation}
	\int_\text{D3} P[C_{(4)}] = \int_{S^3} \mu_2 \, d\mu_2\wedge d\phi_2 \wedge d\phi_3 \int_{S^1} A_\phi^1 \, \sigma_3 = 8\pi^3 \frac{m a}{r_+^2 H_1(r_+)}(c_1 s_2 s_3 - s_1 c_2 c_3) \;.
\end{equation}
The contribution of the tension of the brane is proportional to $\int_\text{D3} \sqrt{-\det(g_\text{D3})}$. The determinant of the induced metric is
\begin{equation}
	\det(g_\text{D3}) = g_{\mu_2\mu_2} \Bigl( g_{\phi\phi} \, g_{\phi_2\phi_2} \, g_{\phi_3\phi_3} - g_{\phi\phi_2}^2 \, g_{\phi_3\phi_3} - g_{\phi\phi_3}^2 \, g_{\phi_2\phi_2} \Bigr) \;.
\end{equation}
Using
\bea
	g_{\phi\phi} &= \frac{\wt{\Lambda \mkern 0mu}^{1/2} f_1}{r^4\HHH^{2/3}} + \sum_{I=1}^{3}\frac{\mu_I^2 \, (A_\phi^I)^2}{\wt\Lambda^{1/2} X^I } \;, &
	g_{\phi\phi_2} &= \frac{\mu_2^2 A_\phi^2}{ \wt\Lambda^{1/2} X^2 } \;, &g_{\phi\phi_3} &=\frac{\mu_3^2 A_\phi^3}{ \wt\Lambda^{1/2} X^3} \;, \\
	g_{\mu_2\mu_2} &= \frac{\mu_2^2 X^2 + \mu_3^2 X^3}{\wt\Lambda^{1/2} \mu_3^2 X^2 X^3} \;,& g_{\phi_2\phi_2} &= \frac{\mu_2^2}{\wt\Lambda^{1/2} X^2} \;, & g_{\phi_3\phi_3} &=  \frac{\mu_3^2}{\wt\Lambda^{1/2} X^3} \;,
\eea
we find
\begin{equation}
	\label{eq: detg 3 charges}
	\sqrt{-\det(g_\text{D3})} = \frac{i\mu_2\sqrt{f_1(r_+)}}{r_+^2H_1(r_+)} \;.
\end{equation} 
Note that the off-diagonal metric terms from the mixing of $\phi_I$ and $A^I$ cancel in the determinant, regardless of the details of the 5d solution (and thus do so also in the single-charge solution discussed in the body of the paper).

Let us begin for simplicity from the case where two of the three charges are equal, namely, $Q_1 \neq Q_2 = Q_3$. Correspondingly we take $\delta_2 = \delta_3$. Using the radial coordinate $R$ defined in \eqref{eq:rToR_coord}, the supersymmetry condition \eqref{eq:susy_cond_3Q}, an additional change of parameters \mbox{$w^2 = 1 + R_+^2 \bigl( 1- e^{4(\delta_1 + \delta_2)} \bigr)$} in order to express $R_+$ in terms of $w$, and some help from Mathematica, we find
\begin{equation}
	m = -\frac{ 2 e^{2 \delta_1+4 \delta_2} (1-w)^2 \left(e^{2 (\delta_1+\delta_2)}-w\right) }{ \bigl( e^{2 (\delta_1+\delta_2)}-1 \bigr)^2 \bigl( e^{2 (\delta_1+\delta_2)}+1 \bigr) \bigl( e^{4 (\delta_1+\delta_2)}-e^{4 \delta_2} (1-w)+e^{2 (\delta_1+\delta_2)} (1-w)-1 \bigr) } \;,
\end{equation}
which reproduces the first branch of \eqref{eq:branches 3 charges def}. The second branch is obtained by the other possible choice of $w$, \ie, by $w \to -w$.

In principle, we need to fix the sign of the $\int_{D3} d^4x \sqrt{-\det(g_\text{D3})}$ like in \eqref{DBI part of D3 action}. However, this proved to be quite technically challenging. Instead, let us initially consider both possible signs, one with a brane and one with an anti-brane, and determine the correct sign by comparing to the 1-charge case.

One can now move forward to compute the action of the brane, and find
\begin{equation}
	\begin{aligned}
		S_\text{D3} &= -\frac{N}{2\pi^2}\int_\text{D3} \biggl( d^4 x \, \pm \sqrt{-\det(g_\text{D3})} \, \pm P \bigl[ C_{(4)} \bigr] \biggr) \\
		&= \mp 4\pi N\left(\frac{i\sqrt{f_1(r_+)}}{2r_+^2H_1(r_+)} + \frac{m a}{r_+^2 H_1(r_+)} \bigl( c_1 s_2 s_3 - s_1 c_2 c_3 \bigr) \right) \\ 
		&= \mp 2\pi N \, \frac{\left(e^{4 \delta _2}+1\right) (w-1)}{e^{4 \left(\delta _1+\delta _2\right)}+e^{4 \delta _2} (w-1)-e^{2 \left(\delta _1+\delta _2\right)} (w-1)-1} \\ 
		&= \mp 2\pi N \, \frac{\Phi_1 - 1}{\Omega - 1} = \mp 2 \pi N \, \frac{\Delta_{\mrg,1}}{\tau_\mrg} \;.
	\end{aligned}
\end{equation}
This agrees with the general expectations described in the main text for a D3-brane and a specific choice of branch for the tension term.

A similar computation can be made for a D3-brane located along $\mu_2 = 0$, instead of $\mu_1 = 0$ (but still keeping the same two charges $Q_2 = Q_3$). The relevant term in $C_{(4)}$ is 
\begin{equation}
	C_{(4)} = \frac{1}{2} \, d\bigl( \mu_1^2 \bigr) \wedge \left(d\phi_1 + A^1\right)\wedge \left(d\phi_3 + A^3\right) \wedge A^2 + \text{other terms} \;,
\end{equation}
while the contribution of the induced metric is now (choosing a branch for the square root as before)
\begin{equation}
	\sqrt{-\det(g_\text{D3})} = -\frac{i\mu_1\sqrt{f_1(r_+)}}{r_+^2H_2(r_+)} \;.
\end{equation}
Using the same change of coordinates and some more help from Mathematica, one finds the action of this brane to be
\begin{equation}
	\begin{aligned}
		S_{D3} &= -\frac{N}{2\pi^2} \int_\text{D3} \biggl( d^4 x \, \sqrt{-\det(g_\text{D3})} \, - P\bigl[ C_{(4)} \bigr] \biggr) \\
		&= 4\pi N \left(\frac{i\sqrt{f_1(r_+)}}{2r_+^2H_2(r_+)} + \frac{m a}{r_+^2 H_2(r_+)} \bigl( c_2 s_1 s_3 - s_2 c_1 c_3 \bigr) \right) \\ 
		&= 2\pi N \, \frac{w-1}{e^{2 (\delta_1+\delta_2)}-1} \\ 
		&= 2\pi N \, \frac{\Phi_2 - 1}{\Omega - 1} = 2\pi N \, \frac{\Delta_{\mrg,2}}{\tau_\mrg} \;.
	\end{aligned}
\end{equation}
Both of these calculations agree with the one performed in Section~\ref{sec: Branes} for the case where the three charges are equal.

We have verified numerically that even when the three charges are different, in a supersymmetric setting the action of a brane like the ones discussed above with $\mu_I = 0$ has action $2\pi N \Delta_{\mrg,I} / \tau_\mrg$, as above. We have not been able to derive this result analytically.

One can also look at another kind of wrapped D3-brane --- one that wraps an $S^3 \subset \text{AdS}_5$ and an $S^1 \subset S^5$, as discussed for equal $U(1)$ charges in Section~\ref{sec:more branes}. For example, let us look at one located at $\mu_1 = 1$, $r = r_+$, and wrapping $\theta, \phi, \psi, \phi_1$. In order to get the second to last term in \eqref{eq:G5 def} and the last term in \eqref{eq:*G5 def}, $C_{(4)}$ must include
\be
C_{(4)} = \frac{1}{2} \, \mu_1^2 \wedge d\phi_1 \wedge \left( \frac{*_5 F_1}{X_1^2} - A_3 \wedge F_2\right) + \text{other terms} \;,
\ee
and the action of the brane (choosing a sign for the square root that reproduces the single-charge solution) is
\begin{equation}
	\begin{aligned}
		S_\mathrm{D3} &= -\frac{N}{2\pi^2} \int_\text{D3} \biggl( d^4 x \, \sqrt{-\det(g_\text{D3})} \, - P\bigl[ C_{(4)} \bigr] \biggr) \\
		&= -4\pi N \left(\frac{i}{2}\sqrt{f_1(r_+)} + m c_1 s_1  \right) \,
	\end{aligned}
\end{equation}
which, for the two charges case $\delta_2 = \delta_3$, results in
\begin{equation}
	\begin{aligned}
		S_\mathrm{D3} 	&= -2\pi N \, \frac{(w-1)^2}{\bigl( e^{2 \left( \delta _1+\delta _2\right)}-1 \bigr){}^2} \\ 
		&= -2\pi N \, \frac{(\Phi_2 - 1)(\Phi_3 - 1)}{(\Omega - 1)^2} = -2\pi N \, \frac{\Delta_{\mrg,2}\Delta_{\mrg,3}}{\tau_\mrg^2} \;.
	\end{aligned}
\end{equation}
This is consistent with (\ref{action new D3-branes equal ang mom}) on the first branch, when $\Delta_{\mg,2} = \Delta_{\mg,3}$.

Similarly for $\mu_3 = 1$ we find 
\be
C_{(4)} = \frac{1}{2} \, \mu_3^2 \wedge d\phi_3 \wedge \left( \frac{*_5 F_3}{X_3^2} - A_2 \wedge F_1\right) + \text{other terms} \;,
\ee
with the action 
\begin{equation}
	\begin{aligned}
		S_\mathrm{D3} &= -\frac{N}{2\pi^2} \int_\text{D3} \biggl( d^4 x \, \sqrt{-\det(g_\text{D3})} \, - P\bigl[ C_{(4)} \bigr] \biggr) \\
		&= -4\pi N \left(\frac{i}{2}\sqrt{f_1(r_+)} + m c_3 s_3  \right) \\ 
		&= -2\pi N \, \frac{ \bigl( e^{4 \delta _2}+1 \bigr) (w-1)^2}{\bigl( e^{2 (\delta _1+\delta _2 )}-1 \bigr) \bigl( e^{4 \left(\delta _1+\delta _2\right)}+e^{4 \delta _2} (w-1)-e^{2 \left(\delta _1+\delta _2\right)} (w-1)-1 \bigr)} \\ 
		&= -2\pi N \, \frac{(\Phi_1 - 1)(\Phi_2 - 1)}{(\Omega - 1)^2} = -2\pi N \, \frac{\Delta_{\mrg,1}\Delta_{\mrg,2}}{\tau_\mrg^2} \;,
	\end{aligned}
\end{equation}
where again we assumed $\delta_2 = \delta_3$ after the second line.

Taking into account the results in this section and in Section~\ref{sec:more branes}, we make the natural conjecture that, in the general case of three unequal charges and two unequal angular momenta, the on-shell action of a supersymmetric Euclidean D3-brane wrapping the $S^3$ horizon at $r=r_+$ along the AdS$_5$ coordinates and wrapping the $S^1$ given by $\mu_a = 1$ inside $S^5$, is
\be
\label{new D3-branes action general}
S_\text{D3} = - 2\pi N \, \frac{\Delta_{\mg,b} \, \Delta_{\mg,c}}{\sigma_\mg \, \tau_\mg} \qquad\text{(\first{} branch)} \;,\qquad
S_\text{D3} = 2\pi N \, \frac{\Delta_{\mg,b} \, \Delta_{\mg,c}}{\sigma_\mg \, \tau_\mg} \qquad\text{(\second{} branch)}
\ee
where $a,b,c$ are all different. We checked this claim numerically in various cases.

\section{Giant gravitons on black hole backgrounds}
\label{app: giant gravitons}

As an aside, we can use the method of Appendix~\ref{app: D3-branes} to exhibit new giant graviton and dual giant graviton solutions on Lorentzian black hole backgrounds. These are supersymmetric Lorentzian D3-brane embeddings in the background of the BPS (\ie, supersymmetric and extremal) black holes of minimal 5d gauged supergravity,%
\footnote{The supersymmetric embeddings we find are valid more generally in the complexified supersymmetric (not necessarily extremal) black hole backgrounds, and are smooth in the extremal limit.}
and they generalize the giant and dual giant graviton embeddings in pure AdS$_5 \times S^5$ \cite{McGreevy:2000cw, Grisaru:2000zn, Hashimoto:2000zp}. Giant and dual giant graviton embeddings in the near-horizon limit of black hole solutions were found in \cite{Sinha:2006sh, Sinha:2007ni}.

\paragraph{Giant gravitons.} Using the orthotoric coordinates (\ref{change to orthotoric coordinates}) in AdS$_5$, consider the embedding
\begin{align}
t &= \sigma^0 \;,\quad& \xi &= \text{const} \;,\quad& \eta &= \text{const} \;,\quad& \Phi &= \text{const} \;,\quad& \Psi &= \text{const} \;, \\
\rho_s &= \text{const} \;,\quad& \theta_s &= \sigma^1 \;,\quad& \varphi_s &= \sigma^2 \;,\quad& \zeta_s &= \sigma^3 \;,\quad& \psi_s &= \sigma^3 + (2\alpha - 3) \sigma^0 \;, \nn
\end{align}
where $\alpha$ is the gauge parameter appearing in the connection in (\ref{CCLP solution I}) and (\ref{gauge potential orthotoric}), while $\sigma^{0,1,2,3}$ are the worldvolume coordinates. This is a Lorentzian D3-brane, sitting at an arbitrary position with constant coordinates (outside or at the black hole horizon) in the spatial slices of AdS$_5$, while wrapping a round $S^3 \subset S^5$ of radius $\sin(\rho_s)$ and orbiting at the speed of light along an orthogonal circle inside $S^5$. Noticing that $f\geq 0$ outside or at the horizon, we compute
\be
\sqrt{-h} = \frac{f \sin(\theta_s)\, \sin^4(\rho_s)}{ 8} \qquad\text{and}\qquad \Theta = - \Gamma^{0789} + \cot(\rho_s) \Bigl( \Gamma^{6789} - \Gamma^{0678} \Bigr) \;.
\ee
Using the projectors (\ref{10d spinor projectors}) we obtain $\Theta \varepsilon = - i \varepsilon$, showing that the embedding is supersymmetric.

We stress that this result is valid for the full family of supersymmetric complex black hole solutions (\ref{CCLP solution I}) with SUSY constraint (\ref{SUSY condition q}), although real Lorentzian metrics that are free of pathologies also satisfy the extremality condition $\wt m=0$.

\paragraph{Dual giant gravitons.} We were able to find dual giant graviton solutions only in the case with equal angular momenta, \ie, with $a=b$. Using orthotoric coordinates (\ref{change to orthotoric coords a=b}) in AdS$_5$, consider the embedding
\begin{align}
t &= \sigma^0 \;,\quad& \wt\xi &= \text{const} \;,\quad& \eta &= \sigma^1 \;,\quad& \wt\Phi &= \sigma^2 \;,\quad& \wt\Psi &= \sigma^3 \;, \\
\rho_s &= \text{const} \;,\quad& \theta_s &= \text{const} \;,\quad& \varphi_s &= \text{const} \;,\quad& \zeta_s &= \text{const} \;,\quad& \psi_s &= (2\alpha - 3) \sigma^0 \;. \nn
\end{align}
This is a Lorentzian D3-brane, wrapping a spatial $S^3$ in AdS$_5$ around the black hole at arbitrary constant radius determined by $\wt\xi$, and orbiting at the speed of light along a maximal circle of $S^5$. For a given choice of sign of the square root, we compute
\be
\sqrt{-h} = \left( f \omega_{\wt\Phi} - \frac23 A_{\wt\Phi} \right) \wt\xi \;,\qquad\qquad \Theta = - \Gamma^{0349} + \sqrt{ \frac{\wt\xi \, \wt\cF}{ -h \, f}} \, \Bigl( \Gamma^{0234} - \Gamma^{2349} \Bigr) \;.
\ee
Using the projectors (\ref{10d spinor projectors}) we obtain $\Theta \varepsilon = - i \varepsilon$, showing that the embedding is supersymmetric. As before, this result is valid for the full family of supersymmetric (not necessarily extremal) complex black hole solutions, and in particular for extremal solutions characterized by $\wt m=0$. It would be interesting to generalize these supersymmetric embeddings to the case $a \neq b$.


\bibliographystyle{ytphys}
\baselineskip=0.97\baselineskip
\bibliography{BHEntropy}

\end{document}